\documentclass[aps,prb,twocolumn,groupedaddress]{revtex4-2}

\usepackage{epsfig,color}
\usepackage{graphicx}
\begin{document}

\title{Non-monotonic dependence of ${T_c}$ on the $c$ axis compression in the HTSC cuprate La$_{2-x}$Sr$_x$CuO$_4$}

\author{I.~A.~Makarov}
\email{maki@iph.krasn.ru}
\author{S.~G.~Ovchinnikov}%
 \email{sgo@iph.krasn.ru}
\affiliation{ Kirensky Institute of Physics, Federal Research Center KSC SB RAS, Akademgorodok 50, bld. 38, 660036 Krasnoyarsk, Russia}
%
%

\date{\today}

\begin{abstract}
The effect of the the $c$ axis compression on the electronic structure and superconducting properties of the HTSC cuprate La$_{2-x}$Sr$_x$CuO$_4$ at different doping is investigated. The electronic structure of quasiparticle excitations is obtained within the effective five-band Hubbard model using the equation of motion method for Green's functions builded on the Hubbard operators. The superconducting gap and ${T_c}$ are calculated taking into account the exchange pairing mechanism involving not only the Zhang-Rice singlet but also excited two-hole triplet and singlet states. The energy of the $a_{1g}$ orbitals increases with increasing compression and the $a_{1g}$ quasiparticle bands begin to strongly interact with the $b_{1g}$ bands at the top of the valence band. The reconstruction of the region of states determining the superconducting properties results in the high density of states near the Fermi level. This mechanism leads to an increase in $T_c$ in the underdoped region with increasing compression. The pairing constants renormalizations under compression results in ${T_c}$ decreasing. The competition of these two effects leads to non-monotonic behavior of $T_c$ under the $c$-axis compression near optimal doping. 
  
\end{abstract}

\maketitle


\section{\label{sec:Intro}Introduction}

In recent years, many unique compounds with high ${T_c}$ up to room temperature have been predicted and discovered. Two important steps to increase ${T_c}$ above the cuprates scale has been done in 2014 by synthesis H$_2$S and H$_3$S with ${T_c}=150$ and $200$ K at the megabar pressure~\cite{Drozdov2014,Drozdov2015,Einaga2016} and the same time by theoretical predictions~\cite{Duan2014,Duan2015}. Further theoretical calculations~\cite{Wang12,Liu2017,Peng2017,Semenok2022} predict that pressure can lead to high ${T_c}$s in hydrides LaH$_{10}$, YH$_{10}$, (La,Y)H$_{10}$ and CaH$_{10}$. In LaH$_{10}$ and (La,Y)H$_{10}$, ${T_c}$s of $250$ K and higher were actually found at pressure $150$ GPa~\cite{Drozdov2019,Somayazulu2019,Semenok2022}. Even higher potential ${T_c}$ ($473$ K) were predicted in calculations of the superconducting properties, crystal and electronic structure as a function of pressure in the ternary hydride Li$_2$MgH$_{16}$~\cite{Sun2019,Zhang2022}. It is seen that the enormous pressure is needed to bring crystal and electronic structure of hydrides into a form favorable for high ${T_c}$ superconductivity.

In most HTSC cuprates, the increase in ${T_c}$ occurs at much lower values of hydrostatic pressure~\cite{Takahashi1995}. The record ${T_c}=164$ and $166$ K among HTSC cuprates were found in HgBa$_2$Ca$_2$Cu$_3$O$_{8+{\delta}}$ and fluorinated one under quasihydrostatic pressure $P=31$~\cite{Gao1994} and $P=23$ GPa~\cite{Monteverde2005}, respectively. The effect of hydrostatic pressure in cuprates has been studied quite well. On the contrary, there are few experimental works devoted to the direct investigation of the influence of uniaxial pressure on ${T_c}$. The increase in ${T_c}$ under hydrostatic pressure and uniaxial pressure along the $a$ and $b$ axes is caused by the reduction of intraplane Cu-O distances which leads to an increase in copper-oxygen hopping integrals $t_{pd}$ and the superexchange pairing interaction constant $J \sim {{t_{pd}^2} \mathord{\left/
 {\vphantom {{t_{pd}^2} U}} \right.
 \kern-\nulldelimiterspace} U}$, where $U$ is the on-site Coulomb repulsion parameter. Existing experiments on the effect of uniaxial pressure along the $c$ axis on ${T_c}$ provide contradictory results although most of them indicate that the greater the distance between the CuO$_2$ plane and the apical oxygens, the greater ${T_c}$. Negative values of the derivative $dT_c/dP^{(c)}$ evidencing that ${T_c}$ increases with increasing distance between the apical oxygens and CuO$_2$ plane were obtained in experiments on uniaxial pressure in La$_{2-x}$Sr$_x$CuO$_4$ (LSCO)~\cite{Motoi1991}, from thermal expansion experiments in LSCO, HgBa$_2$CuO$_{4+{\delta}}$ near optimal doping~\cite{Hardy2010} and Bi$_2$Sr$_2$CaCu$_2$O$_{8+{\delta}}$ (Bi2212)~\cite{Watanabe94,Meingast96}. The authors of the resistivity measurements~\cite{Nakamura1999,Nakamura2000} and the synthesis of thin films LSCO under the Madelung strain~\cite{Butko2009} also came to the same conclusion about the direct correlation between the value of the lattice parameter $c$ and ${T_c}$ value. The effect of stress in Bi2212 and Bi2223 crystals~\cite{Chen91}, the high-resolution thermal expansion data in Y123 crystals~\cite{Meingast90,Meingast91,Welp92} demonstrate very small values $dT_c/dP^{(c)}$. In the other works~\cite{Crommie1989,Budko91,Wuhl91}, uniaxial pressure studies in Y124 showed the opposite behavior, namely, a positive derivative $dT_c/dP^{(c)}$~\cite{Meingast93}. In the work~\cite{Mito2012}, a pressure experiment on Y-124 single crystals via magnetic measurements show that ${T_c}$ increases under ${P\|c}$ contraction. All direct investigations~\cite{Motoi1991,Crommie1989} were performed at low pressure values (up to 0.03 GPa). The study of the region of high uniaxial pressures is not yet possible due to the fragility of the HTSC cuprates. Thermal expansion experiments provide definite derivatives $dT_c/dP^{(c)}$ assuming monotonic behavior regardless of absolute pressure values. In fact, the character of the pressure dependence of ${T_c}$ can change since different mechanisms can be activated in different pressure ranges. Therefore, it is of interest to study the superconducting properties of cuprates in the wide range of uniaxial compression up to those values of hydrostatic pressure at which record ${T_c}$ were reached.
   
The ambiguous behavior of ${T_c}$ under uniaxial pressure in the doping region near $x_{opt}$ which is usually considered in experimental studies is apparently due to the large number of different factors that are affected by pressure and the different contributions of each of these factors in a given compound. All these factors are usually divided into internal caused by changes in the crystal structure and those associated with changes in the number of carriers in the CuO$_2$ planes. In turn, internal factors includes the changes in the intraplane Cu-O distance, the distance between the CuO$_2$ plane and the apical oxygen, the number and types of distortions in the CuO$_2$ plane, and others. The change in the crystal structure can affect superconducting phase by two channels: the electronic structure reconstruction and the renormalization of a superconducting pairing constant which can be easily seen even in the example of the simplest estimate using the BCS theory ${k_B}{T_c} \propto {e^{ - {1 \mathord{\left/
 {\vphantom {1 {\left( {N\left( 0 \right)V} \right)}}} \right.
 \kern-\nulldelimiterspace} {\left( {N\left( 0 \right)V} \right)}}}}$, where ${N\left( 0 \right)}$ is the density of states (DOS) at the Fermi level, $V$ is an coupling constant. The influence of the pressure on superconducting pairing constant explains only the tendency of decreasing ${T_c}$ with increasing the $c$-axis compression $P^{(c)}$. The lattice constants $a$ and $b$ in the CuO$_2$ plane increase with increasing $P^{(c)}$ due to the Poisson effect. If we assume that superconducting pairing occurs through the superexchange interaction, the contraction of the in-plane lattice parameters leads to a decrease in the pairing interaction constant. This was shown in the framework of the t-J model for La$_{2-x}$Sr$_x$CuO$_4$~\cite{Sidorov2016}. 

The influence of the uniaxial pressure on ${T_c}$ through the reconstruction of the electronic structure is more complex, since the formation of electronic states involves states with different numbers of particles, orbital compositions, symmetry, and spins. Pressure may affect states with different values of these characteristics differently, so the resulting effect is not evident. As is known, the only hole in the d$^9$p$^6$ electronic configuration of copper-oxygen layers in undoped HTSC cuprates at normal pressure is located on the ${d_{{x^2} - {y^2}}}$-orbital of the copper atom and doped holes occupy the $p_x$-, $p_y$-orbitals of planar oxygen. Therefore, the electronic structure of low-energy excitations will be formed by orbitals of the $b_{1g}$ symmetry. Obviously, the octahedral symmetry of HTSC cuprates is restored under uniaxial pressure along the $c$ axis, the distance between the apical oxygen and the CuO$_2$ plane decreases, and the Cu-O distances in the CuO$_2$ plane increase. Changes in interatomic distances affect the energies of local electronic states and their orbital composition. The energy gap between $b_{1g}$ and $a_{1g}$ orbitals such as copper $d_z$- and the apical oxygen $p_z$-orbitals should also decrease. It is expected that contribution of a$_{1g}$ symmetry orbitals to the electronic structure of low-energy excitations also will grows. Superconducting pairing in p-type cuprates is realized under hole doping. Two-particle states begin to play an important role in doped compounds. Superconducting pairing at the normal pressure is realized with the participation of the Zhang-Rice singlets with holes in the ${d_{x^2-y^2}}$-orbital of the copper atom and the $p_x$-, $p_y$-orbitals of oxygen. The uniaxial pressure also can increase the contribution to low-energy excitations of two-particle states with symmetry, spin and orbital structure different from the Zhang-Rice singlet.

The present paper is devoted to the study of the effect of uniaxial pressure along the $c$ axis on the electronic structure and concentration dependence of ${T_c}$ in the HTSC cuprate La$_{2-x}$Sr$_x$CuO$_4$. To take into account the orbitals of $a_{1g}$ symmetry, a five-band p-d model is used. HTSC cuprates are the systems in which strong local Coulomb interaction and p-d hybridization significantly influence on structure of the local multiparticle states. We apply the generalized tight-binding (GTB) method~\cite{Ovchinnikov89,Gavrichkov00,Korshunov05,Kresin2021} to construct quasiparticle excitations forming electronic structure on the basis of local multiparticle states which are obtained exactly taking into account strong local interactions. The local multiparticle states and the effective multiband Hubbard model for quasiparticle excitations were obtained in the work~\cite{Makarov2025}. The results of the work~\cite{Makarov2025} revealed that the low-energy electronic structure in the compression range from $0$ to $14$ GPa will be formed by the five quasiparticle excitations. In this work, the electronic structure of quasiparticle excitations and the theory of superconductivity will be obtained within the five-band Hubbard model in the compression $P^{(c)}$ range from $0$ to $10$ GPa to exclude effects related to singlet-triplet crossovers. The role of both the pairing constant renormalization and the electronic structure reconstruction in the formation of superconducting properties under compression will be clarified. In experiments on uniaxial compression in cuprates, the region of low pressure values near the optimal doping was investigated. The monotonic decrease or increase of ${T_c}$ with pressure obtained in these experiments was considered as the main effect of uniaxial compression along the $c$ axis. However, the competition of different mechanisms of pressure influence on the superconducting phase can lead to a different character of the ${T_c}$ behavior in different intervals of pressure and doping. In this paper, it is shown that non-monotonic behavior of ${T_c}$ with pressure is possible near the optimal doping: the effect of uniaxial compression on ${T_c}$ can have a certain character at lower pressure (decrease in ${T_c}$), and the effect can change at higher pressure (increase in ${T_c}$), when the band structure reconstruction play a decisive role.

This paper includes seven sections. Section~\ref{sec:eff_Hub_model} contains the basis of quasiparticle excitations and a general description of the method for calculating the electronic structure and superconducting characteristics within the five-band Hubbard model. Section~\ref{sec:bs_without_pres} is devoted to the band structure of the five quasiparticle excitations in a layer of CuO$_6$ octahedra without pressure in order to understand the features of the electronic structure constructed taking into account the local excited triplet and singlet states, and not just the ground singlet Zhang-Rice state. In the Section~\ref{sec:bs_pres}, the effects of uniaxial pressure on the band structure, DOS, and Fermi contour of quasiparticle excitations are considered. In Section~\ref{sec:Tc_pres}, the concentration dependence of ${T_c}$ at different magnitudes of the $c$-axis compression are given within two-band and five-band Hubbard model. Section~\ref{sec:conclusion} presents a conclusion summarizing the main results. Appendix~\ref{app:model} includes the multiparticle CuO$_6$ cluster eigenstates and the Hamiltonian of the five-band Hubbard model. Appendix~\ref{app:eq_of_motion} provides a detailed description of the equation of motion method for Green's function to obtain the electronic structure and the theory of superconductivity for the five-band Hubbard model.

\section{\label{sec:eff_Hub_model} The method of calculation the electronic structure and superconducting gap}
\begin{figure*}
\includegraphics[width=0.45\linewidth]{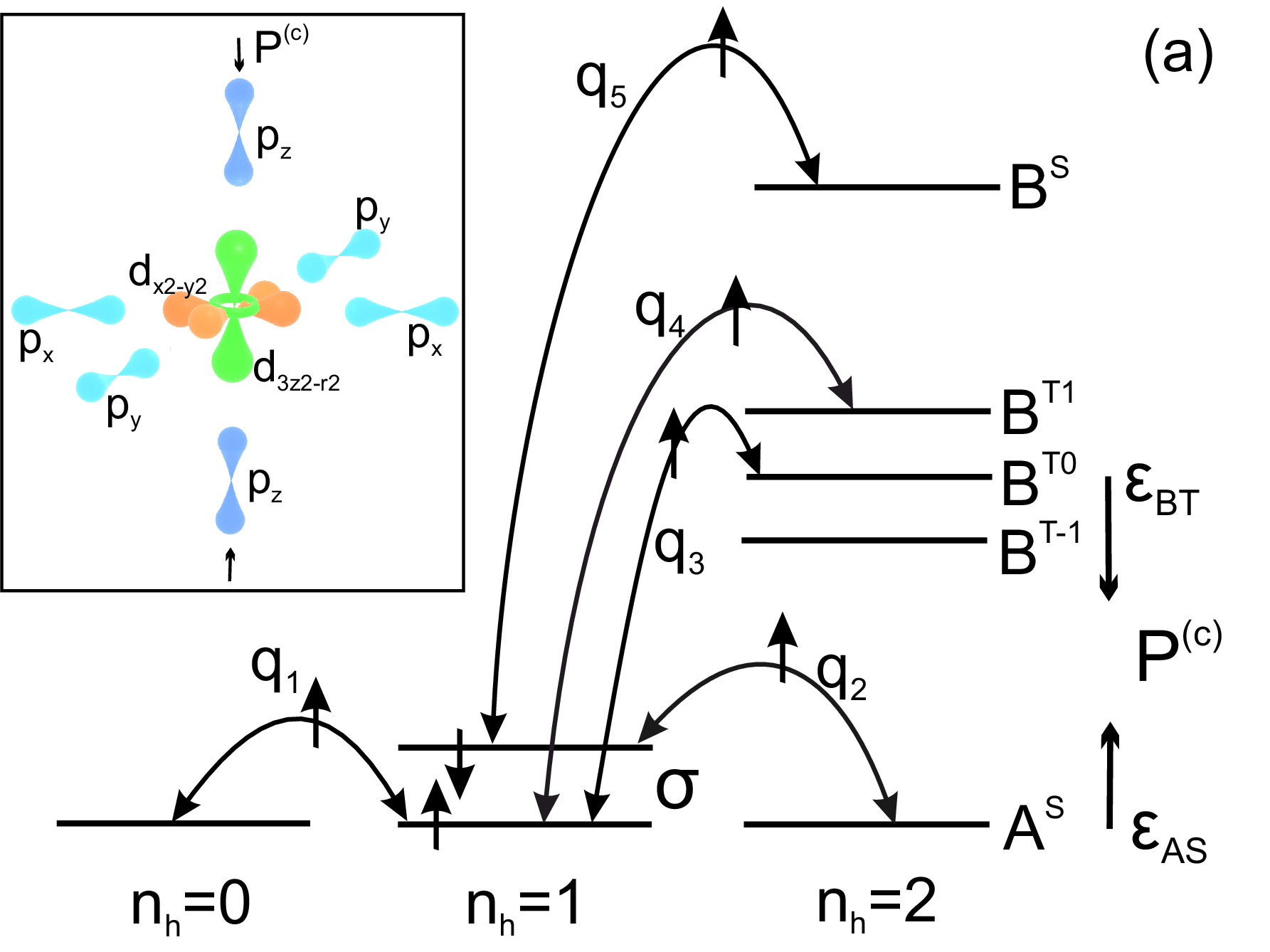}
\includegraphics[width=0.5\linewidth]{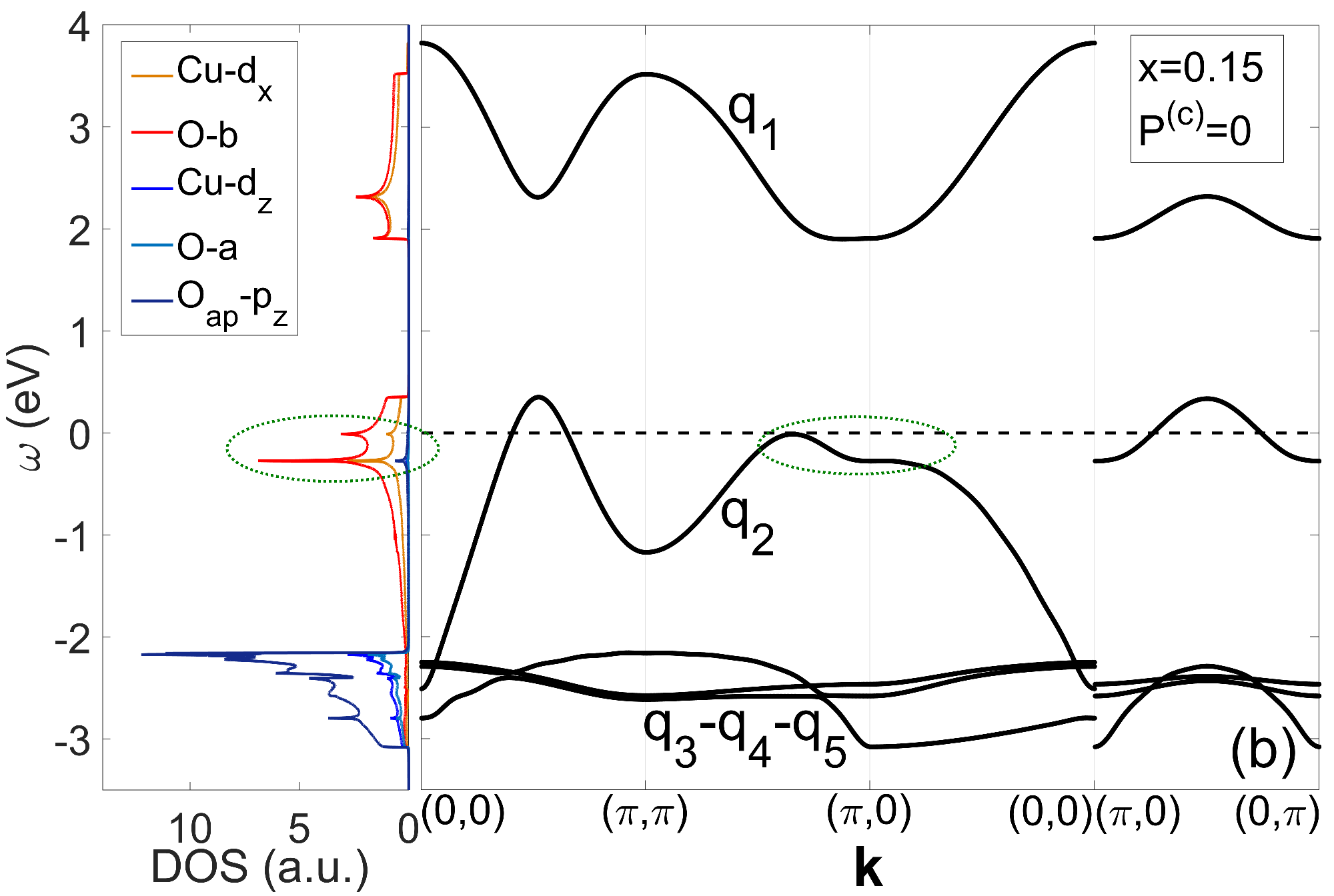}
\caption{\label{fig:quasiparticles} (Color online) (a) The general energy structure of the CuO$_6$ cluster eigenstates with hole number $n_h=0,1,2$ (black horizontal lines are the cluster eigenenergies) and the local quasiparticle excitations $q_1$-$q_5$ with spin projection $ \uparrow $ between them (curves with arrows at the ends) that form the conductivity band and the upper part of the valence band. The arrows with labels $\epsilon_{AS}$ and $\epsilon_{BT}$ indicate the increase and decrease of the corresponding cluster eigenenergies under the $c$-axis pressure $P^{(c)}$. In the inset, the CuO$_6$ cluster including 5 basis atomic orbitals is depicted. (b) The band structure of quasiparticle excitations $q_1$-$q_5$ without spectral weight and partial DOS of the five orbitals $\lambda  = {d_{x^2-y^2}}$, $b$, ${p_z}$, ${d_{3z^2-r^2}}$, $a$ ($b$ and $a$ are the molecular oxygen orbitals formed from $p_x$ and $p_y$ atomic orbitals), at $x=0.15$ and $P^{(c)}=0$~GPa. The dotted green ellipses indicate region of the states that are most actively involved in superconducting pairing and have the strongest effect on ${T_c}$.}
\end{figure*}
 
The effective five-band Hubbard model (Eq.(\ref{eq:H_Hub}) in Appendix~\ref{app:model}) to study electronic structure of La$_{2-x}$Sr$_x$CuO$_4$ was obtained using the GTB method on the basis of the five-band p-d model for the layer of CuO$_6$ octahedra~\cite{Makarov2025}. Each CuO$_6$ octahedra includes five orbitals (inset in Fig.~\ref{fig:quasiparticles}a): ${d_{x^2-y^2}}$-, ${d_{3z^2-r^2}}$-orbitals of copper atoms, $p_x$-, $p_y$-orbitals of planar oxygen atoms and $p_z$-orbitals of apical oxygen atoms. The p-d model Hamiltonian~\cite{Raimondi96,Makarov19} includes the on-site energies ${\varepsilon _{d_{x^2-y^2}}}$, ${\varepsilon _{d_{3z^2-r^2}}}$, ${\varepsilon _p}$, ${\varepsilon _{pz}}$, the $pd$ and $pp$ hopping integrals ${t_{dzpz}}$, ${t_{pd}}$, ${t_{pdz}}$, ${t_{pp}}$, ${t_{ppz}}$, and several types of Coulomb interactions: intraatomic intraorbital ${U_d}$, intraatomic interorbital ${V_d}$ and Hund exchange ${J_d}$ between electrons on ${d_x}$- and ${d_z}$-orbitals of Cu atoms, intraatomic intraorbital ${U_p}$ on O atoms, and interatomic interaction ${V_{pd}}$ between the electrons on copper and oxygen orbitals. The values of the on-site energies and hopping integrals without pressure were calculated in~\cite{Makarov19} using \textit{ab initio} LDA+GTB calculations, their dependencies on the magnitude of $c$-axis uniaxial pressure are obtained in the work~\cite{Makarov2025}. The Coulomb parameters values given in the work~\cite{Makarov2025} were calculated using the local density approximation in the works~\cite{Schluter88,Schluter89,Hybertsen89,Hybertsen90,Mahan90,Grant92}.

The basis of the five-band Hubbard model (Eq.~(\ref{eq:H_Hub})) consists of quasiparticle excitations between the eigenstates of the CuO$_6$ cluster with number of holes $n_h=0,1,2$ (horizontal black lines in Fig.~\ref{fig:quasiparticles}a):
\begin{eqnarray}\label{eq:eigenstates}
&&\left. 1 \right){n_h} = 0\,\,\,\,\,{\left| 0 \right\rangle } \\\nonumber
&&\left. 2 \right){n_h} = 1\,\,\,\,\,\left| {1,\sigma}  \right\rangle ,\,\sigma  =  \downarrow , \uparrow   \\\nonumber
&&\left. 3 \right){n_h} = 2\,\,\,\,\left| {2,L} \right\rangle, L=A^S, B^{T0} ,B^{T1}, B^{T - 1}, B^S .
\end{eqnarray} 
Here $\left| 0 \right\rangle$ is the vacuum hole state or, in electron representation, the completely filled state d$^{10}$p$^6$. $\left| {1,\sigma}  \right\rangle$ is the ground single-hole doublet which corresponds to a mixture of the electronic configurations d$^9$p$^6$ and d$^{10}$p$^5$, it includes the $b_{1g}$ symmetry orbitals ${d_{x^2-y^2}}$ and molecular oxygen orbital $b$ made of the atomic $p_x$- and $p_y$-orbitals. Each of two-hole states $\left| {2,L} \right\rangle $ corresponds to the mixture of the electronic configurations d$^8$p$^6$, d$^{10}$p$^4$, d$^9$p$^5$. The ground two-hole cluster eigenstate is the Zhang-Rice singlet $A^S$ (Eq.~(\ref{eq:A1_singlet})) formed by $b_{1g}$ symmetry orbitals. The first and second excited two-hole cluster eigenstates are the three components of the triplet $B^T$ (Eq.~(\ref{eq:triplet})) and the singlet $B^S$ (Eq.~(\ref{eq:singlet_B2s})) (Fig.~\ref{fig:quasiparticles}a), respectively, which are the mixture of the products of the ${b_{1g}}$ and ${a_{1g}}$ (${d_{3z^2-r^2}}$-, $p_z$- and the molecular oxygen orbital $a$ made of the atomic $p_x$- and $p_y$-orbitals) symmetry orbitals. The CuO$_6$ cluster eigenenergies ${{\varepsilon _m}}$ and their dependence on compression are determined in work~\cite{Makarov2025}, where it is shown that the increase in uniaxial pressure along the $c$ axis results in the increase in the energy of the two-hole singlet $A^S$, while the energies of the triplet $B^T$ and the singlet $B^S$ decrease (illustrated by the vertical arrows with labels $\epsilon_{AS}$ and $\epsilon_{BT}$ in Fig.~\ref{fig:quasiparticles}a). The crossover of the singlet $A^S$ and the triplet $B^T$ occurs at a value of $P_{c1}^{\left( c \right)} \approx 11.8$ GPa, and the crossover of $B^T$ and $B^S$ occurs at $P_{c2}^{\left( c \right)} = 13.4$ GPa. The energy of the single-hole doublet of $b_{1g}$ symmetry increases with pressure increasing, but the crossover with the $a_{1g}$ symmetry doublet occurs at much more higher pressure.

The electronic structure is formed by the quasiparticle excitations with spin projection $ \sigma $ (are depicted by the curves with arrows in Fig.~\ref{fig:quasiparticles}a): 
\begin{eqnarray}\label{eq:q1_q5}
&&{q_1} \equiv \left( {0 \sigma } \right), {q_2} \equiv \left( { {\bar \sigma } A^S} \right), {q_3} \equiv \left( { {\bar \sigma } B^{T0}} \right),  \\\nonumber
&&{q_4} \equiv \left( { \sigma {B^{T2\sigma }}} \right), {q_5} \equiv \left( { {\bar \sigma } B^S} \right),
\end{eqnarray}
The dispersion of quasiparticle excitations is determined by the hopping integrals $t_{\bf{fg}}\left( {{q_i} ,{q_j}} \right)$ with quasiparticle indices $i,j=1-5$ (Eq.~(\ref{eq:X_hoppings})). The band structure and superconducting gap are calculated using the equations of motion for Green's functions method (Appendix~\ref{app:eq_of_motion}). The system of equations (\ref{eq:eq_motion}) for all components of the matrix Green's function built on Hubbard operators $X_{\bf{f}}^{pq} = \left| m \right\rangle \left\langle n \right|$ which describe quasiparticle excitations between the cluster eigenstates $\left| m \right\rangle$ and $\left| n \right\rangle$ is decoupled using the Mori-Zwanzig projection technique~\cite{Plakida2003,Dzebisashvili05} (Appendix~\ref{app:eq_of_motion}). It is assumed that superconducting pairing between quasiparticles occurs due to the superexchange mechanism, and the superconducting gap has d-wave symmetry. We will take into account only the hoppings between nearest clusters $t_{10}\left( {{q_1} ,{q_j}} \right)$ when forming the superexchange interaction. The superexchange mechanism of pairing for the five-band Hubbard model is taken into account in the same way (Appendix~\ref{app:eq_of_motion}) as was used in~\cite{Plakida2003} for the two-band Hubbard model. Compared to the case of the two-band model, the superconducting gap within the five-band Hubbard model has the components associated with the pairing of quasiparticles formed with the participation of excited two-hole states, triplet $B^T$ and singlet $B^S$. The superconducting gap components $\Delta _{0d}^{ij}$ related to pairing between quasiparticles $q_i$ and $q_j$ depends on the effective superexchange interaction between all quasiparticles $J\left( {{q_i},{q_j},{q_{j'}},{q_{j''}}}\right)$ (Eq.~(\ref{eq:J_pm})) and the anomalous averages  ${V_{{\bf{rs}}}}\left( {\sigma L,\sigma 'L'} \right) = \left\langle {X_{\bf{r}}^{\sigma L}X_{\bf{s}}^{\sigma 'L'}} \right\rangle $ (Eqs.~(\ref{eq:d11_q})-(\ref{eq:d21_q})). The anomalous averages ${V_{{\bf{rs}}}}$ characterize the features of the electronic structure of quasiparticle excitations and superconducting gap itself. The system of the equations for all superconducting gap components (\ref{eq:d11_q})-(\ref{eq:d21_q}) is solved self-consistently to find the amplitudes ${\Delta _{0d}^{ij}}$. The superconducting transition temperature ${T_c}$ is determined as the value at which all amplitudes turn to zero. 

\section{\label{sec:bs_without_pres} The band structure of quasiparticle excitations within the five-band Hubbard model without pressure}
\begin{figure*}
\includegraphics[width=0.45\linewidth]{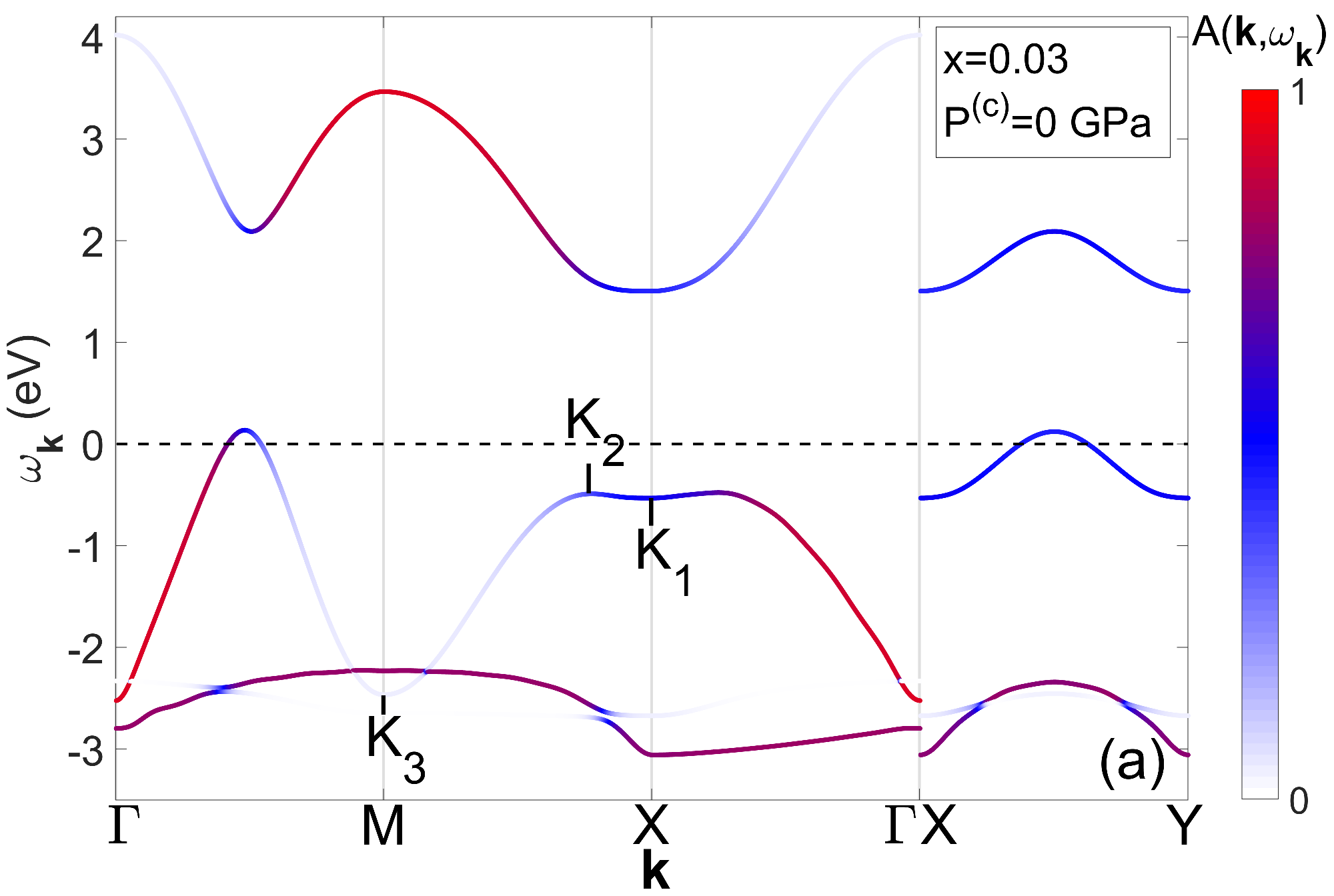}
\includegraphics[width=0.45\linewidth]{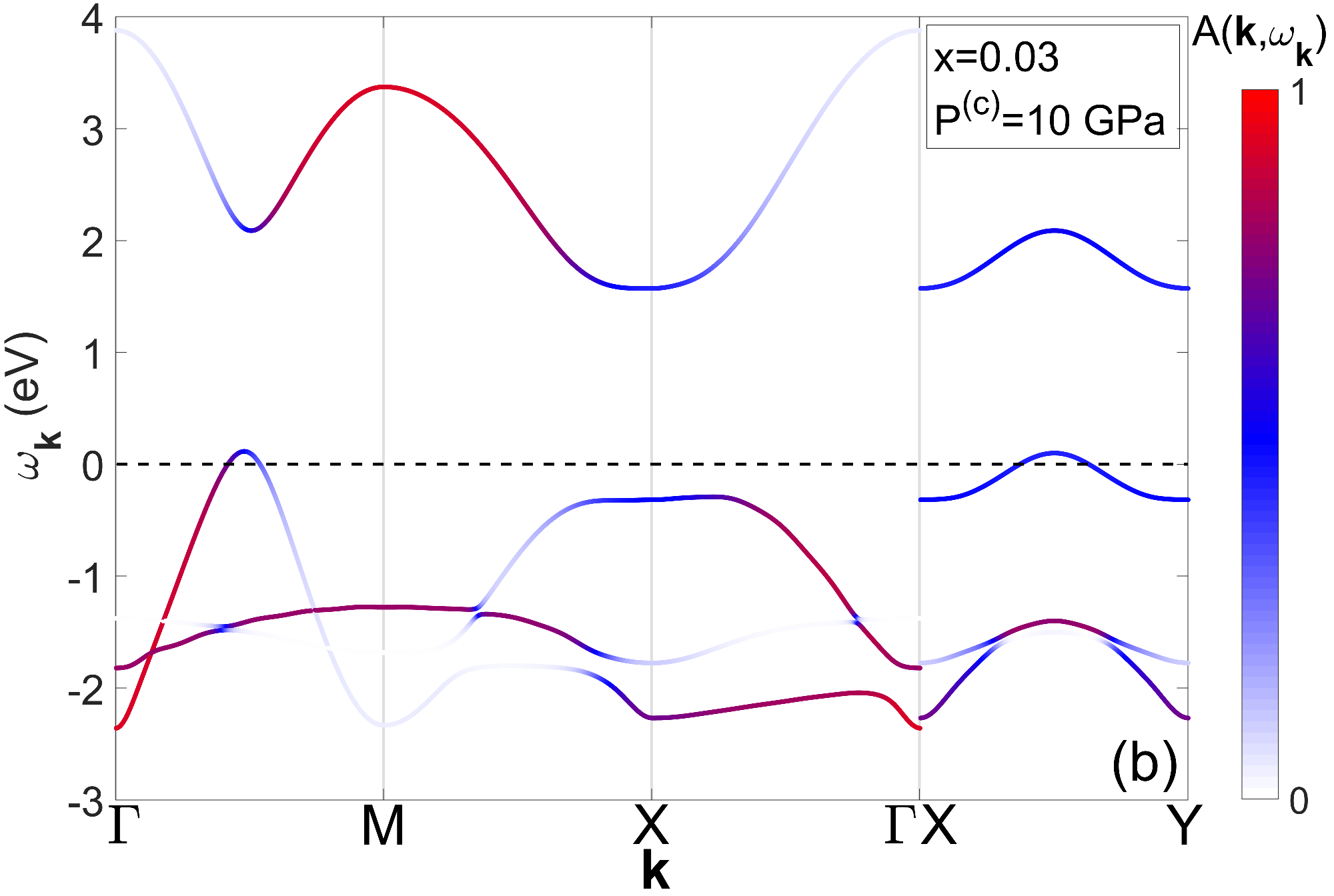}
\includegraphics[width=0.45\linewidth]{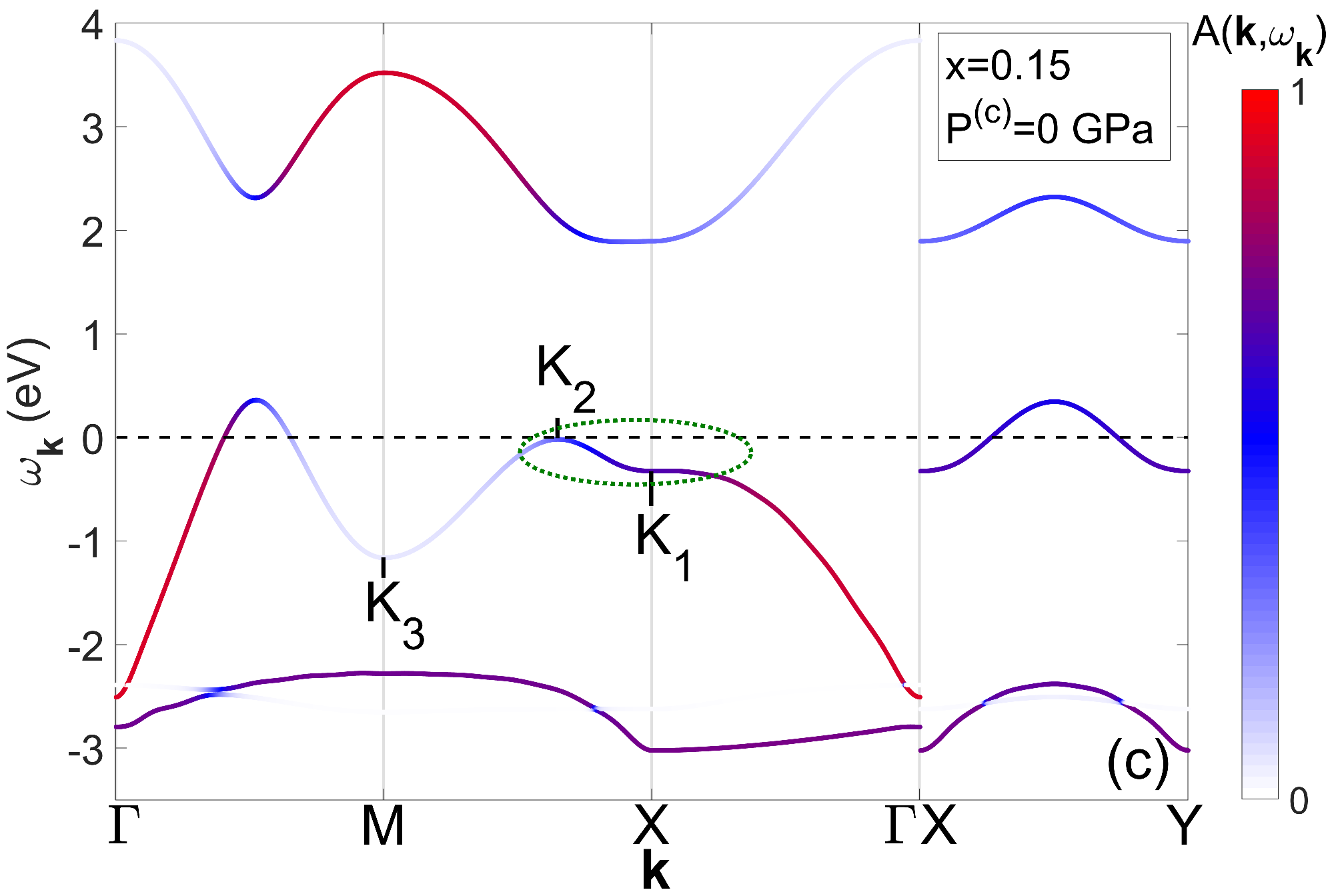}
\includegraphics[width=0.45\linewidth]{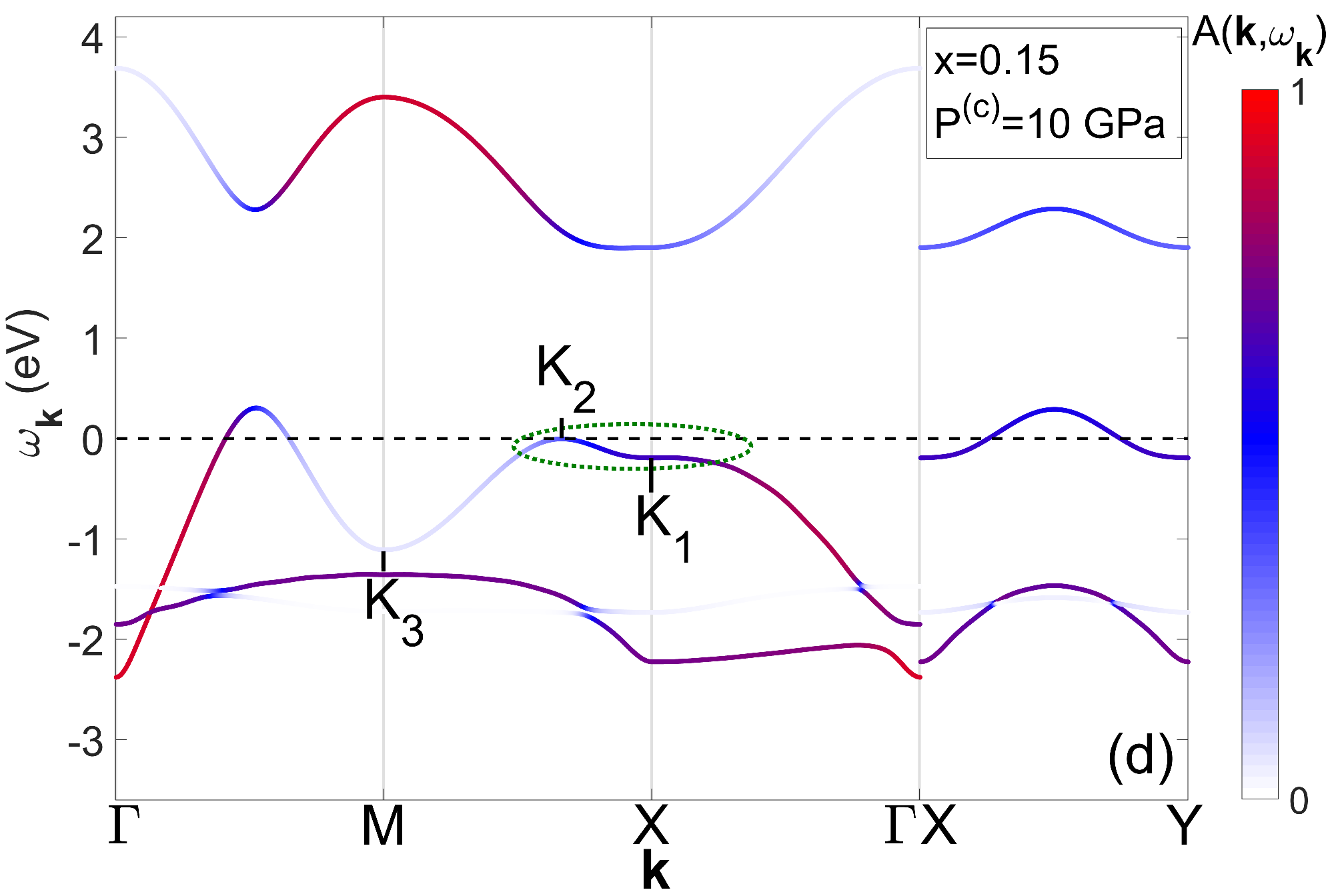}
\includegraphics[width=0.45\linewidth]{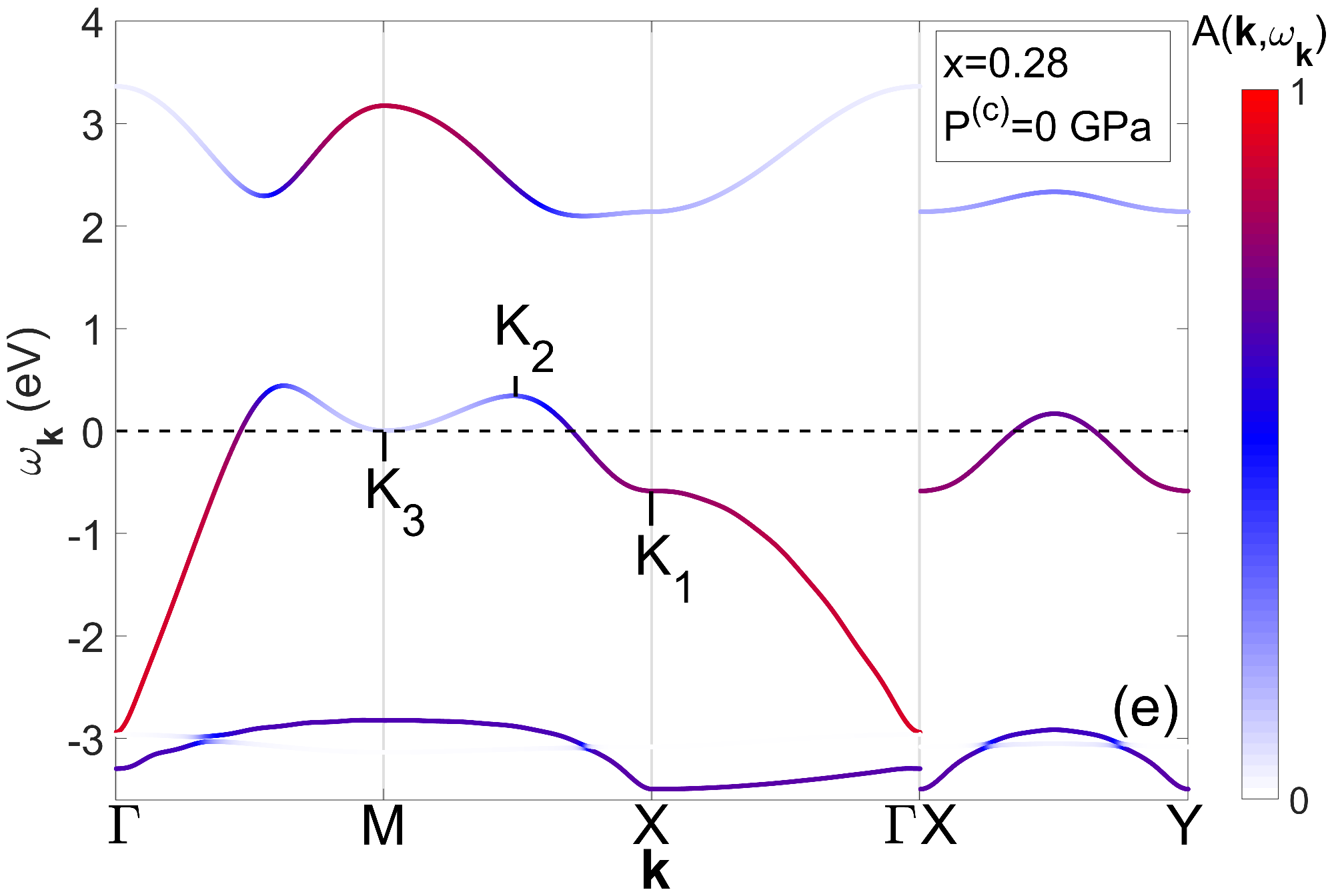}
\includegraphics[width=0.45\linewidth]{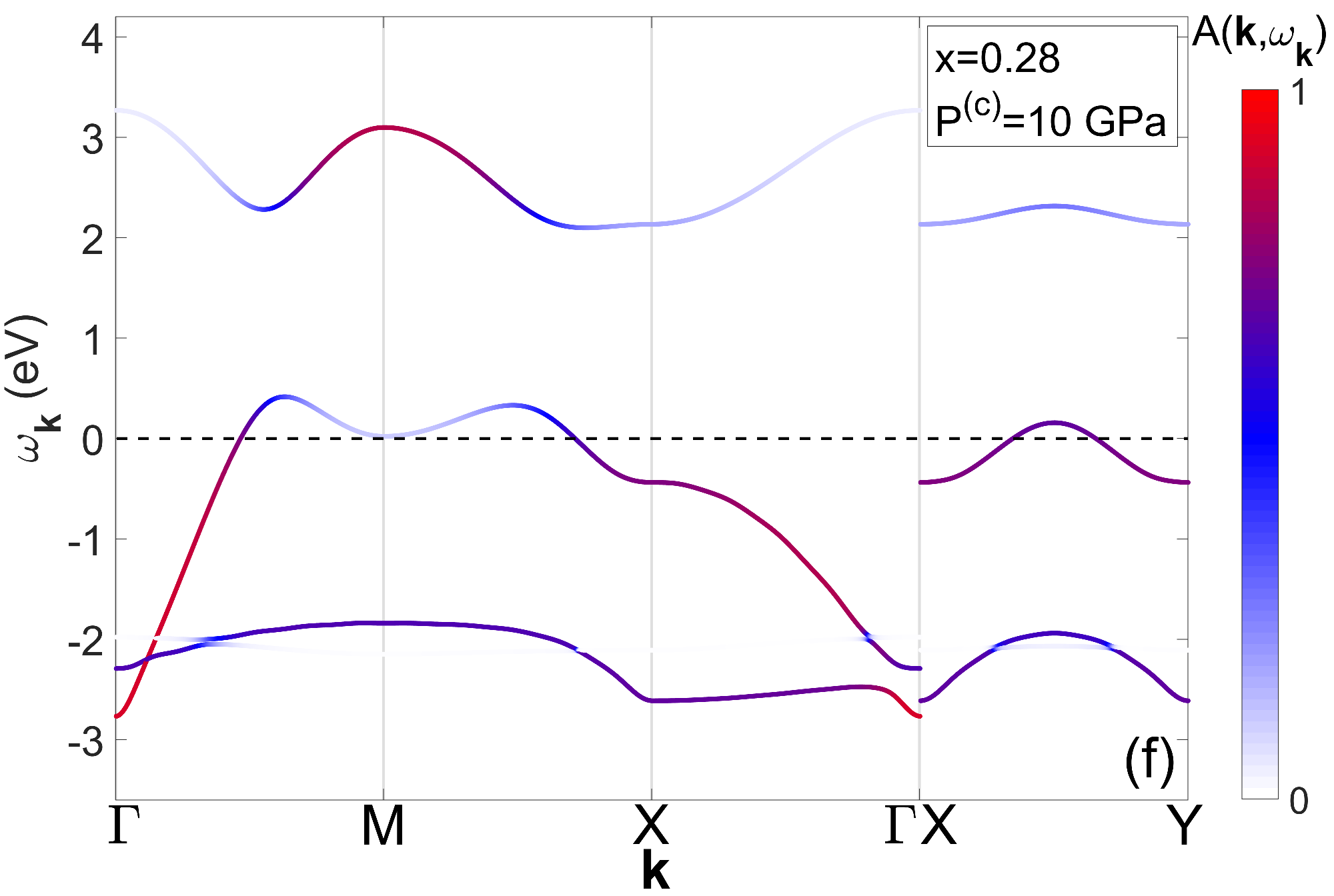}
\caption{\label{fig:bsx} (Color online) The reconstruction of the band structure of quasiparticle excitations under doping $x$ and under uniaxial compression ${P^{(c)}}$. The color of the dots shows the total $A\left( {{\bf{k}},{\omega _{\bf{k}}}} \right)$ spectral density of states with wave vector ${\bf{k}}$ and energy $\omega_{\bf{k}}$. The correspondence between the color and the magnitude of the spectral density is shown on the gradient scale. The dotted green ellipses indicate region of the states that are most actively involved in superconducting pairing and have the strongest effect on ${T_c}$. The following notations are used for k-points: $\Gamma = \left( {0 ,0 } \right)$, ${\rm X}=\left( {\pi ,0} \right)$, ${\rm Y} = \left( {0 ,\pi } \right)$, ${\rm M} = \left( {\pi ,\pi } \right)$.}
\end{figure*}
\begin{figure*}
\includegraphics[width=0.8\linewidth]{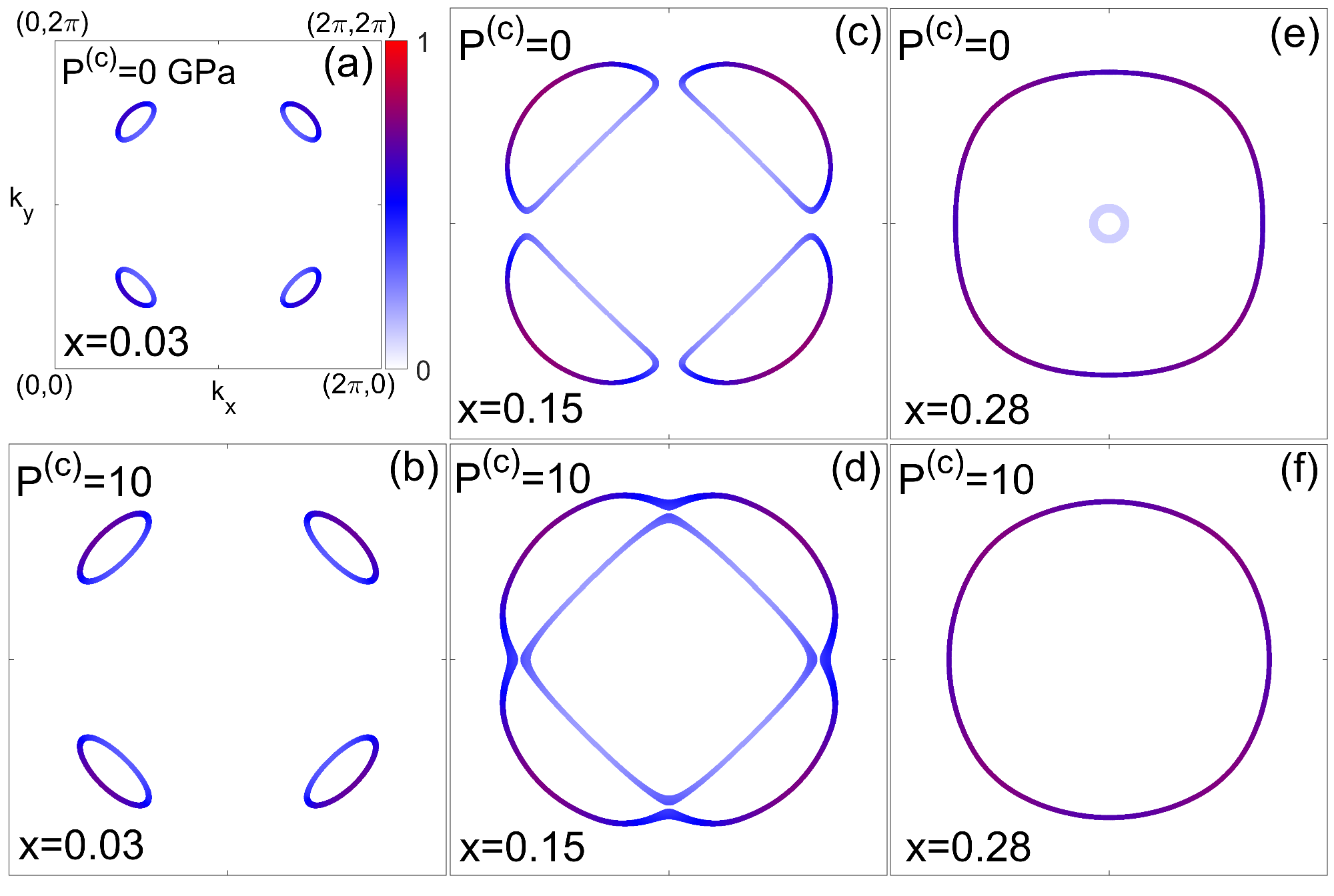}
\caption{\label{fig:FS} (Color online) The reconstruction of the Fermi contour under doping $x$ and under uniaxial compression ${P^{(c)}}$.}
\end{figure*}
\begin{figure}
\includegraphics[width=0.95\linewidth]{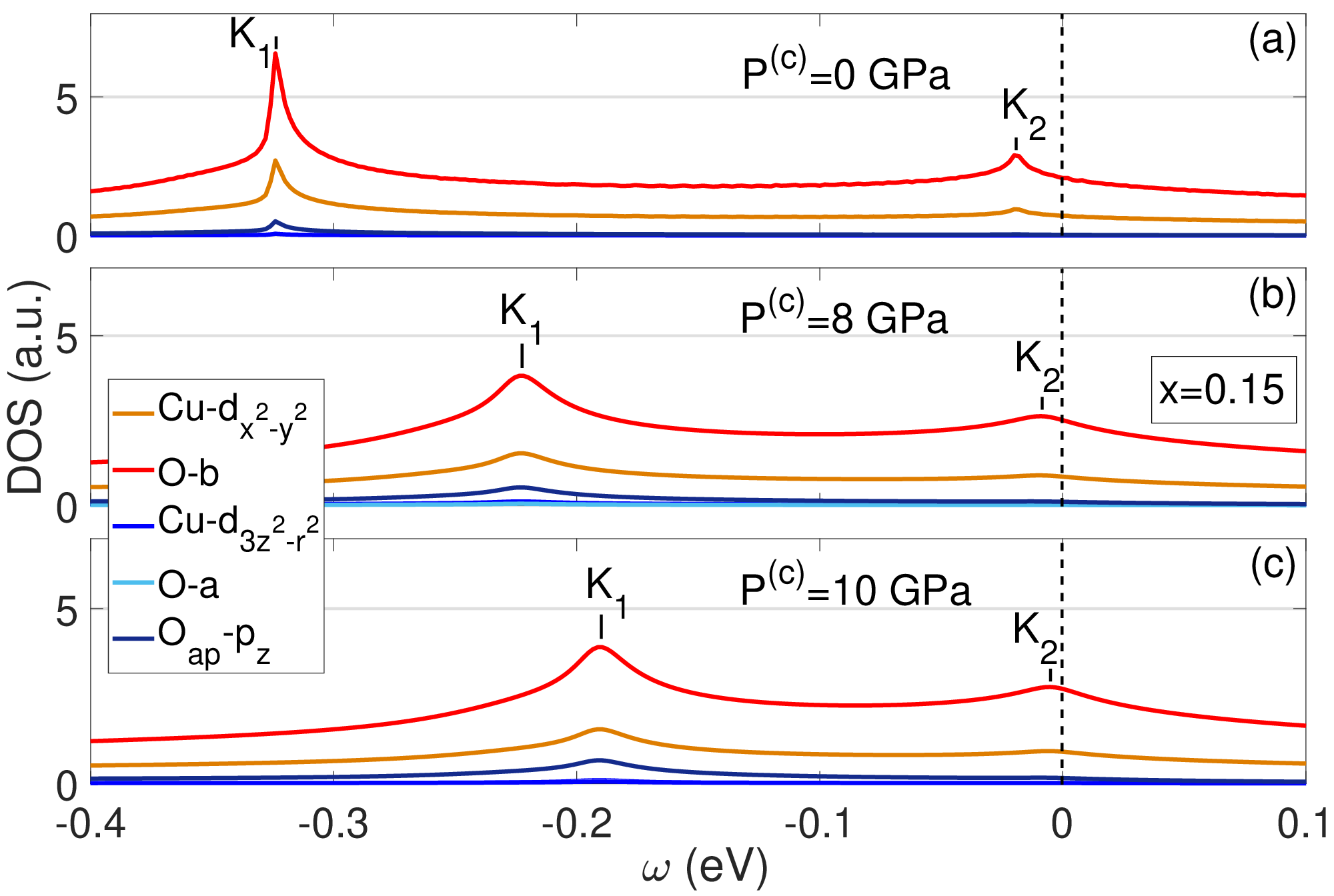}
\caption{\label{fig:DOS} (Color online) Partial DOS for the five orbitals $\lambda  = {d_{x^2-y^2}}$, $b$, ${p_z}$, ${d_{3z^2-r^2}}$, $a$ near Fermi level (dotted line) at different values of the $c$-axis compression ${P^{(c)}}$. Each DOS is calculated at optimal doping ${x_{opt}}$ for a certain compression. The approaching of two peaks corresponding to the states $K_1$ and $K_2$ in the band structure are clearly visible with increasing ${P^{(c)}}$.}
\end{figure}
The band structure within the tight-binding model which is commonly used to interpret the electronic structure of HTSC cuprates in ARPES experiments consists of the wide band with maximum at k-point $M = \left( {\pi ,\pi } \right)$ and with the shoulders at the points ${\rm X}$, ${\tilde {\rm X}}$, ${\rm Y}$, ${\tilde {\rm Y}}$ (${\rm X}=\left( {\pi ,0} \right)$, ${\rm Y} = \left( {0 ,\pi } \right)$, ${\tilde {\rm X}}=\left( {2\pi ,\pi} \right)$, ${\tilde {\rm Y}}=\left( {\pi ,2\pi} \right)$) which lead to the first van Hove singularity in the density of states (DOS) (high-intensity peak inside dotted green ellipse in the left panel of Fig.~\ref{fig:quasiparticles}b). The tight-binding model does not take into account such important features of HTSC cuprates as SEC and magnetic correlations and does not describe the appearance of the band gap. In GTB approach, the single wide band splits into UHB and LHB which correspond to the $q_1$ and $q_2$ subbands, respectively, as a result of taking into account the Coulomb interaction. The quasiparticle ${q_1}$ (Figs.~\ref{fig:quasiparticles}a,b) forms the conductivity band that is in the range from $2$ to $4$ eV (Fig.~\ref{fig:quasiparticles}b, Figs.~\ref{fig:bsx}a,c,e). The band ${q_2}$ lies between $-2.5$ and $0.5$ eV at $P^{\left( c \right)} = 0$ GPa (Figs.~\ref{fig:quasiparticles}a,b, Figs.~\ref{fig:bsx}a,c,e). When we take into account short-range spin-spin correlations, the LHB is reconstructed in such a way that the energy at the point ${\rm M}$ is significantly reduced forming a local minimum $K_3$. Also, a local maximum $K_2$ arises in each of the directions ${\rm X}$-${\rm M}$ (Figs.~\ref{fig:bsx}a,c,e), ${\tilde {\rm X}}$-${\rm M}$, ${\rm Y}$-${\rm M}$, ${\tilde {\rm Y}}$-${\rm M}$ under such a reconstruction. Presence of these maxima results in the appearance of the second van Hove singularity on DOS (low-intensity peak inside dotted green ellipse in the left panel of Fig.~\ref{fig:quasiparticles}b). In underdoped and optimally doped compounds, the states of the band $q_2$ near the k-points ${\rm X}$, ${\tilde {\rm X}}$, ${\rm Y}$, ${\tilde {\rm Y}}$ (states $K_1$) together with the states along directions ${\rm X}$-${\rm M}$, ${\tilde {\rm X}}$-${\rm M}$, ${\rm Y}$-${\rm M}$, ${\tilde {\rm Y}}$-${\rm M}$ up to local maxima $K_2$ form regions (states inside green ellipse in the Fig.~\ref{fig:quasiparticles}b and Figs.~\ref{fig:bsx}c,d) that are responsible for the shape of the superconducting dome, its position and maximal ${T_c}$.

In the effective five-band Hubbard model we use, the LHB (valence band) consists of four bands of the quasiparticles ${q_2}$, ${q_3}$, ${q_4}$, and ${q_5}$ (Fig.~\ref{fig:quasiparticles}b). All quasiparticles contribute to each band but the predominant contribution of a certain quasiparticle can be distinguished for some bands. The band is designated by the quasiparticle name in the sense of the predominant contribution of this quasiparticle to the corresponding band. The maximum of the valence band is at point $\left( {{\pi  \mathord{\left/
 {\vphantom {\pi  2}} \right.
 \kern-\nulldelimiterspace} 2},{\pi  \mathord{\left/
 {\vphantom {\pi  2}} \right.
 \kern-\nulldelimiterspace} 2}} \right)$. The Fermi level crosses the band ${q_2}$ (Figs.~\ref{fig:bsx}a,c,e). The band ${q_2}$ (and also ${q_1}$) is formed predominantly by the $b_{1g}$ symmetry orbitals ${{d_{x^2-y^2}}}$ and ${b}$ (Fig.~\ref{fig:quasiparticles}b, left panel, orange and red partial DOS), the contribution of the $a_{1g}$ symmetry orbitals ${d_{3z^2-r^2}}$, ${a}$, ${p_z}$ is much smaller although the contribution of the ${p_z}$-orbital is noticeable (Fig.~\ref{fig:quasiparticles}b, left panel, dark blue line). The bands of quasiparticles ${q_3}$, ${q_4}$, ${q_5}$ are between $-3.5$ and $-2$ eV at $P^{\left( c \right)} = 0$ GPa (Fig.~\ref{fig:quasiparticles}b, Figs.~\ref{fig:bsx}a,c,e) and thus have a weak effect on the Fermi contour without pressure. The ${q_3}$-${q_4}$-${q_5}$ bands are formed mainly by the $a_{1g}$ symmetry orbitals. The largest contribution to the spectral weight is given by the ${p_z}$-orbital, the contributions of the copper ${d_z}$-orbital and molecular ${a}$-orbital of the planar oxygens are much smaller (Fig.~\ref{fig:quasiparticles}b, left panel, dark blue line). The hybridization of ${q_2}$ and ${q_3}$-${q_4}$-${q_5}$ bands leads to reconstruction of states around $K_1$ and $K_2$ points in comparing with dispersion of the two-band Hubbard model~\cite{Makarov19}: the states of the flat region near $K_1$ rises closer to the chemical potential, and the local maximum $K_2$ becomes wider.
 
 The dispersion and the Fermi contour have different forms for different doping levels $x$. The energy of the states $K_3$ increases with increasing doping (Figs.~\ref{fig:bsx}a,c,e). Another characteristic change is that the maximum $K_2$ becomes more pronounced with higher energy (Figs.~\ref{fig:bsx}a,c,e) when $x$ increases. The Fermi level falls deeper into the valence band as hole doping increases (Figs.~\ref{fig:bsx}a,c,e). The transformation of the Fermi contour under doping is determined by two factors: a dispersion renormalization and a change in the Fermi level position. The shape of the Fermi contour obtained here within the five-band Hubbard model repeats the shape obtained within the standard two-band Hubbard model~\cite{Makarov19}. The Fermi contour at $x = 0.1$ consists of four hole pockets around the points $\left( {{\pi  \mathord{\left/
 {\vphantom {\pi  2}} \right.
 \kern-\nulldelimiterspace} 2},{\pi  \mathord{\left/
 {\vphantom {\pi  2}} \right.
 \kern-\nulldelimiterspace} 2}} \right)$, $\left( {{{3\pi } \mathord{\left/
 {\vphantom {{3\pi } 2}} \right.
 \kern-\nulldelimiterspace} 2},{\pi  \mathord{\left/
 {\vphantom {\pi  2}} \right.
 \kern-\nulldelimiterspace} 2}} \right)$, $\left( {{\pi  \mathord{\left/
 {\vphantom {\pi  2}} \right.
 \kern-\nulldelimiterspace} 2},{{3\pi } \mathord{\left/
 {\vphantom {{3\pi } 2}} \right.
 \kern-\nulldelimiterspace} 2}} \right)$, $\left( {{{3\pi } \mathord{\left/
 {\vphantom {{3\pi } 2}} \right.
 \kern-\nulldelimiterspace} 2},{{3\pi } \mathord{\left/
 {\vphantom {{3\pi } 2}} \right.
 \kern-\nulldelimiterspace} 2}} \right)$ (Fig.~\ref{fig:FS}a). It is seen from Figs.~\ref{fig:FS}a,c,e that the spectral weight is inhomogeneously distributed along the Fermi contour. The inhomogeneous distribution of the spectral weight is caused by interband hoppings which are directly taken into account in the Hubbard model unlike the t-J model~\cite{Makarov19}. The states on one side of the hole pocket have much greater intensity than on the other (Figs.~\ref{fig:FS}a,c). The weak spectral weight of states on one side of the pockets is a possible reason why Fermi arcs rather than pockets are observed in ARPES experiments. Cluster extension of our LDA+GTB approach that takes into account short range magnetic order provides the correct picture with doping and temperature dependent Fermi arcs~\cite{Kuzmin2020}. For the thermodynamical properties discussed in this paper there is no necessity to take into account fluctuations of the magnetic short range order. The Fermi contour at $x = 0.15$ represents four almost connected hole pockets (Fig.~\ref{fig:FS}c). The Fermi contour at $x = 0.28$ consists of the large outer contour and the smaller inner contour around the point ${\rm M}$ (Fig.~\ref{fig:FS}e) which are formed by the merging of the four hole pockets.

The position and dispersion of the bands ${q_2}$-${q_3}$-${q_4}$ for different hole concentrations differ slightly, but no fundamental reconstruction occurs. Only the states of one of these three bands have a significant spectral weight at $P^{\left( c \right)} = 0$ GPa, as can be seen from Figs.~\ref{fig:bsx}a,c,e. The predominant contribution to this band is made by the quasiparticle ${q_4}$; the hybridization with the quasiparticles ${q_3}$ and ${q_5}$ changes its dispersion although the general shape is preserved.
 
\section{\label{sec:bs_pres} Effect of c-axis compression on the electronic structure}

The energy of hole cluster eigenstates having the character of the ${a_{1g}}$ orbitals decreases and the energy of eigenstates formed by the ${b_{1g}}$ orbitals increases under the $c$-axis compression. Therefore, the electron energy of the quasiparticles ${q_3}$, ${q_4}$, ${q_5}$ increases relative to the quasiparticle ${q_2}$ energy under the $c$-axis compression, and the ${q_3}$-${q_4}$-${q_5}$ band rises to the top of the valence band (Figs.~\ref{fig:bsx}a-f). As a result, the hybridization between the quasiparticles ${q_3}$, ${q_4}$, ${q_5}$ and ${q_2}$ becomes stronger. The increase in hybridization leads to the reconstruction of the bands ${q_2}$ and ${q_3}$, ${q_4}$, ${q_5}$ and to the redistribution of the spectral weight between them. The most significant and important for superconducting properties reconstruction occurs in the region near and between the $K_1$ and $K_2$ states marked by the dotted green ellipse in Figs.~\ref{fig:bsx}c,d). The energy of states near the $K_1$ points increases under compression, the dispersion of the whole region of states between and near $K_1$ and $K_2$ points becomes weaker, the wide maximum forms around the points ${\rm X}$, ${\tilde {\rm X}}$, ${\rm Y}$, ${\tilde {\rm Y}}$ (dotted green ellipses in Fig.~\ref{fig:bsx}d). This effect is also appeared as the approaching of two DOS peaks related to the states near $K_1$ and $K_2$ points with increasing pressure: the high-intensity DOS peak $K_1$ in the range from $-0.35$ to $-0.3$ eV at ${P^{\left( c \right)}}=0$ GPa shifts to the range from $-0.2$ to $-0.18$ eV at ${P^{\left( c \right)}}=10$ GPa (Figs.~\ref{fig:DOS}a-c). The height of the high-intensity peak $K_1$ and low-intensity peak $K_2$ in the partial DOS for ${d_x}$-, ${b}$-, ${p_z}$-orbitals as well as the DOS between them increase under $c$-axis compression up to ${P^{\left( c \right)}}=10$ GPa (Figs.~\ref{fig:DOS}a-c). It is clear from Figs.~\ref{fig:DOS}a-c that the states below the Fermi level at a depth of $0.3$ eV are collected in a narrower energy region at a depth of $0.2$ eV with increasing pressure. Thus, a large number of states actively participating in superconducting pairing will arise. The effect of the band $q_2$ reconstruction is most pronounced in the underdoped and optimally doped compounds (Figs.~\ref{fig:bsx}a-d). In the overdoped compounds, this reconstruction is weakly expressed (Figs.~\ref{fig:bsx}e,f) since the merging of the DOS peaks related to the states $K_1$ and $K_2$ under compression does not occur due to the large energy gap between them.

The another region reconstructed under pressure is related to the bands ${q_3}$-${q_4}$-${q_5}$ in the ${\rm X}$-${\rm M}$ (Figs.~\ref{fig:bsx}a,b), ${\tilde {\rm X}}$-${\rm M}$, ${\rm Y}$-${\rm M}$, ${\tilde {\rm Y}}$-${\rm M}$ directions. At $x=0.03$, the splittings are more pronounced between the ${q_2}$ and ${q_3}$-${q_4}$-${q_5}$ bands and within ${q_3}$-${q_4}$-${q_5}$ bands, and also the spectral weight is redistributed between them (Fig.~\ref{fig:bsx}b). This effect is much weaker at $x=0.15$ (Fig.~\ref{fig:bsx}d) and is almost absent at $x=0.28$ (Fig.~\ref{fig:bsx}f).

The degree of influence of the uniaxial pressure on the Fermi contour depends on the position of the Fermi level, i.e. on doping. If the electron system is far from the quantum phase transition at a doping $x_{c}$, then the reconstruction of the Fermi contour with pressure will be insignificant. This can be seen in the example of the underdoped composition with $x=0.03$: the Fermi contour is represented by small hole pockets around the k-points $\left( {{\pi  \mathord{\left/
 {\vphantom {\pi  2}} \right.
 \kern-\nulldelimiterspace} 2},{\pi  \mathord{\left/
 {\vphantom {\pi  2}} \right.
 \kern-\nulldelimiterspace} 2}} \right)$, $\left( {{{3\pi } \mathord{\left/
 {\vphantom {{3\pi } 2}} \right.
 \kern-\nulldelimiterspace} 2},{\pi  \mathord{\left/
 {\vphantom {\pi  2}} \right.
 \kern-\nulldelimiterspace} 2}} \right)$, $\left( {{\pi  \mathord{\left/
 {\vphantom {\pi  2}} \right.
 \kern-\nulldelimiterspace} 2},{{3\pi } \mathord{\left/
 {\vphantom {{3\pi } 2}} \right.
 \kern-\nulldelimiterspace} 2}} \right)$, 
 $\left( {{{3\pi } \mathord{\left/
 {\vphantom {{3\pi } 2}} \right.
 \kern-\nulldelimiterspace} 2},{{3\pi } \mathord{\left/
 {\vphantom {{3\pi } 2}} \right.
 \kern-\nulldelimiterspace} 2}} \right)$ both at ${P^{\left( c \right)}}=0$ and at ${P^{\left( c \right)}}=10$ GPa (Figs.~\ref{fig:FS}a,b). The effect of pressure on a system near a quantum phase transition leads to a fundamental reconstruction of the electronic structure. At $x\approx x_{c1}=0.15$, the Fermi contour transforms from four hole pockets to one large electron and one large hole contour around point ${\rm M}$ with increase in ${P^{\left( c \right)}}$ from $0$ to $10$ GPa (Figs.~\ref{fig:FS}c,d). At $x\approx x_{c2}=0.28$, $c$-axis compression causes the small hole pocket around  ${\rm M}$ to collapse (Figs.~\ref{fig:FS}e,f).
 
\section{\label{sec:Tc_pres} The effect of c-axis compression on T$_c$}
\begin{figure}
\includegraphics[width=0.98\linewidth]{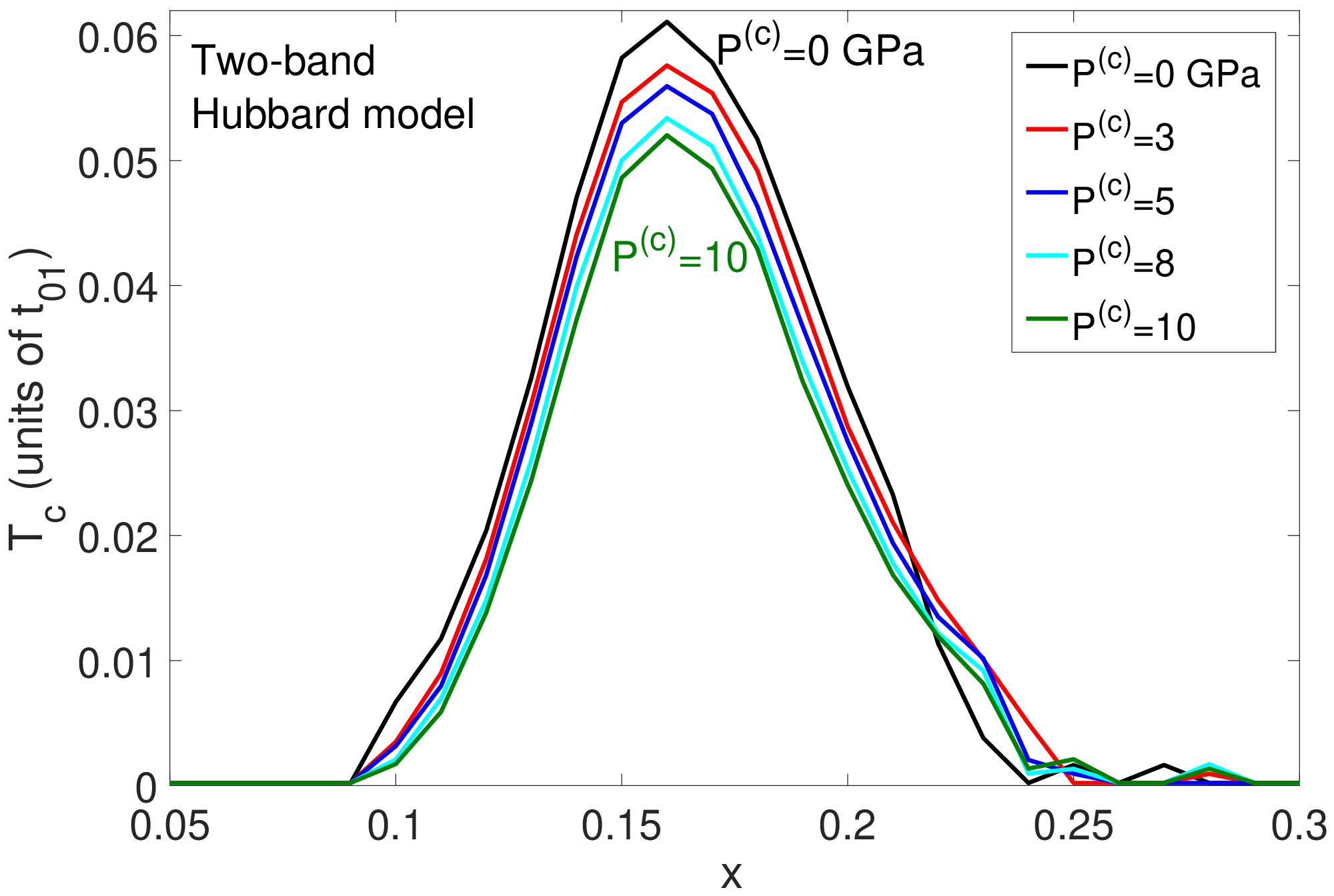}
\caption{\label{fig:Tcx_2band} (Color online) Concentration dependencies of ${T_c}$ (in units of $t_{01}\equiv{t_{10}}(0\sigma ,\bar \sigma A^S)$) at different values of the $c$-axis compression $P^{(c)}$ within the two-band Hubbard model.}
\end{figure}
\begin{figure}
\includegraphics[width=0.98\linewidth]{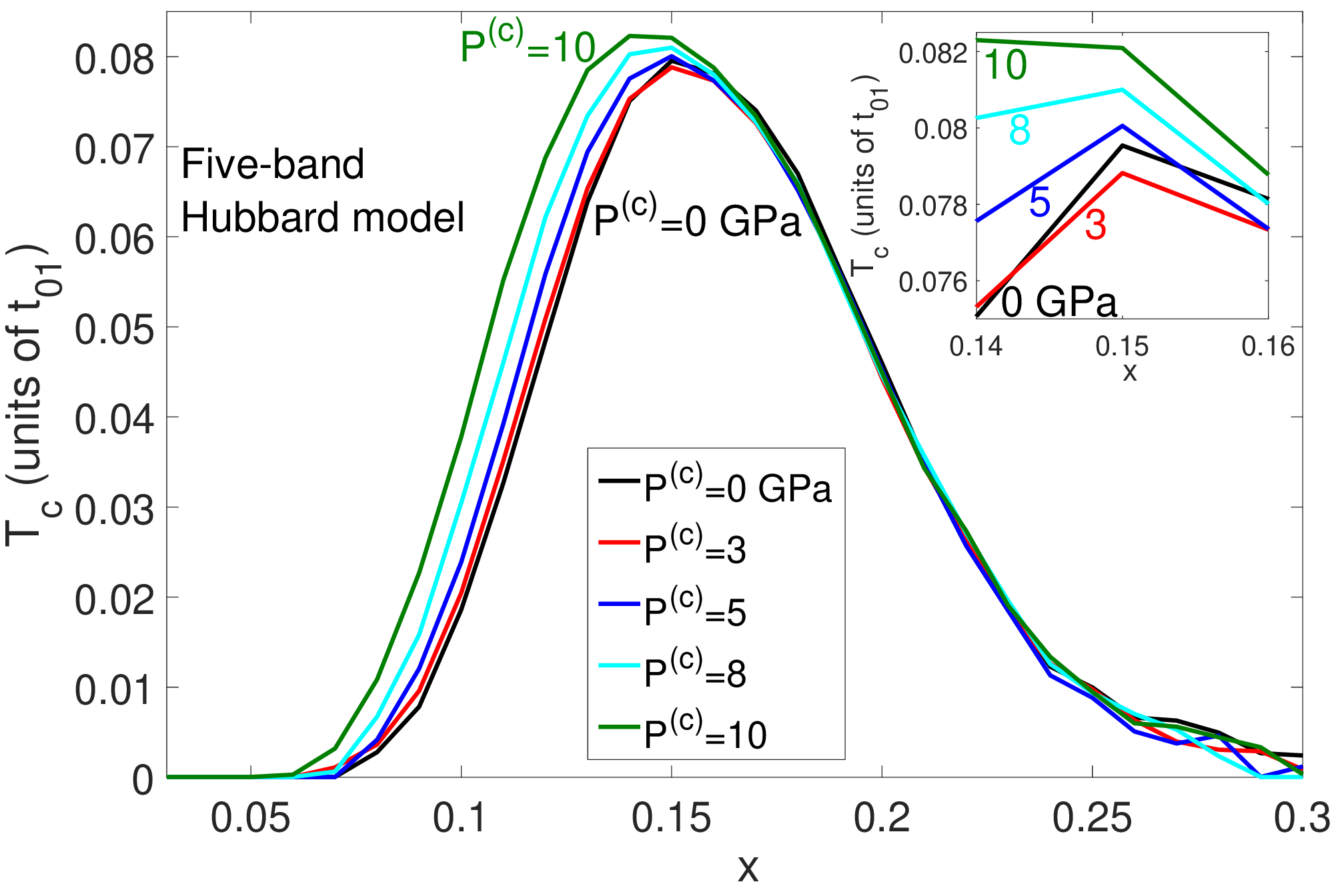}
\caption{\label{fig:Tcx_5band} (Color online) Concentration dependencies of ${T_c}$ (in units of $t_{01}\equiv{t_{10}}(0\sigma ,\bar \sigma A^S)$) at different values of the $c$-axis compression $P^{(c)}$ within the five-band Hubbard model. The inset shows the regions of ${T_c}\left( x \right)$ dependencies near optimal doping $x_{opt}=0.15$.}
\end{figure}
\begin{figure}
\includegraphics[width=0.98\linewidth]{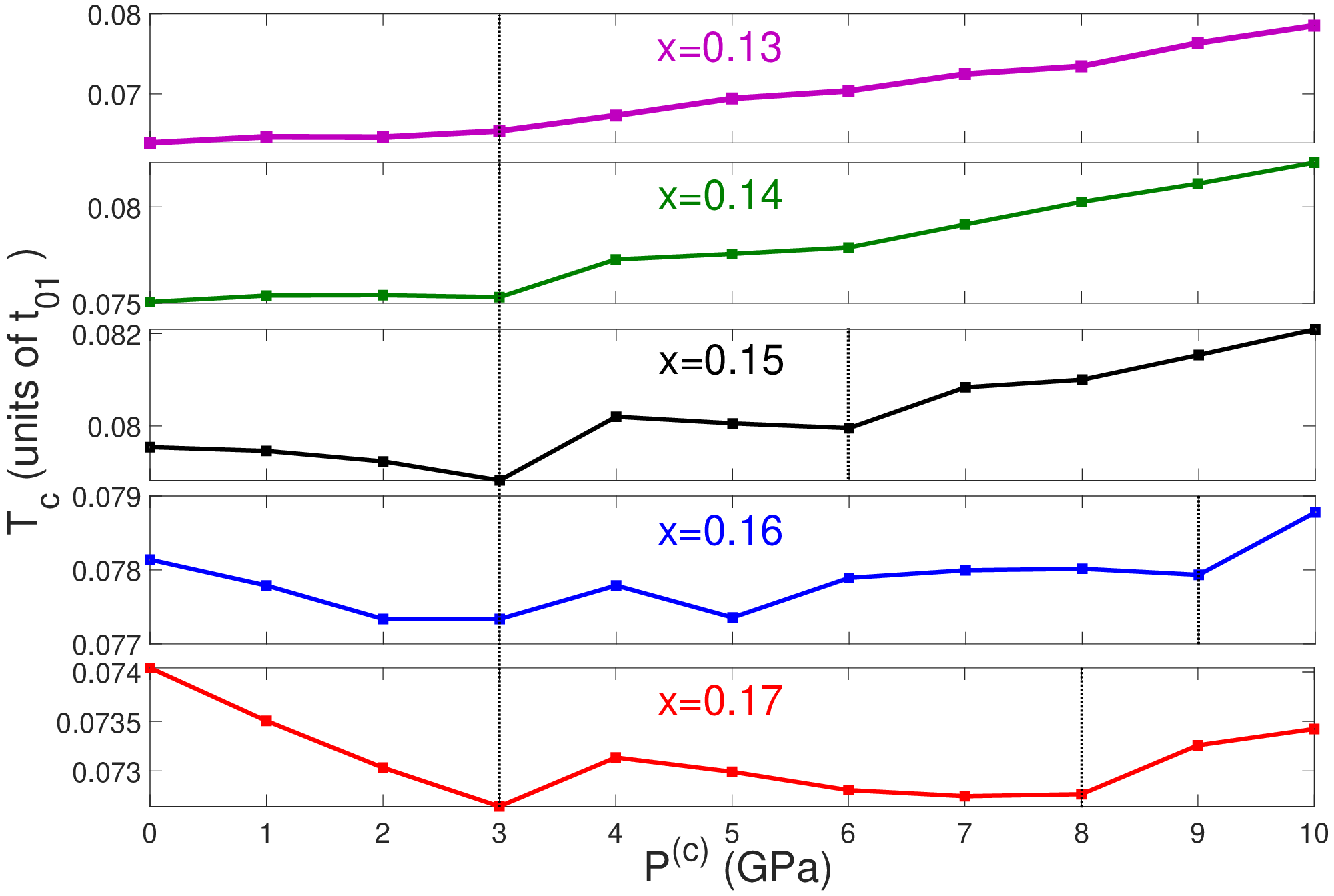}
\caption{\label{fig:Tc_P} (Color online) Dependencies of ${T_c}$ (in units of $t_{01}\equiv{t_{10}}(0\sigma ,\bar \sigma A^S)$) on the $c$-axis compression $P^{(c)}$ at different doping near $x_{opt}=0.15$ within the five-band Hubbard model. The dotted vertical lines separate pressure regions with different $T_c$ behavior.}
\end{figure}
\begin{figure*}
\includegraphics[width=0.8\linewidth]{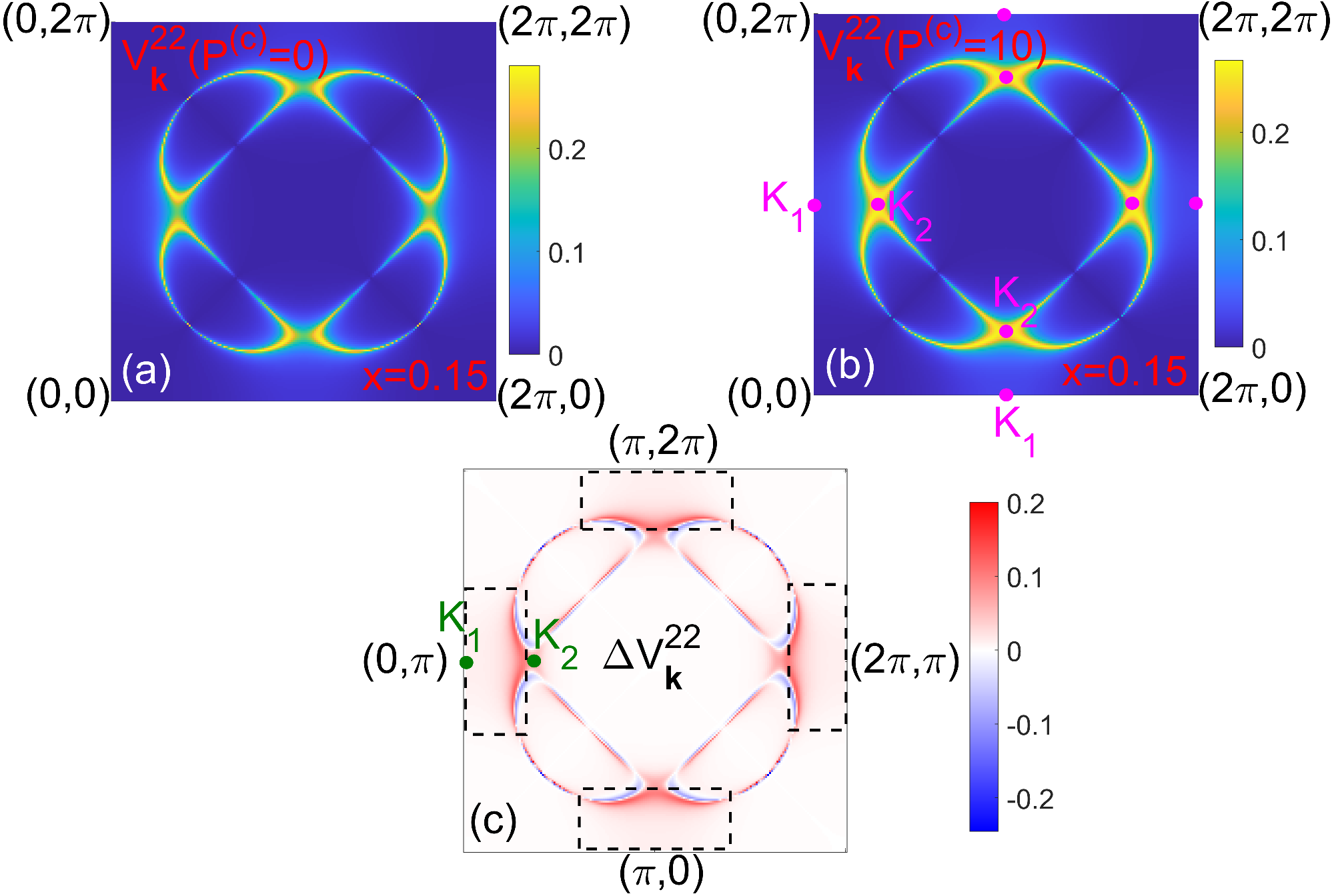}
\caption{\label{fig:an_av} (Color online) Values of the anomalous averages $V_{\bf{k}}^{22}\equiv{V_{{\bf{k}}}}\left( {\bar \sigma A^S,\sigma A^S} \right)$ in the first Brillouin zone at (a) ${P^{\left( c \right)}} = 0$, (b) ${P^{\left( c \right)}} = 10$ GPa and (c) difference between them $\Delta V_{\bf{k}}^{22} = V_{\bf{k}}^{22}\left( {{P^{\left( c \right)}} = 10} \right) - V_{\bf{k}}^{22}\left( {{P^{\left( c \right)}} = 0} \right)$.}
\end{figure*}

First of all, we consider the concentration dependence of ${T_c}$ and its change under uniaxial pressure within the standard two-band Hubbard model. The band structure for the two-band Hubbard model consists of the bands of the quasiparticles ${q_1}$ and ${q_2}$, superconducting pairing is caused by the superexchange interaction constant $J^{22}$. The concentration dependence of ${T_c}$ in the absence of pressure is dome with the maximum at approximately $x=0.16$ (Fig.~\ref{fig:Tcx_2band}, black line). The maximum ${T_c}$ is determined by the hole concentration at which the chemical potential hits the van Hove singularity formed by the states near the $K_2$ points. It is seen from Fig.~\ref{fig:Tcx_2band} that ${T_c}$ within the two-band Hubbard model decreases with increasing the $c$-axis compression $P^{(c)}$ at each $x$ except for the overdoped region $x>0.21$ near which superconductivity disappears. The mechanism (i) for this decrease in ${T_c}\left( x \right)$ is the decrease in the hopping integrals $t_{pd}$, $t_{pp}$ and, consequently, in the superexchange parameter $J^{22}$ with increasing intraplane Cu-O distances under the $c$-axis compression. The coefficient $\Theta_{A^S}$ hardly changes with pressure (Fig.~\ref{fig:paronPc}c). 

The dome shape of the concentration dependence of ${T_c}$ is preserved when it is constructed within the effective five-band Hubbard model (Fig.~\ref{fig:Tcx_5band}). The critical temperatures at each fixed doping and pressure value are higher in the calculations within the five-band Hubbard model (for example, ${T_{cmax}}\sim0.08t_{01}$, where $t_{01}\equiv{t_{10}}(0\sigma ,\bar \sigma A^S)$, at $P^{\left( c \right)} =0$ GPa and $x_{opt}=0.15$, Fig.~\ref{fig:Tcx_5band}) compared with the two-band Hubbard model (${T_{cmax}}\sim0.06t_{01}$ at $P^{\left( c \right)} =0$ GPa, Fig.~\ref{fig:Tcx_2band}). This $T_c$ growth is caused by the dispersion flattening and the DOS increasing at the valence band top due to the hybridization between the $q_2$ and $q_3$-$q_4$-$q_5$ quasiparticles which is taken into account within the five-band model. The behavior of ${T_c}$ under compression within the five-band Hubbard model changes dramatically compared to that for two-band model: ${T_{cmax}}$ increases with pressure and the dependence of ${T_c}$ on the $P^{\left( c \right)}$ magnitude varies strongly at different doping. In the underdoped region $x<0.14$, ${T_c}$ monotonically increases with increasing $P^{\left( c \right)}$ (Fig.~\ref{fig:Tcx_5band}) although the growth rate differs in different pressure regions at $x=0.13$, $0.14$ near optimal doping (Figs.~\ref{fig:Tc_P}a,b). ${T_c}$ does not depend or barely increases with pressure changes in the range from 0 to 3 GPa in the slightly underdoped region (Figs.~\ref{fig:Tc_P}a,b). At $P^{\left( c \right)}>3$ GPa the growth rate increases (Figs.~\ref{fig:Tc_P}a,b). The underdoped side of the superconducting dome together with $x_{opt}$ shifts towards lower concentrations with increasing $P^{\left( c \right)}$. As a result, the larger $P^{\left( c \right)}$ the lower doping values at which superconductivity occurs (Fig.~\ref{fig:Tcx_5band}). The overdoped side of the superconducting dome remains almost unchanged under pressure (Fig.~\ref{fig:Tcx_5band}). 

The different behavior of $T_c$ with pressure in different doping regions is caused by the presence of two competing mechanisms. The first mechanism (i) is the renormalization of superexchange parameters due to a change in interatomic distances. The influence of intraplane interatomic distances on the superexchange pairing parameter $J^{22}$ remains the same as within two-band Hubbard model. There are three more superexchange parameters $J^{33}$, $J^{44}$, $J^{55}$ in addition to $J^{22}$ within the five-band Hubbard model. $J^{33}$, $J^{44}$, $J^{55}$ slightly increase with increasing $c$-axis compression (Fig.~\ref{fig:paronPc}d) since the quasiparticle hopping integrals ${t_{10}}(0\sigma ,\bar \sigma B^T)$, ${t_{10}}(0\sigma ,\bar \sigma B^S)$ increase (Fig.~\ref{fig:paronPc}a). The small degree of change in the superexchange parameters $J^{33}$, $J^{44}$, $J^{55}$ is due to that the coefficients $\Theta_{B^T}$, $\Theta_{B^S}$ in the denominators of (\ref{eq:d11_q})-(\ref{eq:d21_q}) also increase with increasing $c$-axis compression (Fig.~\ref{fig:paronPc}c). The opposite tendencies of change in the parameters $J^{22}$ and $J^{33}$, $J^{44}$, $J^{55}$ under $c$-axis compression compensate each other to a certain extent.

 The second mechanism (ii) is the increase in DOS caused by the flattening of the ${q_2}$ band between $K_1$ and $K_2$ points due to the enhanced hybridization between the ${q_2}$ and $q_3$-$q_4$-$q_5$ bands under compression. Thus, the second mechanism is only possible in the five-band Hubbard model. The analysis of the k-dependence of the largest anomalous average $V_{\bf{k}}^{22}\equiv{B_{{\bf{k}}}}\left( {\bar \sigma A^S,\sigma A^S} \right)$ allows us to find out how the reconstruction of the valence band top influences on the superconducting gap amplitudes (\ref{eq:d11_q})-(\ref{eq:d21_q}). It is seen from Figs.~\ref{fig:an_av}a,b that the dependence of the absolute values $V_{\bf{k}}^{22}$ on wave vectors mostly repeats the topology of the Fermi contour (Fig.~\ref{fig:FS}). The distribution of $V_{\bf{k}}^{22}$ in the k-space changes with increasing compression: the larger value of the compression ${P^{\left( c \right)}}$ the larger number of the significant contributions $V_{\bf{k}}^{22}$ (Figs.~\ref{fig:an_av}a,b). At ${P^{\left( c \right)}}=0$ GPa, the most large values of the anomalous average $V_{\bf{k}}^{22}$ are given by the states between the approaching edges of the hole pockets on the lines $k_x=\pi$ and $k_y=\pi$ (Fig.~\ref{fig:an_av}a). At ${P^{\left( c \right)}}=10$ GPa, the regions of a significant contributions between the hole pockets have larger area, and also a large number of additional contributions appear around k-points ${\rm X}$, ${\tilde {\rm X}}$, ${\rm Y}$, ${\tilde {\rm Y}}$ between points $K_2$ and $K_1$ (Fig.~\ref{fig:an_av}b). These additional contributions are clearly visible on the map of the difference between the anomalous averages at ${P^{\left( c \right)}}=0$ and ${P^{\left( c \right)}}=10$ GPa: $\Delta V_{\bf{k}}^{22} = V_{\bf{k}}^{22}\left(10 \right) - V_{\bf{k}}^{22}\left(0 \right)$ (red regions inside dashed rectangles in Fig.~\ref{fig:an_av}c). Presence of these additional contributions results in the ${T_c}$ growth under the $c$-axis compression.

The continuous growth of ${T_c}$ under compression in the underdoped region (Fig.~\ref{fig:Tcx_5band}) is caused by the predominance of the mechanism (ii) over the mechanism (i). The almost complete absence of dependence of ${T_c}$ on $P^{\left( c \right)}$ in the overdoped region is due to the mutual compensation of the mechanisms (i) and (ii), when the mechanism (ii) is weakened as a result of the absence of any significant effect of reconstruction of the valence band top (Fig.~\ref{fig:bsx}e,f). However, the region of optimal doping is of greatest interest, since it is mainly this region that is studied in experiments. The optimally doped compositions is in transition region between the underdoped region in which $T_c$ increases even at low pressure and the overdoped region in which $T_c$ does not change even at high $P^{\left( c \right)}\approx10$. Therefore, complicated behavior should be expected due to interrelation between the mechanisms (i) and (ii) near optimal doping. Indeed, in the doping region $0.14<x<0.17$, ${T_c}$ changes non-monotonically with increasing $P^{\left( c \right)}$ (inset in Fig.~\ref{fig:Tcx_5band}). Within this region, the character of the non-monotonic pressure dependence of ${T_c}$ differs for different doping (Figs.~\ref{fig:Tc_P}c-e). Three pressure regions for ${T_c}\left( P^{\left( c \right)} \right)$ dependencies at each $0.15<x<0.17$ can be distinguished. At lower pressure values (from 0 to 3 GPa), the mechanism (ii) hardly works since the states of ${a_{1g}}$ symmetry are deep inside the valence band and have little effect on the states at the top of the valence band, and the mechanism (i) dominates. Therefore, ${T_c}$ decreases (Figs.~\ref{fig:Tc_P}c-e) in the pressure range from 0 to 3 GPa. This behavior repeats the pressure dependence of ${T_c}\left( x \right)$ for the two-band Hubbard model and agrees with the most results of uniaxial pressure experiments mentioned in Introduction. The intermediate pressure region is characterized by complex behavior of ${T_c}$ and includes several alternating sections of ${T_c}$ increase and decrease having different lengths at different doping. At $x=0.15$, the intermediate region is from 3 to 6 GPa and is from 3 to 8-9 GPa at $x=0.16$, $0.17$. This variable dependence results from the competition not only between the mechanisms (i) and (ii) but also between the renormalizations of the different types of the superexchange constants ($J^{22}$ and $J^{33}$, $J^{44}$, $J^{55}$). Third region of higher pressure values is where ${T_c}$ increases with increasing ${P^{\left( c \right)}}$ since the influence of bands ${q_3}$, ${q_4}$, ${q_5}$ on the band ${q_2}$ and hybridization between them are strong enough that the increase in the intensity of the van Hove peak compensates for the decrease in ${T_c}$ due to the decrease in superexchange $J^{22}$. This region is from 6 to 10 GPa at $x=0.15$ (Fig.~\ref{fig:Tc_P}c) and from 8-9 to 10 GPa at $x=0.16$, $0.17$ (Figs.~\ref{fig:Tc_P}d,e). It is obvious that the ${T_c}$ behavior with pressure critically depends on the region of applied pressure magnitudes and on the concentration of doped holes. The ambiguous behavior of the uniaxial pressure dependences of ${T_c}$ in the various experiments mentioned in the Introduction may be due to small differences in the doping degree of the CuO$_2$ planes in the various samples.
 
\section{\label{sec:conclusion} Conclusion}

In this work, the effect of the $c$-axis compression on the electronic structure of quasiparticle excitations and the concentration dependence of ${T_c}$ in the HTSC cuprate La$_{2-x}$Sr$_x$CuO$_4$ is studied within the framework of the effective five-band Hubbard model. At zero compression, the top of the valence band is formed by the single band with the character of the $b_{1g}$ orbitals, the $a_{1g}$ quasiparticle bands are located deeper into valence band.  Uniaxial compression restores the octahedral symmetry of CuO$_6$ octahedra and leads to an increase in the energy of the $a_{1g}$ quasiparticle excitations. The hybridization between the $b_{1g}$ and $a_{1g}$ bands leads to the flattening of the dispersion on the top of the valence band and to the DOS growth under the $c$-axis compression. An increase in the number of states involved in pairing in the regions near the k-points $\left( {\pi ,0} \right)$, $ \left( {0 ,\pi } \right)$, $\left( {2\pi ,\pi} \right)$, $\left( {\pi ,2\pi} \right)$ leads to an increase in ${T_c}$. This electronic structure reconstruction is one (ii) of the two mechanisms of the $c$-axis compression influence on the superconducting gap and dome-shaped concentration dependence of ${T_c}$. The mechanism (ii) results in the $T_c$ increase with increasing compression $P^{(c)}$ in the underdoped region. The other mechanism (i) is the renormalization of the superconducting pairing constants. The $c$-axis compression reduces the largest superexchange constant for pairing within the $b_{1g}$ band and thus results in the ${T_c}$ decrease. Near the optimal doping, it is shown that $T_c$ either decreases (at the optimal and slightly overdoped region) or increases insignificantly (in the slightly underdoped region) at pressures less than 3 GPa, when the mechanism (i) prevails. This agrees with the experimental results for this pressure region. Moreover, the behavior of $T_c$ with pressure changes significantly with small changes in doping. The ambiguity of the results of various measurements can be related to their sensitivity to the number of doped carriers in the CuO$_2$ planes which is difficult to control experimentally. Our calculations demonstrate that ${T_{cmax}}$ can be increased by the applying the uniaxial $c$-axis compression $\sim10$ GPa due to activation of the mechanism (ii) even near the optimal doping.

\section*{Acknowledgments}
 The present work was carried out within the state assignment of Kirensky Institute of Physics.

\appendix
\section{\label{app:model} The effective five-band Hubbard model of quasiparticle excitations}
\begin{figure*}
\includegraphics[width=0.45\linewidth]{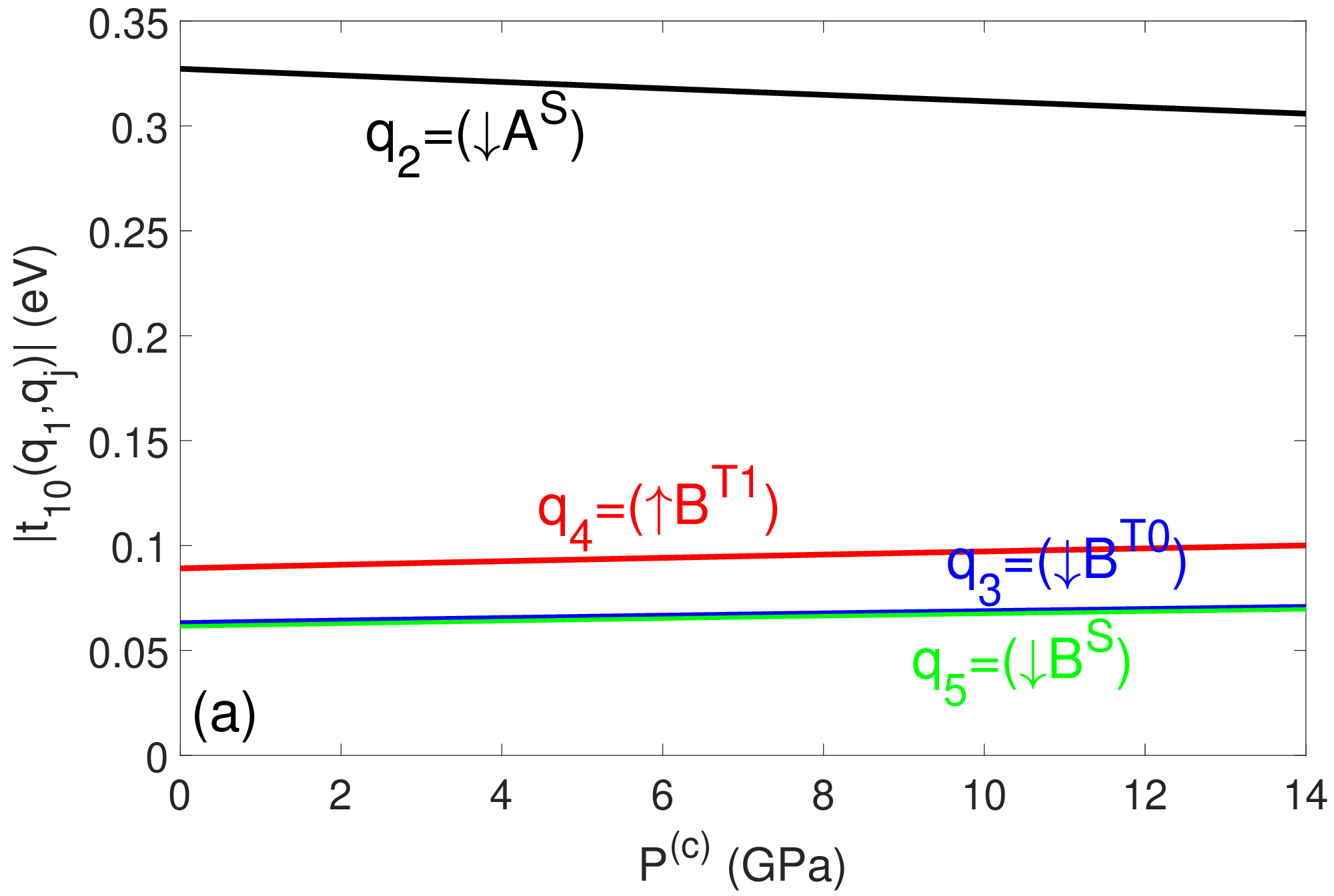}
\includegraphics[width=0.45\linewidth]{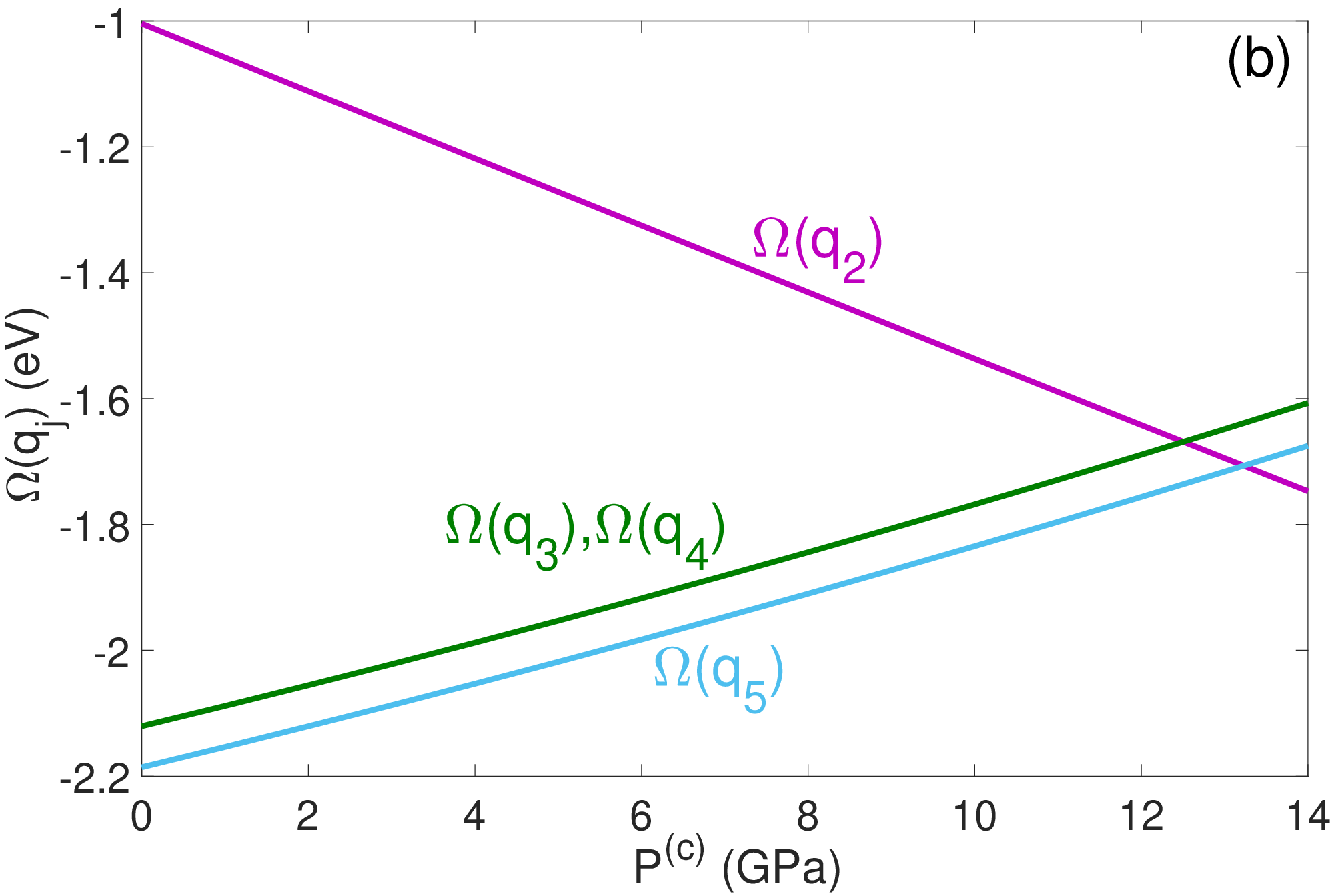}
\includegraphics[width=0.45\linewidth]{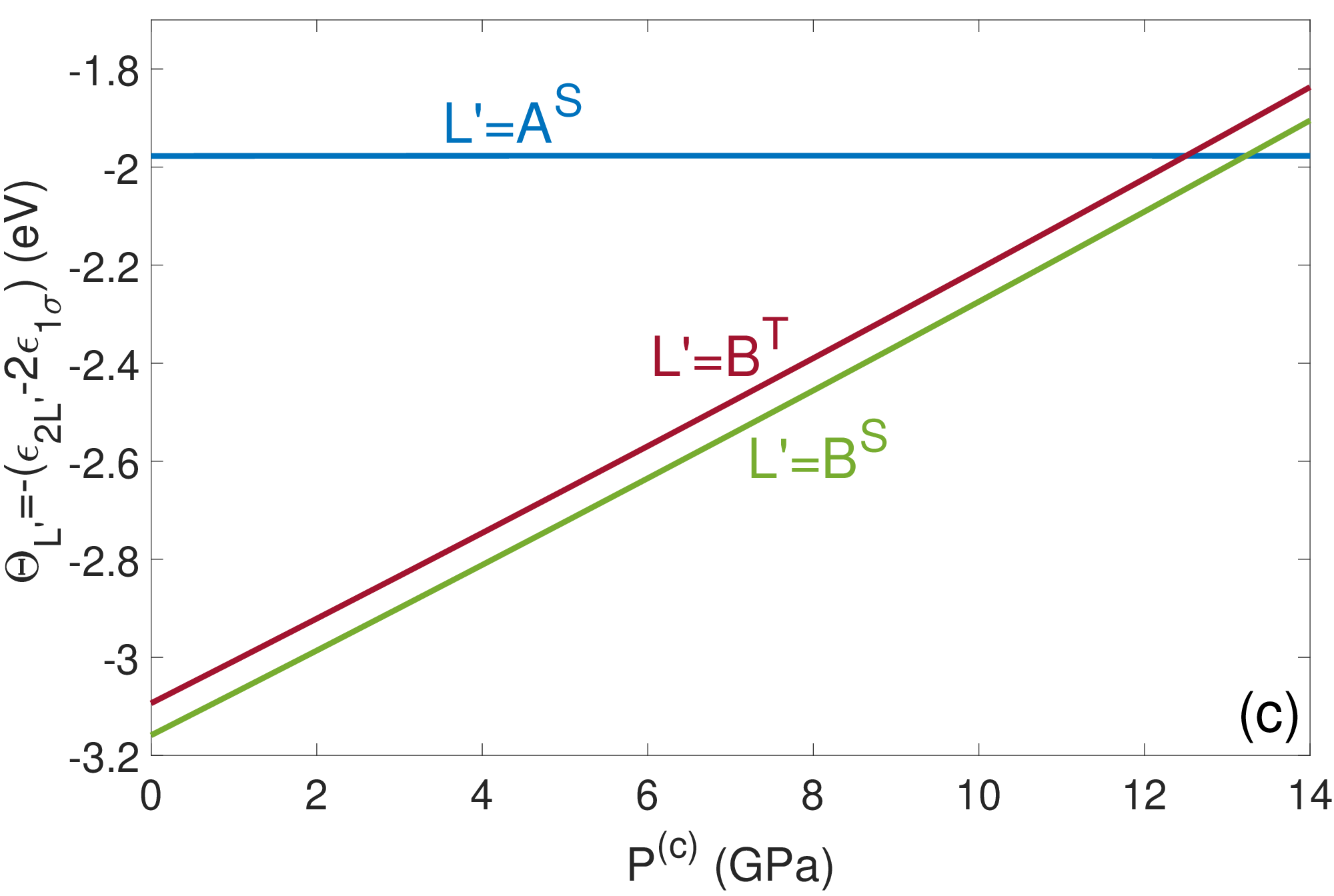}
\includegraphics[width=0.45\linewidth]{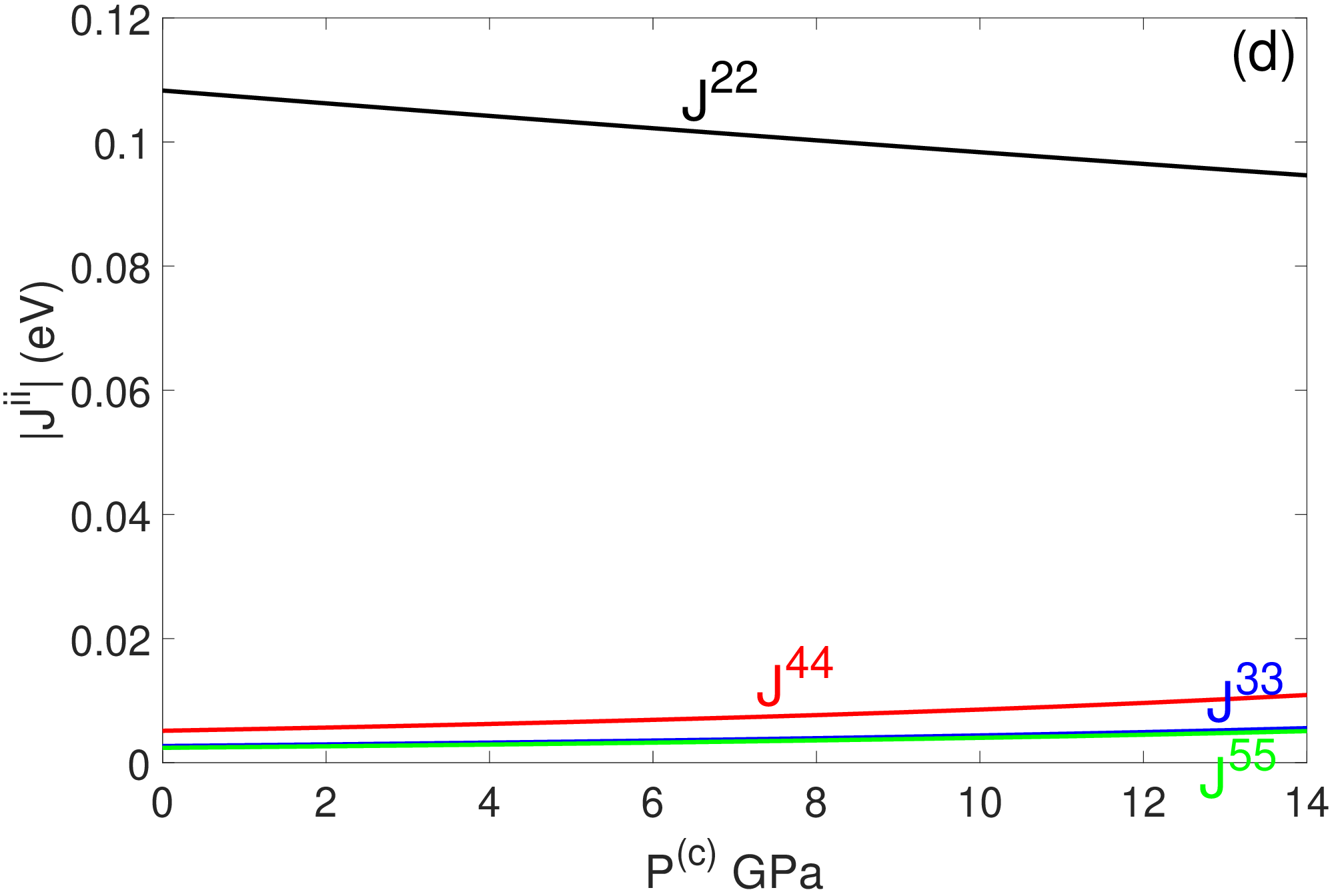}
\caption{\label{fig:paronPc} (Color online) Dependencies of (a) absolute values of hopping integrals $\left| {{t_{10}}\left( {{q_1 },{q_j} } \right)} \right|$ between quasiparticles in the nearest CuO$_6$ clusters,(b) energies $\Omega \left( {q_j} \right)$,  (c) parameters $\Theta_{L'}=-\left( {\epsilon_{2L'} - 2\epsilon_{1\sigma}} \right)$ and (d) superexchange constants $\left| {{J^{jj}}} \right|$ for the quasiparticles $q_2$-$q_5$ on the $c$-axis compression $P^{(c)}$.}
\end{figure*}

In the GTB method, the crystal lattice is divided into clusters, the full Hamiltonian is represented as a sum of the intracluster Hamiltonian and the Hamiltonian of intercluster interactions. The multiparticle cluster eigenstates are obtained using exact diagonalization. The ground single-hole doublet is  $\left| {1,\sigma}  \right\rangle  \equiv \left| {1,{b_{1\sigma }}} \right\rangle  = {c_{dx}}\left| {{d_{x\sigma }}} \right\rangle  + {c_b}\left| {{b_\sigma }} \right\rangle $. The molecular oxygen orbitals $b$- and $a$ are defined in the k-space as a combinations of $p_{x}$- and $p_{y}$-orbitals according with the Shastry procedure~\cite{Shastry89,Raimondi96}, these orbitals are orthogonal in adjacent clusters. The Zhang-Rice singlet $A^S$
\begin{eqnarray}\label{eq:A1_singlet}
&&\left| {2,A^S} \right\rangle  =\\\nonumber
&& {L_{dxb}}\left( {\left| {{d_{x \downarrow }}{b_ \uparrow }} \right\rangle  - \left| {{d_{x \uparrow }}{b_ \downarrow }} \right\rangle } \right) + {L_{dxdx}}\left| {{d_{x \downarrow }}{d_{x \uparrow }}} \right\rangle  +{L_{bb}}\left| {{b_ \downarrow }{b_ \uparrow }} \right\rangle
\end{eqnarray}
is the ground two-hole state. The three components of the triplet $\left| {2,B^T} \right\rangle  = \left\{ {\left| {2,B^{T0}} \right\rangle } \right.$, $\left| {2,B^{T1}} \right\rangle $, $\left. {\left| {2,B^{T - 1}} \right\rangle } \right\}$(Fig.~\ref{fig:quasiparticles}a) have the form:
\begin{eqnarray}\label{eq:triplet} 
\left| {2,B^{T0}} \right\rangle &=& {L_{dxdz}}\left( {\left| {{d_{x \downarrow }}{d_{z \uparrow }}} \right\rangle  + \left| {{d_{x \uparrow }}{d_{z \downarrow }}} \right\rangle } \right) +\\\nonumber
&& {L_{dxa}}\left( {\left| {{d_{x \downarrow }}{a_ \uparrow }} \right\rangle  + \left| {{d_{x \uparrow }}{a_ \downarrow }} \right\rangle } \right) +\\\nonumber
&& {L_{dxpz}}\left( {\left| {{d_{x \downarrow }}{p_{z \uparrow }}} \right\rangle  + \left| {{d_{x \uparrow }}{p_{z \downarrow }}} \right\rangle } \right) + \cdots,\\\nonumber
\left| {2,B^{T1}} \right\rangle &=& {L'_{dxdz}}\left| {{d_{x \uparrow }}{d_{z \uparrow }}} \right\rangle  + {L'_{dxa}}\left| {{d_{x \uparrow }}{a_ \uparrow }} \right\rangle  +\\\nonumber
&& {L'_{dxpz}}\left| {{d_{x \uparrow }}{p_{z \uparrow }}} \right\rangle  + \cdots,\\\nonumber
\left| {2,B^{T - 1}} \right\rangle &=& {L'_{dxdz}}\left| {{d_{x \downarrow }}{d_{z \downarrow }}} \right\rangle  + {L'_{dxa}}\left| {{d_{x \downarrow }}{a_ \downarrow }} \right\rangle  +\\\nonumber
&& {L'_{dxpz}}\left| {{d_{x \downarrow }}{p_{z \downarrow }}} \right\rangle  + \cdots.
\end{eqnarray}
The singlet state $\left| {2,B^S} \right\rangle$ 
\begin{eqnarray}\label{eq:singlet_B2s} 
&&\left| {2,B^S} \right\rangle ={{L''}_{dxdz}}\left( {\left| {{d_{x \downarrow }}{d_{z \uparrow }}} \right\rangle  - \left| {{d_{x \uparrow }}{d_{z \downarrow }}} \right\rangle } \right) +\\\nonumber
&&{{L''}_{dxa}}\left( {\left| {{d_{x \downarrow }}{a_ \uparrow }} \right\rangle  - \left| {{d_{x \uparrow }}{a_ \downarrow }} \right\rangle } \right) +\\\nonumber
&&{{L''}_{dxpz}}\left( {\left| {{d_{x \downarrow }}{p_{z \uparrow }}} \right\rangle  - \left| {{d_{x \uparrow }}{p_{z \downarrow }}} \right\rangle } \right) +\\\nonumber
&&{{L''}_{bdz}}\left( {\left| {{b_ \downarrow }{d_{z \uparrow }}} \right\rangle  - \left| {{b_ \uparrow }{d_{z \downarrow }}} \right\rangle } \right) +\\\nonumber
&& {{L''}_{ba}}\left( {\left| {{b_ \downarrow }{a_ \uparrow }} \right\rangle  - \left| {{b_ \uparrow }{a_ \downarrow }} \right\rangle } \right) + {{L''}_{bpz}}\left( {\left| {{b_ \downarrow }{p_{z \uparrow }}} \right\rangle  - \left| {{b_ \uparrow }{p_{z \downarrow }}} \right\rangle } \right)
\end{eqnarray}
is formed by the same orbitals as the triplet $\left| {2,B^T} \right\rangle$ (Fig.~\ref{fig:quasiparticles}a).

Each quasiparticle excitation $\left( {mm'} \right)$ between the initial $\left| {n,m'} \right\rangle $ and final $\left| {n \pm 1,m} \right\rangle $ multiparticle cluster eigenstates is described by the Hubbard operator $X_{\bf{f}}^{mm'} = \left| {n \pm 1,m} \right\rangle \left\langle {n,m'} \right|$. Electron (or hole) can be expressed as a sum of quasiparticle excitations by the relation:
\begin{equation}\label{eq:a_X}
{a_{\lambda {\bf{f}}\sigma }} = \sum\limits_{mm'} {{\gamma _{\lambda \sigma }}\left( {mm'} \right)X_{\bf{f}}^{mm'}} 
\end{equation}
where the matrix elements ${\gamma _{\lambda \sigma }}$ are determined by the relation ${\gamma _{\lambda \sigma }}\left( mm'  \right) = \left\langle {n - 1,m} \right|{a_{\lambda {\bf{f}}\sigma }}\left| {n,m'} \right\rangle$~\cite{Ovchinnikov89,Gavrichkov00,Korshunov05}. The intracluster and intercluster parts of the five-band p-d model Hamiltonian are expressed in terms of the Hubbard operators within the framework of the cluster form of perturbation theory. The full Hamiltonian of the effective five-band Hubbard model in the hole representation has the form:
\begin{eqnarray}\label{eq:H_Hub}
&&H = \sum\limits_{{\bf{f}}m} {{\varepsilon _m}X_{\bf{f}}^{mm}}  + \sum\limits_{{\bf{fg}}{\beta _\sigma }{\beta _\sigma }^\prime \sigma } {{t_{{\bf{fg}}}}\left( {{\beta _\sigma },{\beta _\sigma }^\prime } \right)\mathop {X_{\bf{f}}^{{\beta _\sigma }}}\limits^\dag X_{\bf{g}}^{{\beta _\sigma }^\prime }}  
\end{eqnarray}
where ${\bf{f}}$, ${\bf{g}}$ - cluster coordinates, the index $m$ enumerates eight multiparticle cluster eigenstates (\ref{eq:eigenstates}) with the energies ${{\varepsilon _m}}$. The root vector ${\beta _\sigma }$ runs over the five quasiparticle excitations ${q_j}$ with spin projection $\sigma$, where $j=1-5$. The first term of (\ref{eq:H_Hub}) is the intracluster interaction including energies of the cluster hole eigenstates. The second term consists of quasiparticle hoppings between different clusters. The hopping integrals are defined by the expression:
\begin{eqnarray}\label{eq:X_hoppings}
&&{t_{{\bf{fg}}}}\left( {\beta ,\beta '} \right) = t_{{\bf{fg}}}^{bb}\left( {\beta ,\beta '} \right) + t_{{\bf{fg}}}^{aa}\left( {\beta ,\beta '} \right) + t_{{\bf{fg}}}^{ba}\left( {\beta ,\beta '} \right),\\\nonumber
&&t_{{\bf{fg}}}^{bb}\left( {\beta ,\beta '} \right) =  - 2{t_{pd}}{\mu _{{\bf{fg}}}}\left( {\gamma _{dx}^ * \left( \beta  \right){\gamma _b}\left( {\beta '} \right) + \gamma _b^ * \left( \beta  \right){\gamma _{dx}}\left( {\beta '} \right)} \right) - \\\nonumber
&&2{t_{pp}}{\nu _{{\bf{fg}}}}\gamma _b^ * \left( \beta  \right){\gamma _b}\left( {\beta '} \right),\\\nonumber
&&t_{{\bf{fg}}}^{aa}\left( {\beta ,\beta '} \right) =  - 2{t_{pdz}}{\lambda _{{\bf{fg}}}}\left( {\gamma _{dz}^ * \left( \beta  \right){\gamma _a}\left( {\beta '} \right) + \gamma _a^ * \left( \beta  \right){\gamma _{dz}}\left( {\beta '} \right)} \right) - \\\nonumber
&&2{t_{ppz}}{\lambda _{{\bf{fg}}}}\left( {\gamma _a^ * \left( \beta  \right){\gamma _{pz}}\left( {\beta '} \right) + \gamma _{pz}^ * \left( \beta  \right){\gamma _a}\left( {\beta '} \right)} \right) + \\\nonumber
&&2{t_{pp}}{\nu _{{\bf{fg}}}}\gamma _a^ * \left( \beta  \right){\gamma _a}\left( {\beta '} \right),\\\nonumber
&&t_{{\bf{fg}}}^{ba}\left( {\beta ,\beta '} \right) = 2{t_{pdz}}{\xi _{{\bf{fg}}}}\left( {\gamma _{dz}^ * \left( \beta  \right){\gamma _b}\left( {\beta '} \right) + \gamma _b^ * \left( \beta  \right){\gamma _{dz}}\left( {\beta '} \right)} \right) - \\\nonumber
&&2{t_{pp}}{\chi _{{\bf{fg}}}}\left( {\gamma _b^ * \left( \beta  \right){\gamma _a}\left( {\beta '} \right) + \gamma _a^ * \left( \beta  \right){\gamma _b}\left( {\beta '} \right)} \right) - \\\nonumber
&& - 2{t_{ppz}}{\xi _{{\bf{fg}}}}\left( {\gamma _b^ * \left( \beta  \right){\gamma _{pz}}\left( {\beta '} \right) + \gamma _{pz}^ * \left( \beta  \right){\gamma _b}\left( {\beta '} \right)} \right),
\end{eqnarray}
where the structural factors ${\nu _{{\bf{fg}}}}$, ${\lambda _{{\bf{fg}}}}$, ${\xi _{{\bf{fg}}}}$, ${\chi _{{\bf{fg}}}}$ are determined in $k$-space in~\cite{Raimondi96}. The hoppings $t_{10}\left( {{q_1} ,{q_2}} \right)$ and $t_{10}\left( {{q_1} ,{q_3}} \right)$, $t_{10}\left( {{q_1} ,{q_4}} \right)$, $t_{10}\left( {{q_1} ,{q_5}} \right)$ have different dependencies on pressure. The Cu-O and O-O distances in CuO$_2$ plane increase under the $c$-axis compression and, consequently, the hoppings $t_{pd}$ and $t_{pp}$ within this plane decrease. Both quasiparticles $q_1$ and $q_2$ have the character of ${b_{1g}}$ symmetry orbitals therefore the only non-zero contribution to the quasiparticle hopping integral ${t_{10}}\left( {q_1 ,q_2} \right)$ is given by the component $t_{10}^{bb}$. Therefore the hopping integral ${t_{10}}(q_1,q_2)$ decrease under pressure (Fig.~\ref{fig:paronPc}a). The Cu-O$_{ap}$ and O$_{pl}$-O$_{ap}$ distances decrease under the $c$-axis compression, the hoppings between CuO$_2$ plane orbitals and $p_z$-orbitals $t_{dzpz}$ and $t_{ppz}$ increase. The quasiparticles $q_3$, $q_4$, $q_5$ have the character of ${a_{1g}}$ symmetry orbitals so the hopping integrals ${t_{10}}(q_1,q_3)$, ${t_{10}}(q_1,q_4)$, ${t_{10}}(q_1,q_5)$ are determined by the component $t_{10}^{ba}$. These three hopping integrals increase with increasing pressure (Fig.~\ref{fig:paronPc}a).

\section{\label{app:eq_of_motion} Theory of superconductivity within the five-band Hubbard model}
The method of equations of motion for the Green's functions of quasiparticle excitations is used to obtain the electronic structure and superconducting properties. The Green's functions built on the creation and annihilation Fermi operators is expressed in terms of the Green's functions constructed on the Hubbard operators by the following expression:
\begin{eqnarray}\label{eq:anih_Hub}
&&\left\langle {\left\langle {{{a_{\lambda {\bf{r}}\sigma }}}}
 \mathrel{\left | {\vphantom {{{a_{\lambda {\bf{r}}\sigma }}} {a_{\lambda '{\bf{r'}}\sigma }^\dag }}}
 \right. \kern-\nulldelimiterspace}
 {{a_{\lambda '{\bf{r'}}\sigma }^\dag }} \right\rangle } \right\rangle  =\\\nonumber 
 &&\sum\limits_{\beta \beta '} {{\gamma _{\lambda \sigma }}\left( \beta  \right)\gamma _{\lambda '\sigma }^ * \left( {\beta '} \right)\left\langle {\left\langle {{X_{\bf{r}}^\beta }}
 \mathrel{\left | {\vphantom {{X_{\bf{r}}^\beta } {\mathop {X_{{\bf{r'}}}^{\beta '}}\limits^\dag  }}}
 \right. \kern-\nulldelimiterspace}
 {{\mathop {X_{{\bf{r'}}}^{\beta '}}\limits^\dag  }} \right\rangle } \right\rangle } 
\end{eqnarray}
To describe the superconducting phase, we consider the retarded Green's function ${\hat D}_\sigma\left( {{\bf{r}},{\bf{r'}};t} \right) = \left\langle {\left\langle {{{{\hat X}_{\bf{r}}}\left( t \right)}}
 \mathrel{\left | {\vphantom {{{{\hat X}_{\bf{r}}}\left( t \right)} {\hat X_{{\bf{r'}}}^\dag \left( 0 \right)}}}
 \right. \kern-\nulldelimiterspace}
 {{\hat X_{{\bf{r'}}}^\dag \left( 0 \right)}} \right\rangle } \right\rangle $ built on the ten-component Nambu operator
\begin{eqnarray}\label{eq:Nambu}
&&{{\hat X}_{\bf{r}}} = \left\{ {{{\hat X}_{\sigma {\bf{r}}}},\hat X_{\bar \sigma {\bf{r}}}^\dag } \right\},\\
&&{{\hat X}_{\sigma {\bf{r}}}} = \left\{ {X_{\bf{r}}^{0\sigma },X_{\bf{r}}^{\bar \sigma {A^S}},X_{\bf{r}}^{\bar \sigma {B^{T0}}},X_{\bf{r}}^{\sigma {B^{T2\sigma }}},X_{\bf{r}}^{\bar \sigma {B^S}}} \right\}.
\end{eqnarray}
The Green's function ${\hat D}_\sigma\left( {{\bf{r}},{\bf{r'}};\omega } \right)$ can be represented as a supermatrix $2\times2$
\begin{eqnarray}\label{eq:matrixD}
&&{{\hat D}_\sigma }\left( {{\bf{r}},{\bf{r'}};\omega } \right) = \left[ {\begin{array}{*{20}{c}}
{{{\hat G}_\sigma }\left( {{\bf{r}},{\bf{r'}};\omega } \right)}&{{{\hat W}_\sigma }\left( {{\bf{r}},{\bf{r'}};\omega } \right)}\\
{\hat W_\sigma ^\dag \left( {{\bf{r}},{\bf{r'}};\omega } \right)}&{\hat G_{\bar \sigma }^\dag \left( {{\bf{r}},{\bf{r'}};\omega } \right)}
\end{array}} \right].
\end{eqnarray}
formed by the matrices $5\times5$ of normal components ${{\hat G}_\sigma }\left( {{\bf{r}},{\bf{r'}};\omega } \right)= \left\langle {\left\langle {{{{\hat X}_{\sigma {\bf{r}}}}}}
 \mathrel{\left | {\vphantom {{{{\hat X}_{\sigma {\bf{r}}}}} {\hat X_{\sigma {\bf{r'}}}^\dag }}}
 \right. \kern-\nulldelimiterspace}
 {{\hat X_{\sigma {\bf{r'}}}^\dag }} \right\rangle } \right\rangle $ and $\hat G_{\bar \sigma }^\dag \left( {{\bf{r}},{\bf{r'}};\omega } \right) = \left\langle {\left\langle {{\hat X_{\bar \sigma {\bf{r}}}^\dag }}
 \mathrel{\left | {\vphantom {{\hat X_{\bar \sigma {\bf{r}}}^\dag } {{{\hat X}_{\bar \sigma {\bf{r'}}}}}}}
 \right. \kern-\nulldelimiterspace}
 {{{{\hat X}_{\bar \sigma {\bf{r'}}}}}} \right\rangle } \right\rangle$ and also the matrices ${{\hat W}_\sigma }\left( {{\bf{r}},{\bf{r'}};\omega } \right) = \left\langle {\left\langle {{{{\hat X}_{\sigma {\bf{r}}}}}}
 \mathrel{\left | {\vphantom {{{{\hat X}_{\sigma {\bf{r}}}}} {{{\hat X}_{\bar \sigma {\bf{r'}}}}}}}
 \right. \kern-\nulldelimiterspace}
 {{{{\hat X}_{\bar \sigma {\bf{r'}}}}}} \right\rangle } \right\rangle $ and $\hat W_\sigma ^\dag \left( {{\bf{r}},{\bf{r'}};\omega } \right) = \left\langle {\left\langle {{\hat X_{\bar \sigma {\bf{r}}}^\dag }}
 \mathrel{\left | {\vphantom {{\hat X_{\bar \sigma {\bf{r}}}^\dag } {\hat X_{\sigma {\bf{r'}}}^\dag }}}
 \right. \kern-\nulldelimiterspace}
 {{\hat X_{\sigma {\bf{r'}}}^\dag }} \right\rangle } \right\rangle$ of anomalous components. The equation of motion for each component of the Green's function ${\hat D}_\sigma\left( {{\bf{r}},{\bf{r'}};\omega } \right)$ has the general form:
\begin{eqnarray}\label{eq:eq_motion}
&&\omega {D^{\left( {pm} \right)\left( {m'p'} \right)}}\left( {{\bf{r}},{\bf{r'}};\omega } \right) = {\delta _{{\bf{rr'}}}}{\delta _{pp'}}{\delta _{mm'}}F\left( {pm} \right) + \\\nonumber
&&\Omega \left( {pm} \right){D^{\left( {pm} \right)\left( {m'p'} \right)}}\left( {{\bf{r}},{\bf{r'}};\omega } \right) + {Z^{\left( {pm} \right)\left( {m'p'} \right)}}\left( {{\bf{r}},{\bf{r'}};\omega } \right),
\end{eqnarray}
where $F\left( {pm} \right) = \left\langle {{X^{pp}}} \right\rangle  + \left\langle {{X^{mm}}} \right\rangle $ and $\Omega \left( {pm} \right) = {\varepsilon _p} - {\varepsilon _m}$ are the filling factor and energy, respectively, of the quasiparticle excitation $\left( {pm} \right)$. The filling numbers $\left\langle {{X^{mm}}} \right\rangle$ and $\left\langle {{X^{pp}}} \right\rangle$ of the initial $m$ and final $p$ eigenstates for the quasiparticle $\left( {pm} \right)$ are calculated self-consistently with the chemical potential and kinematic correlation functions~\cite{Makarov22}. The term ${Z^{\left( {pm} \right)\left( {m'p'} \right)}}\left( {{\bf{r}},{\bf{r'}};\omega } \right) = {\left\langle {\left\langle {{Z_{{\bf{r}}}^{\left( {pm} \right)}}}
 \mathrel{\left | {\vphantom {{Z_{\bf{r}}^{\left( {pm} \right)}} {X_{{\bf{r'}}}^{\left( {m'p'} \right)}}}}
 \right. \kern-\nulldelimiterspace}
 {{X_{{\bf{r'}}}^{\left( {m'p'} \right)}}} \right\rangle } \right\rangle _\omega }$ includes the high-order Green's functions. The dependence of quasiparticle energies $\Omega \left( {q_j} \right)$ on the $c$-axis pressure magnitude $P^{(c)}$ is shown in Fig.~\ref{fig:paronPc}b: the energy of the quasiparticle $q_2$ increases and of the quasiparticles $q_3$, $q_4$, $q_5$ energies decrease with increasing pressure.   
 
 The system of equations (\ref{eq:eq_motion}) for all components of the matrix Green's function ${\hat D}_\sigma\left( {{\bf{r}},{\bf{r'}};\omega } \right)$ is decoupled within the generalized mean-field approximation using the Mori-Zwanzig projection technique~\cite{Plakida2003,Dzebisashvili05}. In this decoupling method, the operators $Z_{\bf{r}}^{\left( {pm} \right)}$ are represented as the sum of reducible and irreducible parts. The reducible part can be linearized using the X-operators for the quasiparticle basis (\ref{eq:q1_q5}):
\begin{eqnarray}\label{eq:red_Z}
&&Z_{{\bf{r}}}^{\left( {pm} \right)} =  \\\nonumber
&&\sum\limits_{{\bf{s}}ln} {\left( {{T_{{\bf{rs}}}}\left( {pm,nl} \right)X_{{\bf{s}}}^{ln} + {\Delta _{{\bf{rs}}}}\left( {pm,ln} \right)\mathop {X_{{\bf{s}}}^{ln}}\limits^\dag  } \right)}+Z_{{\bf{r}}}^{\left( {pm} \right)\left( {irr} \right)}
\end{eqnarray}
where ${T_{{\bf{rs}}}}\left( {pm,nl} \right) = {{\left\langle {\left\{ {Z_{\bf{r}}^{\left( {pm} \right)},X_{\bf{s}}^{nl}} \right\}} \right\rangle } \mathord{\left/
 {\vphantom {{\left\langle {\left\{ {Z_{\bf{r}}^{\left( {pm} \right)},X_{\bf{s}}^{nl}} \right\}} \right\rangle } {\left\langle {\left\{ {{\rm X}_{\bf{s}}^{ln},X_{\bf{s}}^{nl}} \right\}} \right\rangle }}} \right.
 \kern-\nulldelimiterspace} {\left\langle {\left\{ {{\rm X}_{\bf{s}}^{ln},X_{\bf{s}}^{nl}} \right\}} \right\rangle }}$ - normal, ${\Delta _{{\bf{rs}}}}\left( {pm,ln} \right) = {{\left\langle {\left\{ {Z_{\bf{r}}^{\left( {pm} \right)},X_{\bf{s}}^{ln}} \right\}} \right\rangle } \mathord{\left/
 {\vphantom {{\left\langle {\left\{ {Z_{\bf{r}}^{\left( {pm} \right)},X_{\bf{s}}^{ln}} \right\}} \right\rangle } {\left\langle {\left\{ {X_{\bf{s}}^{ln},{\rm X}_{\bf{s}}^{nl}} \right\}} \right\rangle }}} \right.
 \kern-\nulldelimiterspace} {\left\langle {\left\{ {X_{\bf{s}}^{ln},{\rm X}_{\bf{s}}^{nl}} \right\}} \right\rangle }}$ - anomalous linearization coefficients. Thus, the Green's function $\hat D\left( {{\bf{r}},{\bf{r'}};\omega } \right)$ is expressed in terms of the normal and anomalous Green's functions. The terms ${T_{{\bf{rs}}}}\left( {pm,nl} \right)$ define the self-energy operator, they contain hoppings, kinematic and spin-spin correlation functions:
\begin{widetext}
\begin{eqnarray}
&&{T_{\sigma {\bf{k}}}}\left( {\sigma 0,0\sigma } \right) = F\left( {0\sigma } \right){t_{\bf{k}}}\left( {\sigma 0,0\sigma } \right) +\\\nonumber
&&\frac{1}{{NF\left( {0\sigma } \right)}}\sum\limits_{\bf{q}} {\left[ {{t_{\bf{q}}}\left( {\bar \sigma 0,0\bar \sigma } \right){K_{\bf{q}}}\left( {\bar \sigma 0,0\bar \sigma } \right) + \sum\limits_{\sigma ''L''} {{t_{\bf{q}}}\left( {\bar \sigma 0,\sigma ''L''} \right){K_{\bf{q}}}\left( {\sigma '0,\sigma ''L''} \right)} } \right]}  + \\\nonumber
&&\frac{1}{{NF\left( {0\sigma } \right)}}\sum\limits_{{\bf{q}}\sigma 'L''} {\left[ {\sum\limits_{L'} {{t_{\bf{q}}}\left( {L'\sigma ',\sigma L''} \right){K_{\bf{q}}}\left( {L'\sigma ',\sigma L''} \right)}  - {t_{\bf{q}}}\left( {\sigma '0,\sigma L''} \right){K_{\bf{q}}}\left( {\sigma '0,\sigma L''} \right)} \right]}  + \\\nonumber
&&\frac{1}{N}\sum\limits_{\bf{q}} {\frac{1}{{F\left( {0\sigma } \right)}}{t_{{\bf{k}} - {\bf{q}}}}\left( {\bar \sigma 0,0\bar \sigma } \right){C_{\bf{q}}}\left( {\bar \sigma \sigma ,\sigma \bar \sigma } \right)},\\
\,\,  \\\nonumber
&&{T_{\sigma {\bf{k}}}}\left( {\sigma 0,{\sigma _1}{L_1}} \right) = F\left( {0\sigma } \right){t_{\bf{k}}}\left( {\sigma 0,{\sigma _1}{L_1}} \right) +\\\nonumber
&& \frac{1}{{NF\left( {{\sigma _1}{L_1}} \right)}}\sum\limits_{\bf{q}} {\left\{ {{t_{\bf{q}}}\left( {{\sigma _1}0,0{\sigma _1}} \right){K_{\bf{q}}}\left( {{L_1}\sigma ,0{\sigma _1}} \right) - \sum\limits_{\sigma 'L'} {{t_{\bf{q}}}\left( {L'\sigma ',\sigma {L_1}} \right){K_{\bf{q}}}\left( {L'\sigma ',0{\sigma _1}} \right)} } \right.}  + \\\nonumber
&& \sum\limits_{\sigma 'L''} {{t_{\bf{q}}}\left( {\sigma '0,\sigma L''} \right){K_{\bf{q}}}\left( {\sigma '0,0{\sigma _1}} \right)}  + \sum\limits_{\sigma ''L''} {{t_{\bf{q}}}\left( {{\sigma _1}0,\sigma ''L''} \right){K_{\bf{q}}}\left( {{L_1}\sigma ,\sigma ''L''} \right)}  + \\\nonumber
&& \left. {\sum\limits_{\sigma ' \ne \sigma ,\sigma '' \ne {\sigma _1}} {{t_{{\bf{k}} - {\bf{q}}}}\left( {\sigma '0,\sigma ''{L_1}} \right){C_{\bf{q}}}\left( {\sigma '\sigma ,\sigma ''{\sigma _1}} \right)} } \right\},\\
\,\, \\\nonumber
&&{T_{\sigma {\bf{k}}}}\left( {L{\sigma _0},0{\sigma _1}} \right) = F\left( {{\sigma _0}L} \right){t_{\bf{k}}}\left( {L{\sigma _0},0{\sigma _1}} \right) + \\\nonumber
&&\frac{1}{{NF\left( {0{\sigma _1}} \right)}}\sum\limits_{\bf{q}} {\left\{ {{t_{\bf{q}}}\left( {{\sigma _0}0,0{\sigma _0}} \right){K_{\bf{q}}}\left( {{\sigma _0}0,{\sigma _1}L} \right) - \sum\limits_{\sigma ''L''} {{t_{\bf{q}}}\left( {L{\sigma _1},\sigma ''L''} \right){K_{\bf{q}}}\left( {{\sigma _0}0,\sigma ''L''} \right)} } \right.}  + \\\nonumber
&& \sum\limits_{\sigma 'L'} {{t_{\bf{q}}}\left( {L'\sigma ',0{\sigma _0}} \right){K_{\bf{q}}}\left( {L'\sigma ',{\sigma _1}L} \right)}  - \sum\limits_{\sigma ''} {{t_{\bf{q}}}\left( {L{\sigma _1},0\sigma ''} \right){K_{\bf{q}}}\left( {{\sigma _0}0,0\sigma ''} \right)}  + \\\nonumber
&&\left. {\sum\limits_{\sigma ' \ne {\sigma _0},\sigma '' \ne {\sigma _1}} {{t_{{\bf{k}} - {\bf{q}}}}\left( {L\sigma ',0\sigma ''} \right){C_{\bf{q}}}\left( {{\sigma _0}\sigma ',{\sigma _1}\sigma ''} \right)} } \right\},\\
\,\,  \\\nonumber
&&{T_{\sigma {\bf{k}}}}\left( {L{\sigma _0},{\sigma _1}{L_1}} \right) = F\left( {{\sigma _0}L} \right){t_{\bf{k}}}\left( {L{\sigma _0},{\sigma _1}{L_1}} \right) + \\\nonumber
&&\frac{1}{{NF\left( {{\sigma _1}{L_1}} \right)}}\sum\limits_{\bf{q}} {\left\{ {{\delta _{{\sigma _0}{\sigma _1}}}\sum\limits_{\sigma '\sigma ''} {\left[ {\sum\limits_{L''} {{t_{\bf{q}}}\left( {L\sigma ',\sigma ''L''} \right){K_{\bf{q}}}\left( {{L_1}\sigma ',\sigma ''L''} \right)}  - {t_{\bf{q}}}\left( {L\sigma ',0\sigma ''} \right){K_{\bf{q}}}\left( {{L_1}\sigma ',0\sigma ''} \right)} \right]} } \right.}  - \\\nonumber
&& {\delta _{L{L_1}}}\sum\limits_{L'} {\left[ {\sum\limits_{\sigma ''L''} {{t_{\bf{q}}}\left( {L'{\sigma _0},\sigma ''L''} \right){K_{\bf{q}}}\left( {L'{\sigma _1},\sigma ''L''} \right)}  + \sum\limits_{\sigma 'L'} {{t_{\bf{q}}}\left( {L'\sigma ',0{\sigma _0}} \right){K_{\bf{q}}}\left( {L'\sigma ',0{\sigma _1}} \right)} } \right]}  + \\\nonumber
&& {\delta _{L{L_1}}}\sum\limits_{L'\sigma ''} {{t_{\bf{q}}}\left( {L'{\sigma _0},0\sigma ''} \right){K_{\bf{q}}}\left( {L'{\sigma _1},0\sigma ''} \right)}  - {t_{\bf{q}}}\left( {{\sigma _0}0,0{\sigma _0}} \right){K_{\bf{q}}}\left( {{\sigma _0}0,0{\sigma _1}} \right) + \\\nonumber
&& \,\left. {\sum\limits_{\sigma ' \ne {\sigma _0},\sigma '' \ne {\sigma _1}} {{t_{{\bf{k}} - {\bf{q}}}}\left( {L\sigma ',\sigma ''{L_1}} \right){C_{\bf{q}}}\left( {{\sigma _0}\sigma ',\sigma ''{\sigma _1}} \right)} } \right\}.
\label{eq:kinenergy2}
\end{eqnarray}
\end{widetext}
The kinematic correlation functions ${K_{{\bf{fg}}}} = \left\langle {X_{\bf{f}}^{\sigma 0}X_{\bf{g}}^{0\sigma }} \right\rangle $, $\left\langle {X_{\bf{f}}^{\sigma 0}X_{\bf{g}}^{\sigma 'L}} \right\rangle $, $\left\langle {X_{\bf{f}}^{L\sigma '}X_{\bf{g}}^{0\sigma }} \right\rangle $, $\left\langle {X_{\bf{f}}^{L\sigma }X_{\bf{g}}^{\sigma 'L}} \right\rangle $, where $L$ runs over the five two-hole eigenstates in (\ref{eq:eigenstates}), are obtained from the corresponding Green's functions according with the spectral theorem. The values of the spin-spin correlation functions ${C_{{\bf{fg}}}} = \left\langle {X_{\bf{f}}^{\sigma \bar \sigma }X_{\bf{g}}^{\bar \sigma \sigma }} \right\rangle $ were calculated within the framework of the t-J* model in~\cite{Dzebisashvili05,Korshunov07} taking into account only ground two-hole singlet state $\left| {A^S} \right\rangle $. The spin-spin correlator values may change if the triplet states are taken into account. It was shown in~\cite{Gavrichkov2008} that the overwhelming contribution to the effective exchange interaction in the CuO$_2$ plane comes from hoppings of the excitations involving the singlet state $\left| {A^S} \right\rangle $; this contribution is of the antiferromagnetic type. The contribution to the superexchange interaction that is caused by the hoppings of the quasiparticles involving the triplet states is only 3$\%$. Accordingly, the influence of the exchange interaction caused by the triplet states on spin correlations will be small. These conclusions are valid even if the triplet state becomes the ground state~\cite{Gavrichkov2008}.

We obtain the Gor'kov equations in $k$-space neglecting the irreducible operator $Z_{\bf{r}}^{\left( {pm} \right)\left( {irr} \right)}$:
\begin{eqnarray}\label{eq:Gorkov} 
&&\left[ {\begin{array}{*{20}{c}}
{\omega \hat I}&0\\
0&{\omega \hat I}
\end{array}} \right]{\hat D}_\sigma\left( {{\bf{k}};\omega } \right) = \left[ {\begin{array}{*{20}{c}}
{{{\hat F}_\sigma }}&0\\
0&{{{\hat F}_{\bar \sigma }}}
\end{array}} \right] + \\\nonumber
&&\left[ {\begin{array}{*{20}{c}}
{{{\hat \Omega }_\sigma }}&0\\
0&{ - {{\hat \Omega }_{\bar \sigma }}}
\end{array}} \right]{{\hat D}_\sigma}\left( {{\bf{k}};\omega } \right) + \left[ {\begin{array}{*{20}{c}}
{{{\hat T}_{\sigma {\bf{k}}}}}&{{{\hat \Delta }_{\bf{k}}}}\\
{\hat \Delta _\sigma^\dag }&{ - {{\hat T}_{\bar \sigma {\bf{k}}}}}
\end{array}} \right]{{\hat D}_\sigma}\left( {{\bf{k}};\omega } \right)
\end{eqnarray}
where ${\hat I}$ is the identity $5\times5$ matrix,
\begin{eqnarray}\label{eq:Omega}
&&\hat \Omega_\sigma  = \\\nonumber
&&diag\left( {\Omega \left( {0\sigma } \right),\Omega \left( {\bar \sigma A^S} \right),\Omega \left( {\bar \sigma B^{T0}} \right),\Omega \left( {\sigma B^{T2\sigma}} \right),\Omega \left( {\bar \sigma B^S} \right)} \right)
\end{eqnarray}
is the diagonal matrix of quasiparticle energies,
\begin{eqnarray}\label{eq:matr_F} 
&&{\hat F_\sigma } = \\\nonumber
&&diag\left( {F\left( {0\sigma } \right),F\left( {\bar \sigma A^S} \right),F\left( {\bar \sigma B^{T0}} \right),F\left( {\sigma B^{T2\sigma}} \right),F\left( {\bar \sigma B^S} \right)} \right)
\end{eqnarray}
is the diagonal matrix of filling factors, and ${{\hat T}_{\sigma {\bf{k}}}}$ is the matrix form of the self-energy operator with the elements $T_{\sigma {\bf{k}}}\left( {pm,m'p'} \right)$. The dispersion of quasipaticle excitations is defined by the poles of the Green's function ${\hat G_\sigma }\left( {{\bf{k}};\omega } \right)$ calculated from the Dyson equation:
\begin{equation}\label{eq:Dyson} 
{{\hat G}_\sigma }\left( {{\bf{k}};\omega } \right) = {\left( {\omega \hat I - {{\hat \Omega }_\sigma } - {{\hat F}_\sigma }{{\hat T}_{\sigma {\bf{k}}}}} \right)^{ - 1}},
\end{equation}
and their spectral weight $A_\sigma\left( {{\bf{k}},\omega  } \right) = \sum\limits_{\lambda  } {{A_{\lambda \sigma }}\left( {{\bf{k}},\omega } \right)}$ is obtained using spectral theorem:
\begin{eqnarray}\label{eq:Akwlam} 
&&{A_{\lambda \sigma }}\left( {{\bf{k}},\omega  + i\delta } \right) = \left( { - \frac{1}{\pi }} \right) \times \\\nonumber
&&\sum\limits_{pp'mm'} {{\gamma _{\lambda \sigma }}\left( {pm} \right)\gamma _{\lambda \sigma }^ * \left( {m'p'} \right){\mathop{\rm Im}\nolimits} {{\left\langle {\left\langle {{X_{\bf{k}}^{pm}}}
 \mathrel{\left | {\vphantom {{X_{\bf{k}}^{pm}} {\mathop {X_{\bf{k}}^{m'p'}}\limits^\dag  }}}
 \right. \kern-\nulldelimiterspace}
 {{\mathop {X_{\bf{k}}^{m'p'}}\limits^\dag  }} \right\rangle } \right\rangle }_{\omega  + i\delta }}} , 
\end{eqnarray}
where ${{A_{\lambda \sigma }}}$ is the contribution from certain orbital $\lambda  = {d_x}$, $b$, ${p_z}$, ${d_z}$, $a$.

The linearization coefficients ${\Delta _{{\bf{rs}}}}\left( {pm,nl} \right)$ determine the components of the superconducting gap. Among the components of the superconducting gap, we can distinguish those that are caused by pairing of quasiparticles

1) within the upper Hubbard band (UHB)
\begin{eqnarray}\label{eq:gap11}
&&\Delta _{{\bf{rs}}}^{11} = {\Delta _{{\bf{rs}}}}\left( {0\sigma ,0\bar \sigma } \right) =\\\nonumber
&& \frac{1}{{{F}\left( {0\bar \sigma } \right)}}\left[ {\sum\limits_{L'} {{t_{{\bf{rs}}}}\left( {\sigma 0,\bar \sigma L'} \right)\left\langle {\left( {X_{\bf{r}}^{00} + X_{\bf{r}}^{\sigma \sigma }} \right)X_{\bf{s}}^{0L'}} \right\rangle } } \right. - \\\nonumber
&&\left. {\sum\limits_{L'} {{t_{{\bf{sr}}}}\left( {\bar \sigma 0,\sigma L'} \right)\left\langle {\left( {X_{\bf{s}}^{00} + X_{\bf{s}}^{\bar \sigma \bar \sigma }} \right)X_{\bf{r}}^{0L'}} \right\rangle } } \right]
\end{eqnarray}

2) within the lower Hubbard band (LHB)
\begin{eqnarray}\label{eq:gap22}
&&\Delta _{{\bf{rs}}}^{jj'} = {\Delta _{{\bf{rs}}}}\left( {{\sigma _0}L,{\sigma _1}{L_1}} \right) = \\\nonumber
&&\frac{1}{{{F}\left( {{\sigma _1}{L_1}} \right)}}\left[ {{t_{{\bf{js}}}}\left( {L{\sigma _0},0{\sigma _1}} \right)\left\langle {\left( {X_{\bf{r}}^{{\sigma _0}{\sigma _0}} + X_{\bf{r}}^{LL}} \right)X_{\bf{s}}^{0{L_1}}} \right\rangle } \right. - \\\nonumber
&&\left. {{t_{{\bf{sr}}}}\left( {{L_1}{\sigma _1},0{\sigma _0}} \right)\left\langle {\left( {X_{\bf{s}}^{{\sigma _1}{\sigma _1}} + X_{\bf{s}}^{{L_1}{L_1}}} \right)X_{\bf{r}}^{0L}} \right\rangle } \right]
\end{eqnarray}
and

3) between the UHB and LHB
\begin{eqnarray}\label{eq:gap12}
&&\Delta _{{\bf{rs}}}^{1j'} = {\Delta _{{\bf{rs}}}}\left( {0\sigma ,{\sigma _1}{L_1}} \right) = \\\nonumber
&&\frac{1}{{{F}\left( {\sigma {L_1}} \right)}}\left[ {{t_{{\bf{rs}}}}\left( {\sigma 0,0\sigma } \right)\left\langle {\left( {X_{\bf{r}}^{00} + X_{\bf{r}}^{\sigma \sigma }} \right)X_{\bf{s}}^{0{L_1}}} \right\rangle } \right. - \\\nonumber
&&\left. {\sum\limits_{L'} {{t_{{\bf{sr}}}}\left( {{L_1}{\sigma _1},\sigma L'} \right)\left\langle {\left( {X_{\bf{s}}^{{\sigma _1}{\sigma _1}} + X_{\bf{s}}^{{L_1}{L_1}}} \right)X_{\bf{r}}^{0L'}} \right\rangle } } \right],
\end{eqnarray}

\begin{eqnarray}\label{eq:gap21}
&&\Delta _{{\bf{rs}}}^{j1} = {\Delta _{{\bf{rs}}}}\left( {{\sigma _0}L,0\bar \sigma } \right) = \\\nonumber
&&\frac{1}{{{F}\left( {0\bar \sigma } \right)}}\left[ {\sum\limits_{L'} {{t_{{\bf{rs}}}}\left( {L{\sigma _0},\bar \sigma L'} \right)\left\langle {\left( {X_{\bf{r}}^{{\sigma _0}{\sigma _0}} + X_{\bf{r}}^{LL}} \right)X_{\bf{s}}^{0L'}} \right\rangle } } \right. - \\\nonumber
&&\left. {\frac{1}{{{F_{\bf{s}}}\left( {0{\sigma _0}} \right)}}{t_{{\bf{sr}}}}\left( {{\sigma _0}0,0{\sigma _0}} \right)\left\langle {\left( {X_{\bf{s}}^{00} + X_{\bf{s}}^{{\sigma _0}{\sigma _0}}} \right)X_{\bf{r}}^{0L}} \right\rangle } \right]
\end{eqnarray}
where 
\begin{eqnarray}\label{eq:sL_sL}
&&\left( {{\sigma _0}L} \right) = \left\{ {\bar \sigma A_1^S,\bar \sigma B_1^{T0},\sigma B_1^{T2\sigma},\bar \sigma B_1^S} \right\},\\\nonumber
&&\left( {{\sigma _1}{L_1}} \right) = \left\{ {\sigma A_1^S,\sigma B_1^{T0},\bar \sigma B_1^{T2{\bar \sigma}},\sigma B_1^S} \right\}
\end{eqnarray}
and index $j\left( {j'} \right)=2-5$ enumerates quasiparticle excitations with spin projection ${\sigma}$(${\bar \sigma}$) between single-hole and two-hole eigenstates. In order to calculate averages $\left\langle {X_{\bf{r}}^{pm}X_{\bf{s}}^{ln}} \right\rangle $, we use the same approach which was applied in the work~\cite{Plakida2003} to take into account the exchange pairing mechanism for the two-band Hubbard model. This approach consists of the next steps: (i) to write the equation of motion for the corresponding Green's functions, for example, $\left\langle {\left\langle {{X_{\bf{s}}^{0L'}}}
 \mathrel{\left | {\vphantom {{X_{\bf{s}}^{0L'}} {\left( {X_{\bf{r}}^{00} + X_{\bf{r}}^{\sigma \sigma }} \right)}}}
 \right. \kern-\nulldelimiterspace}
 {{\left( {X_{\bf{r}}^{00} + X_{\bf{r}}^{\sigma \sigma }} \right)}} \right\rangle } \right\rangle $, (ii) to neglect interband transitions and pairing within UHB, (iii) to use the two-site approximation to express an average of the type $\left\langle {\left( {X_{\bf{r}}^{00} + X_{\bf{r}}^{\sigma \sigma }} \right)X_{\bf{s}}^{0L'}} \right\rangle $ and $\left\langle {\left( {X_{\bf{r}}^{{\sigma _0}{\sigma _0}} + X_{\bf{r}}^{LL}} \right)X_{\bf{s}}^{0L'}} \right\rangle $  in terms of the averages for the quasiparticle pairing within the LHB ${V_{{\bf{rs}}}}\left( {\sigma L'',\sigma ''L'} \right) = \left\langle {X_{\bf{r}}^{\sigma L''}X_{\bf{s}}^{\sigma ''L'}} \right\rangle $: 
\begin{eqnarray}\label{eq:averages}
&&\left\langle {\left( {X_{\bf{r}}^{00} + X_{\bf{r}}^{\sigma \sigma }} \right)X_{\bf{s}}^{0L'}} \right\rangle  = \\\nonumber
&&\frac{1}{{{\Theta _{L'}}}}\sum\limits_{{\sigma ''}{L''}} {{t_{{\bf{sr}}}}\left( {{\sigma ''}0,\sigma {L''}} \right)\left\langle {X_{\bf{r}}^{\sigma {L''}}X_{\bf{s}}^{{\sigma ''}L'}} \right\rangle } 
\end{eqnarray}
\begin{eqnarray}\label{eq:averages2}
&&\left\langle {\left( {X_{\bf{r}}^{{\sigma _0}{\sigma _0}} + X_{\bf{r}}^{LL}} \right)X_{\bf{s}}^{0L'}} \right\rangle = \\\nonumber
&&\frac{1}{{{\Theta _{L'}}}}\sum\limits_{\sigma ''L''} {{t_{{\bf{sr}}}}\left( {\sigma ''0,{\sigma _0}L''} \right)\left\langle {X_{\bf{r}}^{{\sigma _0}L''}X_{\bf{s}}^{\sigma ''L'}} \right\rangle }
\end{eqnarray}
where the parameter ${\Theta_{L'}} =  - \left( {{\varepsilon _{2L'}} - 2{\varepsilon _{1\sigma }}} \right)$ (the minus sign in front of the whole expression appears in the electron representation while the energies ${\varepsilon _{m}}$ are in the hole representation) plays a role of the Coulomb interaction in the exchange constant of the t-J model\cite{Chao77}. Since the overwhelming majority of experiments indicate that the superconducting gap in HTSC cuprates has d-symmetry~\cite{VanHarlingen95,Tsuei00,Damascelli2003}, we consider the following form of the superconducting gap components $ij$:
\begin{eqnarray}\label{eq:gen_sc_gap} 
\Delta_{\bf{k}}^{ij}  = {\Delta _{0d}^{ij}}\left( {\cos \left( {{k_x}} \right) - \cos \left( {{k_y}} \right)} \right)
\end{eqnarray}

The expressions (\ref{eq:gap11})-(\ref{eq:gap21}) are obtained when we neglect the terms caused by kinematic pairing mechanism~\cite{Zaitsev87,Zaitsev88,Zaitsev88_2} which contribute to the s-wave symmetry gap and do not contribute to the d-wave gap~\cite{Plakida2003}.
The expressions (\ref{eq:gap11})-(\ref{eq:gap21}) can be simplified to calculate the gap amplitudes ${\Delta _{0d}^{ij}}$. We multiply both parts of each equation (\ref{eq:gap11})-(\ref{eq:gap21}) by $\left( {\cos \left( {{k_x}} \right) - \cos \left( {{k_y}} \right)} \right)$ and integrate $\int\limits_0^{2\pi } {\int\limits_0^{2\pi } {d{k_x}d{k_y}} } $. The expressions for the superconducting gap components in the nearest neighbors approximation after the Fourier transformation will take the form:
\widetext
\begin{eqnarray}\label{eq:d11_q}
&&\Delta _{0d}^{11}\equiv\Delta _{0d}\left( {0\sigma ,\bar \sigma 0} \right) = \frac{2}{N}\sum\limits_{{\bf{q}}\sigma 'L'\sigma ''L''} {\left( {{\delta _{\sigma '\sigma }} - {\delta _{\sigma '{\sigma _1}}}} \right)\frac{{{V_{\bf{q}}}\left( {\sigma 'L',\sigma ''L''} \right)}}{{F\left( {0{\sigma _1}} \right)}} \times } \\\nonumber
&&\left[ {J_{10}^{\left(  +  \right)}\left( {\sigma '0,{\sigma _1}L'';\sigma ''0,\sigma 'L'} \right)\left( {\cos \left( {{q_x}} \right) - \cos \left( {{q_y}} \right)} \right) + J_{10}^{\left(  -  \right)}\left( {\sigma '0,{\sigma _1}L'';\sigma ''0,\sigma 'L'} \right)\left( {\cos \left( {{q_x}} \right) + \cos \left( {{q_y}} \right)} \right)} \right],
\end{eqnarray}

\begin{eqnarray}\label{eq:d22_q}
&&\Delta _{0d}^{jj'}\equiv\Delta _{0d}\left( {{\sigma _0}L,{\sigma _1}{L_1}} \right) = \\\nonumber
&&\frac{2}{N}\sum\limits_{{\bf{q}}\sigma 'L'\sigma ''L''\sigma '''L'''} {\left( {{\delta _{L''L}}{\delta _{L'''{L_1}}}{\delta _{\sigma '{\sigma _1}}}{\delta _{\sigma ''{\sigma _0}}} - {\delta _{L''{L_1}}}{\delta _{L'''L}}{\delta _{\sigma '{\sigma _0}}}{\delta _{\sigma ''{\sigma _1}}}} \right)\frac{{{V_{\bf{q}}}\left( {\sigma ''L',\sigma '''L'''} \right)}}{{F\left( {{\sigma _1}{L_1}} \right)}}}  \times  \\\nonumber
&&\left[ {J_{10}^{\left(  +  \right)}\left( {0\sigma ',\sigma ''L'';0\sigma ''',\sigma ''L'} \right)\left( {\cos \left( {{q_x}} \right) - \cos \left( {{q_y}} \right)} \right) + J_{10}^{\left(  -  \right)}\left( {0\sigma ',\sigma ''L'';0\sigma ''',\sigma ''L'} \right)\left( {\cos \left( {{q_x}} \right) + \cos \left( {{q_y}} \right)} \right)} \right],
\end{eqnarray}

\begin{eqnarray}\label{eq:d12_q}
&&\Delta _{0d}^{1j'} \equiv {\Delta _{0d}}\left( {0\sigma ,{\sigma _1}{L_1}} \right) = \frac{1}{N}\sum\limits_{{\bf{q}}\sigma ''L'} {\left\{ {{\delta _{\sigma {\sigma _1}}}\frac{{{B_{\bf{q}}}\left( {\sigma L',\sigma ''{L_1}} \right)}}{{{F_{\bf{s}}}\left( {{\sigma _1}{L_1}} \right)}} \times } \right.} \\\nonumber
&&\left[ {J_{10}^{\left(  +  \right)}\left( {\sigma 0,0\sigma ;\sigma ''0,\sigma L'} \right)\left( {\cos \left( {{q_x}} \right) - \cos \left( {{q_q}} \right)} \right) + J_{10}^{\left(  -  \right)}\left( {\sigma 0,0\sigma ;\sigma ''0,\sigma L'} \right)\left( {\cos \left( {{q_x}} \right) + \cos \left( {{q_q}} \right)} \right)} \right] - \\\nonumber
&&\sum\limits_{L''} {\frac{{{B_{\bf{q}}}\left( {{\sigma _1}L',\sigma ''L''} \right)}}{{{F_{\bf{s}}}\left( {{\sigma _1}{L_1}} \right)}}}  \times \\\nonumber
&&\left. {\left[ {J_{10}^{\left(  +  \right)}\left( {{\sigma _1}{L_1},\sigma L'';\sigma ''0,{\sigma _1}L'} \right)\left( {\cos \left( {{q_x}} \right) - \cos \left( {{q_y}} \right)} \right) + J_{10}^{\left(  -  \right)}\left( {{\sigma _1}{L_1},\sigma L'';\sigma ''0,{\sigma _1}L'} \right)\left( {\cos \left( {{q_x}} \right) + \cos \left( {{q_y}} \right)} \right)} \right]} \right\},
\end{eqnarray}

\begin{eqnarray}\label{eq:d21_q}
&&\Delta _{0d}^{j'1} = {\Delta _{0d}}\left( {{\sigma _0}L,0{\sigma _1}} \right) = \frac{2}{N}\sum\limits_{{\bf{q}}\sigma 'L'} {\left\{ {\sum\limits_{L''} {\frac{{{B_{\bf{q}}}\left( {{\sigma _0}L',\sigma 'L''} \right)}}{{F\left( {0{\sigma _1}} \right)}}} } \right.}  \times \\\nonumber
&&\left[ {J_{10}^{\left(  +  \right)}\left( {{\sigma _0}L,{\sigma _1}L'';\sigma '0,{\sigma _0}L'} \right)\left( {\cos \left( {{q_x}} \right) - \cos \left( {{q_q}} \right)} \right) + J_{10}^{\left(  -  \right)}\left( {{\sigma _0}L,{\sigma _1}L'';\sigma '0,{\sigma _0}L'} \right)\left( {\cos \left( {{q_x}} \right) + \cos \left( {{q_q}} \right)} \right)} \right] - \\\nonumber
&&{\delta _{{\sigma _1}{\sigma _0}}}\frac{{{B_{\bf{q}}}\left( {{\sigma _0}L',\sigma 'L} \right)}}{{F\left( {0{\sigma _1}} \right)}} \times \\\nonumber
&&\left. {\left[ {J_{10}^{\left(  +  \right)}\left( {{\sigma _0}0,0{\sigma _0};\sigma '0,{\sigma _0}L'} \right)\left( {\cos \left( {{q_x}} \right) - \cos \left( {{q_y}} \right)} \right) + J_{10}^{\left(  -  \right)}\left( {{\sigma _0}0,0{\sigma _0};\sigma '0,{\sigma _0}L'} \right)\left( {\cos \left( {{q_x}} \right) + \cos \left( {{q_y}} \right)} \right)} \right]} \right\},
\end{eqnarray}
\endwidetext
where the terms 
\begin{eqnarray}
&&J_{10}^{\left(  \pm  \right)}\left( {{\beta _1},{\beta _2};{\beta _3},{\beta _4}} \right) = \frac{1}{{{\Theta _{L'}}}} \times \\\nonumber
&&\left( {t_{10}^{\left(  +  \right)}\left( {{\beta _1},{\beta _2}} \right)t_{10}^{\left(  +  \right)}\left( {{\beta _3},{\beta _4}} \right) \pm t_{10}^{\left(  -  \right)}\left( {{\beta _1},{\beta _2}} \right)t_{10}^{\left(  -  \right)}\left( {{\beta _3},{\beta _4}} \right)} \right)
\label{eq:J_pm}
\end{eqnarray}
are the superexchange parameters which are defined as the ratio between the product of quasiparticle hopping integrals and the parameter $\Theta_L'$. The parameter ${\Theta_{L'}} =  - \left( {{\varepsilon _{2L'}} - 2{\varepsilon _{1\sigma }}} \right)$ (the minus sign in front of the whole expression appears in the electron representation while the energies ${\varepsilon _{m}}$ are in the hole representation) plays a role of the Coulomb interaction in the exchange constant of t-J model\cite{Chao77}. The hopping integrals between nearest neighbors $t_{10}^{\left(  \pm  \right)}\left( {\beta ,\beta '} \right)$ have the form:
\begin{equation}
t_{10}^{\left(  \pm  \right)}\left( {\beta ,\beta '} \right) = t_{10}^{bb}\left( {\beta ,\beta '} \right) + t_{10}^{aa}\left( {\beta ,\beta '} \right) \pm t_{10}^{ba}\left( {\beta ,\beta '} \right).
\label{eq:tba_pm}
\end{equation}
The magnitudes of the hopping integrals of quasiparticles ${q_1}$ and ${q_2}$ having the nature of orbitals of ${b_{1g}}$ symmetry are 6-15 times greater than those of quasiparticles ${q_3}$, ${q_4}$, ${q_5}$ with the nature of orbitals of ${a_{1g}}$ symmetry. The characteristic intraband exchange parameters having the largest magnitudes are
\begin{eqnarray}\label{eq:J22}
&&{J^{22}} \equiv J_{10}^{\left(  +  \right)}\left( {0\sigma ,\bar \sigma A_1^S;0\bar \sigma ,\sigma A_1^S} \right) \sim 0.1,\\\nonumber
&&{J^{33}} \equiv J_{10}^{\left(  +  \right)}\left( {0\sigma ,\bar \sigma B_1^{T0};0\bar \sigma ,\sigma B_1^{T0}} \right) \sim 0.01,\\\nonumber
&&{J^{44}} \equiv J_{10}^{\left(  +  \right)}\left( {0\sigma ,\bar \sigma B_1^{T2\bar \sigma };0\bar \sigma ,\sigma B_1^{T2\sigma }} \right) \sim {\rm{0}}{\rm{.006}},\\\nonumber
&&{J^{55}} \equiv J_{10}^{\left(  +  \right)}\left( {0\sigma ,\bar \sigma B_1^S;0\bar \sigma ,\sigma B_1^{TS}} \right) \sim {\rm{0}}{\rm{.006}}.
\end{eqnarray}

\bibliography{refers_crossover_3}

\begin{thebibliography}{60}%
\makeatletter
\providecommand \@ifxundefined [1]{%
 \@ifx{#1\undefined}
}%
\providecommand \@ifnum [1]{%
 \ifnum #1\expandafter \@firstoftwo
 \else \expandafter \@secondoftwo
 \fi
}%
\providecommand \@ifx [1]{%
 \ifx #1\expandafter \@firstoftwo
 \else \expandafter \@secondoftwo
 \fi
}%
\providecommand \natexlab [1]{#1}%
\providecommand \enquote  [1]{``#1''}%
\providecommand \bibnamefont  [1]{#1}%
\providecommand \bibfnamefont [1]{#1}%
\providecommand \citenamefont [1]{#1}%
\providecommand \href@noop [0]{\@secondoftwo}%
\providecommand \href [0]{\begingroup \@sanitize@url \@href}%
\providecommand \@href[1]{\@@startlink{#1}\@@href}%
\providecommand \@@href[1]{\endgroup#1\@@endlink}%
\providecommand \@sanitize@url [0]{\catcode `\\12\catcode `\$12\catcode
  `\&12\catcode `\#12\catcode `\^12\catcode `\_12\catcode `\%12\relax}%
\providecommand \@@startlink[1]{}%
\providecommand \@@endlink[0]{}%
\providecommand \url  [0]{\begingroup\@sanitize@url \@url }%
\providecommand \@url [1]{\endgroup\@href {#1}{\urlprefix }}%
\providecommand \urlprefix  [0]{URL }%
\providecommand \Eprint [0]{\href }%
\providecommand \doibase [0]{https://doi.org/}%
\providecommand \selectlanguage [0]{\@gobble}%
\providecommand \bibinfo  [0]{\@secondoftwo}%
\providecommand \bibfield  [0]{\@secondoftwo}%
\providecommand \translation [1]{[#1]}%
\providecommand \BibitemOpen [0]{}%
\providecommand \bibitemStop [0]{}%
\providecommand \bibitemNoStop [0]{.\EOS\space}%
\providecommand \EOS [0]{\spacefactor3000\relax}%
\providecommand \BibitemShut  [1]{\csname bibitem#1\endcsname}%
\let\auto@bib@innerbib\@empty
\bibitem [{\citenamefont {Drozdov}\ \emph {et~al.}(2014)\citenamefont
  {Drozdov}, \citenamefont {Eremets},\ and\ \citenamefont
  {Troyan}}]{Drozdov2014}%
  \BibitemOpen
  \bibfield  {author} {\bibinfo {author} {\bibfnamefont {A.}~\bibnamefont
  {Drozdov}}, \bibinfo {author} {\bibfnamefont {M.}~\bibnamefont {Eremets}},\
  and\ \bibinfo {author} {\bibfnamefont {I.}~\bibnamefont {Troyan}},\
  }\href@noop {} {\bibfield  {journal} {\bibinfo  {journal} {Preprint at
  http://arXiv.org/abs/1412.0460}\ } (\bibinfo {year} {2014})}\BibitemShut
  {NoStop}%
\bibitem [{\citenamefont {Drozdov}\ \emph {et~al.}(2015)\citenamefont
  {Drozdov}, \citenamefont {Eremets}, \citenamefont {Troyan}, \citenamefont
  {Ksenofontov},\ and\ \citenamefont {Shylin}}]{Drozdov2015}%
  \BibitemOpen
  \bibfield  {author} {\bibinfo {author} {\bibfnamefont {A.}~\bibnamefont
  {Drozdov}}, \bibinfo {author} {\bibfnamefont {M.}~\bibnamefont {Eremets}},
  \bibinfo {author} {\bibfnamefont {I.}~\bibnamefont {Troyan}}, \bibinfo
  {author} {\bibfnamefont {V.}~\bibnamefont {Ksenofontov}},\ and\ \bibinfo
  {author} {\bibfnamefont {S.}~\bibnamefont {Shylin}},\ }\href
  {https://doi.org/10.1038/nature14964} {\bibfield  {journal} {\bibinfo
  {journal} {Nature}\ }\textbf {\bibinfo {volume} {525}},\ \bibinfo {pages}
  {73} (\bibinfo {year} {2015})}\BibitemShut {NoStop}%
\bibitem [{\citenamefont {Einaga}\ \emph {et~al.}(2016)\citenamefont {Einaga},
  \citenamefont {Sakata}, \citenamefont {Ishikawa}, \citenamefont {Shimizu},
  \citenamefont {Eremets}, \citenamefont {Drozdov}, \citenamefont {Troyan},
  \citenamefont {Hirao},\ and\ \citenamefont {Ohishi}}]{Einaga2016}%
  \BibitemOpen
  \bibfield  {author} {\bibinfo {author} {\bibfnamefont {M.}~\bibnamefont
  {Einaga}}, \bibinfo {author} {\bibfnamefont {M.}~\bibnamefont {Sakata}},
  \bibinfo {author} {\bibfnamefont {T.}~\bibnamefont {Ishikawa}}, \bibinfo
  {author} {\bibfnamefont {K.}~\bibnamefont {Shimizu}}, \bibinfo {author}
  {\bibfnamefont {M.}~\bibnamefont {Eremets}}, \bibinfo {author} {\bibfnamefont
  {A.}~\bibnamefont {Drozdov}}, \bibinfo {author} {\bibfnamefont
  {I.}~\bibnamefont {Troyan}}, \bibinfo {author} {\bibfnamefont
  {N.}~\bibnamefont {Hirao}},\ and\ \bibinfo {author} {\bibfnamefont
  {Y.}~\bibnamefont {Ohishi}},\ }\href {https://doi.org/10.1038/nphys3760}
  {\bibfield  {journal} {\bibinfo  {journal} {Nature Phys.}\ }\textbf {\bibinfo
  {volume} {12}},\ \bibinfo {pages} {835} (\bibinfo {year} {2016})}\BibitemShut
  {NoStop}%
\bibitem [{\citenamefont {Duan}\ \emph {et~al.}(2014)\citenamefont {Duan},
  \citenamefont {Liu}, \citenamefont {Tian}, \citenamefont {Li}, \citenamefont
  {Huang}, \citenamefont {Zhao}, \citenamefont {Yu}, \citenamefont {Liu},
  \citenamefont {Tian},\ and\ \citenamefont {Cui}}]{Duan2014}%
  \BibitemOpen
  \bibfield  {author} {\bibinfo {author} {\bibfnamefont {D.}~\bibnamefont
  {Duan}}, \bibinfo {author} {\bibfnamefont {Y.}~\bibnamefont {Liu}}, \bibinfo
  {author} {\bibfnamefont {F.}~\bibnamefont {Tian}}, \bibinfo {author}
  {\bibfnamefont {D.}~\bibnamefont {Li}}, \bibinfo {author} {\bibfnamefont
  {X.}~\bibnamefont {Huang}}, \bibinfo {author} {\bibfnamefont
  {Z.}~\bibnamefont {Zhao}}, \bibinfo {author} {\bibfnamefont {H.}~\bibnamefont
  {Yu}}, \bibinfo {author} {\bibfnamefont {B.}~\bibnamefont {Liu}}, \bibinfo
  {author} {\bibfnamefont {W.}~\bibnamefont {Tian}},\ and\ \bibinfo {author}
  {\bibfnamefont {T.}~\bibnamefont {Cui}},\ }\href
  {https://doi.org/10.1038/srep06968} {\bibfield  {journal} {\bibinfo
  {journal} {Sci. Rep.}\ }\textbf {\bibinfo {volume} {4}},\ \bibinfo {pages}
  {6968} (\bibinfo {year} {2014})}\BibitemShut {NoStop}%
\bibitem [{\citenamefont {Duan}\ \emph {et~al.}(2015)\citenamefont {Duan},
  \citenamefont {Huang}, \citenamefont {Tian}, \citenamefont {Li},
  \citenamefont {Yu}, \citenamefont {Liu}, \citenamefont {Ma}, \citenamefont
  {Liu},\ and\ \citenamefont {Cui}}]{Duan2015}%
  \BibitemOpen
  \bibfield  {author} {\bibinfo {author} {\bibfnamefont {D.}~\bibnamefont
  {Duan}}, \bibinfo {author} {\bibfnamefont {X.}~\bibnamefont {Huang}},
  \bibinfo {author} {\bibfnamefont {F.}~\bibnamefont {Tian}}, \bibinfo {author}
  {\bibfnamefont {D.}~\bibnamefont {Li}}, \bibinfo {author} {\bibfnamefont
  {H.}~\bibnamefont {Yu}}, \bibinfo {author} {\bibfnamefont {Y.}~\bibnamefont
  {Liu}}, \bibinfo {author} {\bibfnamefont {Y.}~\bibnamefont {Ma}}, \bibinfo
  {author} {\bibfnamefont {B.}~\bibnamefont {Liu}},\ and\ \bibinfo {author}
  {\bibfnamefont {T.}~\bibnamefont {Cui}},\ }\href
  {https://doi.org/10.1103/physrevb.91.180502} {\bibfield  {journal} {\bibinfo
  {journal} {Phys. Rev. B}\ }\textbf {\bibinfo {volume} {91}},\ \bibinfo
  {pages} {180502} (\bibinfo {year} {2015})}\BibitemShut {NoStop}%
\bibitem [{\citenamefont {Wang}\ \emph {et~al.}(2012)\citenamefont {Wang},
  \citenamefont {Tse}, \citenamefont {Tanaka}, \citenamefont {Iitaka},\ and\
  \citenamefont {Ma}}]{Wang12}%
  \BibitemOpen
  \bibfield  {author} {\bibinfo {author} {\bibfnamefont {H.}~\bibnamefont
  {Wang}}, \bibinfo {author} {\bibfnamefont {J.}~\bibnamefont {Tse}}, \bibinfo
  {author} {\bibfnamefont {K.}~\bibnamefont {Tanaka}}, \bibinfo {author}
  {\bibfnamefont {T.}~\bibnamefont {Iitaka}},\ and\ \bibinfo {author}
  {\bibfnamefont {Y.}~\bibnamefont {Ma}},\ }\href
  {https://doi.org/10.1073/pnas.1118168109} {\bibfield  {journal} {\bibinfo
  {journal} {Proc. Natl. Acad. Sci. USA}\ }\textbf {\bibinfo {volume} {109}},\
  \bibinfo {pages} {6463} (\bibinfo {year} {2012})}\BibitemShut {NoStop}%
\bibitem [{\citenamefont {Liu}\ \emph {et~al.}(2017)\citenamefont {Liu},
  \citenamefont {Naumov}, \citenamefont {Hoffmann}, \citenamefont {Ashcroft},\
  and\ \citenamefont {Hemley}}]{Liu2017}%
  \BibitemOpen
  \bibfield  {author} {\bibinfo {author} {\bibfnamefont {H.}~\bibnamefont
  {Liu}}, \bibinfo {author} {\bibfnamefont {I.}~\bibnamefont {Naumov}},
  \bibinfo {author} {\bibfnamefont {R.}~\bibnamefont {Hoffmann}}, \bibinfo
  {author} {\bibfnamefont {N.}~\bibnamefont {Ashcroft}},\ and\ \bibinfo
  {author} {\bibfnamefont {R.}~\bibnamefont {Hemley}},\ }\href
  {https://doi.org/10.1073/pnas.1704505114} {\bibfield  {journal} {\bibinfo
  {journal} {Proc. Natl. Acad. Sci. USA}\ }\textbf {\bibinfo {volume} {114}},\
  \bibinfo {pages} {6990} (\bibinfo {year} {2017})}\BibitemShut {NoStop}%
\bibitem [{\citenamefont {Peng}\ \emph {et~al.}(2017)\citenamefont {Peng},
  \citenamefont {Sun}, \citenamefont {Pickard}, \citenamefont {Needs},
  \citenamefont {Wu},\ and\ \citenamefont {Ma}}]{Peng2017}%
  \BibitemOpen
  \bibfield  {author} {\bibinfo {author} {\bibfnamefont {F.}~\bibnamefont
  {Peng}}, \bibinfo {author} {\bibfnamefont {Y.}~\bibnamefont {Sun}}, \bibinfo
  {author} {\bibfnamefont {C.}~\bibnamefont {Pickard}}, \bibinfo {author}
  {\bibfnamefont {R.}~\bibnamefont {Needs}}, \bibinfo {author} {\bibfnamefont
  {Q.}~\bibnamefont {Wu}},\ and\ \bibinfo {author} {\bibfnamefont
  {Y.}~\bibnamefont {Ma}},\ }\href
  {https://doi.org/10.1103/PhysRevLett.119.107001} {\bibfield  {journal}
  {\bibinfo  {journal} {Phys. Rev. Lett.}\ }\textbf {\bibinfo {volume} {119}},\
  \bibinfo {pages} {107001} (\bibinfo {year} {2017})}\BibitemShut {NoStop}%
\bibitem [{\citenamefont {Semenok}\ \emph {et~al.}(2021)\citenamefont
  {Semenok}, \citenamefont {Troyan}, \citenamefont {Ivanova}, \citenamefont
  {Kvashnin}, \citenamefont {Kruglov}, \citenamefont {Hanfland}, \citenamefont
  {Sadakov}, \citenamefont {Sobolevskiy}, \citenamefont {Pervakov},
  \citenamefont {Lyubutin}, \citenamefont {Glazyrin}, \citenamefont {Giordano},
  \citenamefont {Karimov}, \citenamefont {Vasiliev}, \citenamefont {Akashi},
  \citenamefont {Pudalov},\ and\ \citenamefont {Oganov}}]{Semenok2022}%
  \BibitemOpen
  \bibfield  {author} {\bibinfo {author} {\bibfnamefont {D.}~\bibnamefont
  {Semenok}}, \bibinfo {author} {\bibfnamefont {I.}~\bibnamefont {Troyan}},
  \bibinfo {author} {\bibfnamefont {A.}~\bibnamefont {Ivanova}}, \bibinfo
  {author} {\bibfnamefont {A.}~\bibnamefont {Kvashnin}}, \bibinfo {author}
  {\bibfnamefont {I.}~\bibnamefont {Kruglov}}, \bibinfo {author} {\bibfnamefont
  {M.}~\bibnamefont {Hanfland}}, \bibinfo {author} {\bibfnamefont
  {A.}~\bibnamefont {Sadakov}}, \bibinfo {author} {\bibfnamefont
  {O.}~\bibnamefont {Sobolevskiy}}, \bibinfo {author} {\bibfnamefont
  {K.}~\bibnamefont {Pervakov}}, \bibinfo {author} {\bibfnamefont
  {I.}~\bibnamefont {Lyubutin}}, \bibinfo {author} {\bibfnamefont
  {K.}~\bibnamefont {Glazyrin}}, \bibinfo {author} {\bibfnamefont
  {N.}~\bibnamefont {Giordano}}, \bibinfo {author} {\bibfnamefont {D.~N.}\
  \bibnamefont {Karimov}}, \bibinfo {author} {\bibfnamefont {A.}~\bibnamefont
  {Vasiliev}}, \bibinfo {author} {\bibfnamefont {R.}~\bibnamefont {Akashi}},
  \bibinfo {author} {\bibfnamefont {V.}~\bibnamefont {Pudalov}},\ and\ \bibinfo
  {author} {\bibfnamefont {A.}~\bibnamefont {Oganov}},\ }\href
  {https://doi.org/10.1016/j.mattod.2021.03.025} {\bibfield  {journal}
  {\bibinfo  {journal} {Mater. Today}\ }\textbf {\bibinfo {volume} {48}},\
  \bibinfo {pages} {18} (\bibinfo {year} {2021})}\BibitemShut {NoStop}%
\bibitem [{\citenamefont {Drozdov}\ \emph {et~al.}(2019)\citenamefont
  {Drozdov}, \citenamefont {Kong}, \citenamefont {Minkov}, \citenamefont
  {Besedin}, \citenamefont {Kuzovnikov}, \citenamefont {Mozaffari},
  \citenamefont {Balicas}, \citenamefont {Balakirev}, \citenamefont {Graf},
  \citenamefont {Prakapenka}, \citenamefont {Greenberg}, \citenamefont
  {Knyazev}, \citenamefont {Tkacz},\ and\ \citenamefont
  {Eremets}}]{Drozdov2019}%
  \BibitemOpen
  \bibfield  {author} {\bibinfo {author} {\bibfnamefont {A.}~\bibnamefont
  {Drozdov}}, \bibinfo {author} {\bibfnamefont {P.}~\bibnamefont {Kong}},
  \bibinfo {author} {\bibfnamefont {V.}~\bibnamefont {Minkov}}, \bibinfo
  {author} {\bibfnamefont {S.}~\bibnamefont {Besedin}}, \bibinfo {author}
  {\bibfnamefont {M.}~\bibnamefont {Kuzovnikov}}, \bibinfo {author}
  {\bibfnamefont {S.}~\bibnamefont {Mozaffari}}, \bibinfo {author}
  {\bibfnamefont {L.}~\bibnamefont {Balicas}}, \bibinfo {author} {\bibfnamefont
  {F.}~\bibnamefont {Balakirev}}, \bibinfo {author} {\bibfnamefont
  {E.}~\bibnamefont {Graf}}, \bibinfo {author} {\bibfnamefont {V.}~\bibnamefont
  {Prakapenka}}, \bibinfo {author} {\bibfnamefont {E.}~\bibnamefont
  {Greenberg}}, \bibinfo {author} {\bibfnamefont {D.}~\bibnamefont {Knyazev}},
  \bibinfo {author} {\bibfnamefont {M.}~\bibnamefont {Tkacz}},\ and\ \bibinfo
  {author} {\bibfnamefont {M.}~\bibnamefont {Eremets}},\ }\href
  {https://doi.org/10.1038/s41586-019-1201-8} {\bibfield  {journal} {\bibinfo
  {journal} {Nature}\ }\textbf {\bibinfo {volume} {569}},\ \bibinfo {pages}
  {528} (\bibinfo {year} {2019})}\BibitemShut {NoStop}%
\bibitem [{\citenamefont {Somayazulu}\ \emph {et~al.}(2019)\citenamefont
  {Somayazulu}, \citenamefont {Ahart}, \citenamefont {Mishra}, \citenamefont
  {Geballe}, \citenamefont {Baldini}, \citenamefont {Meng}, \citenamefont
  {Struzhkin},\ and\ \citenamefont {Hemley}}]{Somayazulu2019}%
  \BibitemOpen
  \bibfield  {author} {\bibinfo {author} {\bibfnamefont {M.}~\bibnamefont
  {Somayazulu}}, \bibinfo {author} {\bibfnamefont {M.}~\bibnamefont {Ahart}},
  \bibinfo {author} {\bibfnamefont {A.}~\bibnamefont {Mishra}}, \bibinfo
  {author} {\bibfnamefont {Z.}~\bibnamefont {Geballe}}, \bibinfo {author}
  {\bibfnamefont {M.}~\bibnamefont {Baldini}}, \bibinfo {author} {\bibfnamefont
  {Y.}~\bibnamefont {Meng}}, \bibinfo {author} {\bibfnamefont {V.}~\bibnamefont
  {Struzhkin}},\ and\ \bibinfo {author} {\bibfnamefont {R.}~\bibnamefont
  {Hemley}},\ }\href {https://doi.org/10.1103/PhysRevLett.122.027001}
  {\bibfield  {journal} {\bibinfo  {journal} {Phys. Rev. Lett.}\ }\textbf
  {\bibinfo {volume} {122}},\ \bibinfo {pages} {027001} (\bibinfo {year}
  {2019})}\BibitemShut {NoStop}%
\bibitem [{\citenamefont {Sun}\ \emph {et~al.}(2019)\citenamefont {Sun},
  \citenamefont {Lv}, \citenamefont {Xie}, \citenamefont {Liu},\ and\
  \citenamefont {Ma}}]{Sun2019}%
  \BibitemOpen
  \bibfield  {author} {\bibinfo {author} {\bibfnamefont {Y.}~\bibnamefont
  {Sun}}, \bibinfo {author} {\bibfnamefont {J.}~\bibnamefont {Lv}}, \bibinfo
  {author} {\bibfnamefont {Y.}~\bibnamefont {Xie}}, \bibinfo {author}
  {\bibfnamefont {H.}~\bibnamefont {Liu}},\ and\ \bibinfo {author}
  {\bibfnamefont {Y.}~\bibnamefont {Ma}},\ }\href
  {https://doi.org/10.1103/PhysRevLett.123.097001} {\bibfield  {journal}
  {\bibinfo  {journal} {Phys. Rev. Lett.}\ }\textbf {\bibinfo {volume} {123}},\
  \bibinfo {pages} {097001} (\bibinfo {year} {2019})}\BibitemShut {NoStop}%
\bibitem [{\citenamefont {Zhang}\ \emph {et~al.}(2022)\citenamefont {Zhang},
  \citenamefont {Zhao},\ and\ \citenamefont {Yang}}]{Zhang2022}%
  \BibitemOpen
  \bibfield  {author} {\bibinfo {author} {\bibfnamefont {X.}~\bibnamefont
  {Zhang}}, \bibinfo {author} {\bibfnamefont {Y.}~\bibnamefont {Zhao}},\ and\
  \bibinfo {author} {\bibfnamefont {G.}~\bibnamefont {Yang}},\ }\href
  {https://doi.org/10.1002/wcms.1582} {\bibfield  {journal} {\bibinfo
  {journal} {WIREs Comput Mol Sci}\ }\textbf {\bibinfo {volume} {12}},\
  \bibinfo {pages} {e1582} (\bibinfo {year} {2022})}\BibitemShut {NoStop}%
\bibitem [{\citenamefont {Takahashi}\ and\ \citenamefont
  {M{\^o}ri}(1995)}]{Takahashi1995}%
  \BibitemOpen
  \bibfield  {author} {\bibinfo {author} {\bibfnamefont {H.}~\bibnamefont
  {Takahashi}}\ and\ \bibinfo {author} {\bibfnamefont {N.}~\bibnamefont
  {M{\^o}ri}},\ }\href@noop {} {\emph {\bibinfo {title} {Studies of High
  Temperature Superconductors}}}\ (\bibinfo  {publisher} {Nova Science
  Publishers, Inc., New York},\ \bibinfo {year} {1995})\BibitemShut {NoStop}%
\bibitem [{\citenamefont {Gao}\ \emph {et~al.}(1994)\citenamefont {Gao},
  \citenamefont {Xue}, \citenamefont {Chen}, \citenamefont {Xiong},
  \citenamefont {Meng}, \citenamefont {Ramirez}, \citenamefont {Chu},
  \citenamefont {Eggert},\ and\ \citenamefont {Mao}}]{Gao1994}%
  \BibitemOpen
  \bibfield  {author} {\bibinfo {author} {\bibfnamefont {L.}~\bibnamefont
  {Gao}}, \bibinfo {author} {\bibfnamefont {Y.}~\bibnamefont {Xue}}, \bibinfo
  {author} {\bibfnamefont {F.}~\bibnamefont {Chen}}, \bibinfo {author}
  {\bibfnamefont {Q.}~\bibnamefont {Xiong}}, \bibinfo {author} {\bibfnamefont
  {R.}~\bibnamefont {Meng}}, \bibinfo {author} {\bibfnamefont {D.}~\bibnamefont
  {Ramirez}}, \bibinfo {author} {\bibfnamefont {C.}~\bibnamefont {Chu}},
  \bibinfo {author} {\bibfnamefont {J.}~\bibnamefont {Eggert}},\ and\ \bibinfo
  {author} {\bibfnamefont {H.}~\bibnamefont {Mao}},\ }\href
  {https://doi.org/10.1103/PhysRevB.50.4260} {\bibfield  {journal} {\bibinfo
  {journal} {Phys. Rev. B}\ }\textbf {\bibinfo {volume} {50}},\ \bibinfo
  {pages} {4260} (\bibinfo {year} {1994})}\BibitemShut {NoStop}%
\bibitem [{\citenamefont {Monteverde}\ \emph {et~al.}(2005)\citenamefont
  {Monteverde}, \citenamefont {Acha}, \citenamefont {N\'{u}{\~n}ez-Regueiro},
  \citenamefont {Pavlov}, \citenamefont {Lokshin}, \citenamefont {Putilin},\
  and\ \citenamefont {Antipov}}]{Monteverde2005}%
  \BibitemOpen
  \bibfield  {author} {\bibinfo {author} {\bibfnamefont {M.}~\bibnamefont
  {Monteverde}}, \bibinfo {author} {\bibfnamefont {C.}~\bibnamefont {Acha}},
  \bibinfo {author} {\bibfnamefont {M.}~\bibnamefont {N\'{u}{\~n}ez-Regueiro}},
  \bibinfo {author} {\bibfnamefont {D.}~\bibnamefont {Pavlov}}, \bibinfo
  {author} {\bibfnamefont {K.}~\bibnamefont {Lokshin}}, \bibinfo {author}
  {\bibfnamefont {S.}~\bibnamefont {Putilin}},\ and\ \bibinfo {author}
  {\bibfnamefont {E.}~\bibnamefont {Antipov}},\ }\href
  {https://doi.org/10.1209/epl/i2005-10247-3} {\bibfield  {journal} {\bibinfo
  {journal} {Europhys. Lett.}\ }\textbf {\bibinfo {volume} {72}},\ \bibinfo
  {pages} {458} (\bibinfo {year} {2005})}\BibitemShut {NoStop}%
\bibitem [{\citenamefont {Motoi}\ \emph {et~al.}(1991)\citenamefont {Motoi},
  \citenamefont {Fujimoto}, \citenamefont {Uwe},\ and\ \citenamefont
  {Sakudo}}]{Motoi1991}%
  \BibitemOpen
  \bibfield  {author} {\bibinfo {author} {\bibfnamefont {Y.}~\bibnamefont
  {Motoi}}, \bibinfo {author} {\bibfnamefont {K.}~\bibnamefont {Fujimoto}},
  \bibinfo {author} {\bibfnamefont {H.}~\bibnamefont {Uwe}},\ and\ \bibinfo
  {author} {\bibfnamefont {T.}~\bibnamefont {Sakudo}},\ }\href
  {https://doi.org/10.1143/jpsj.60.384} {\bibfield  {journal} {\bibinfo
  {journal} {J. Phys. Soc. Jpn.}\ }\textbf {\bibinfo {volume} {60}},\ \bibinfo
  {pages} {384} (\bibinfo {year} {1991})}\BibitemShut {NoStop}%
\bibitem [{\citenamefont {Hardy}\ \emph {et~al.}(2010)\citenamefont {Hardy},
  \citenamefont {Hillier}, \citenamefont {Meingast}, \citenamefont {Colson},
  \citenamefont {Li}, \citenamefont {Bari\v{s}i\'{c}}, \citenamefont {Yu},
  \citenamefont {Zhao}, \citenamefont {Greven},\ and\ \citenamefont
  {Schilling}}]{Hardy2010}%
  \BibitemOpen
  \bibfield  {author} {\bibinfo {author} {\bibfnamefont {F.}~\bibnamefont
  {Hardy}}, \bibinfo {author} {\bibfnamefont {N.}~\bibnamefont {Hillier}},
  \bibinfo {author} {\bibfnamefont {C.}~\bibnamefont {Meingast}}, \bibinfo
  {author} {\bibfnamefont {D.}~\bibnamefont {Colson}}, \bibinfo {author}
  {\bibfnamefont {Y.}~\bibnamefont {Li}}, \bibinfo {author} {\bibfnamefont
  {N.}~\bibnamefont {Bari\v{s}i\'{c}}}, \bibinfo {author} {\bibfnamefont
  {G.}~\bibnamefont {Yu}}, \bibinfo {author} {\bibfnamefont {X.}~\bibnamefont
  {Zhao}}, \bibinfo {author} {\bibfnamefont {M.}~\bibnamefont {Greven}},\ and\
  \bibinfo {author} {\bibfnamefont {J.}~\bibnamefont {Schilling}},\ }\href
  {https://doi.org/10.1103/PhysRevLett.105.167002} {\bibfield  {journal}
  {\bibinfo  {journal} {Phys. Rev. Lett.}\ }\textbf {\bibinfo {volume} {105}},\
  \bibinfo {pages} {167002} (\bibinfo {year} {2010})}\BibitemShut {NoStop}%
\bibitem [{\citenamefont {Watanabe}\ \emph {et~al.}(1994)\citenamefont
  {Watanabe}, \citenamefont {Fukamachi}, \citenamefont {Ueda}, \citenamefont
  {Tsushima}, \citenamefont {Balbashov}, \citenamefont {Nakanishi},\ and\
  \citenamefont {M{\^o}ri}}]{Watanabe94}%
  \BibitemOpen
  \bibfield  {author} {\bibinfo {author} {\bibfnamefont {N.}~\bibnamefont
  {Watanabe}}, \bibinfo {author} {\bibfnamefont {K.}~\bibnamefont {Fukamachi}},
  \bibinfo {author} {\bibfnamefont {Y.}~\bibnamefont {Ueda}}, \bibinfo {author}
  {\bibfnamefont {K.}~\bibnamefont {Tsushima}}, \bibinfo {author}
  {\bibfnamefont {A.}~\bibnamefont {Balbashov}}, \bibinfo {author}
  {\bibfnamefont {T.}~\bibnamefont {Nakanishi}},\ and\ \bibinfo {author}
  {\bibfnamefont {N.}~\bibnamefont {M{\^o}ri}},\ }\href
  {https://doi.org/10.1016/0921-4534(94)91879-1} {\bibfield  {journal}
  {\bibinfo  {journal} {Physica C}\ }\textbf {\bibinfo {volume} {235-240}},\
  \bibinfo {pages} {1309} (\bibinfo {year} {1994})}\BibitemShut {NoStop}%
\bibitem [{\citenamefont {Meingast}\ \emph {et~al.}(1996)\citenamefont
  {Meingast}, \citenamefont {Junod},\ and\ \citenamefont
  {Walker}}]{Meingast96}%
  \BibitemOpen
  \bibfield  {author} {\bibinfo {author} {\bibfnamefont {C.}~\bibnamefont
  {Meingast}}, \bibinfo {author} {\bibfnamefont {A.}~\bibnamefont {Junod}},\
  and\ \bibinfo {author} {\bibfnamefont {E.}~\bibnamefont {Walker}},\ }\href
  {https://doi.org/10.1016/S0921-4534(96)00589-8} {\bibfield  {journal}
  {\bibinfo  {journal} {Physica C}\ }\textbf {\bibinfo {volume} {272}},\
  \bibinfo {pages} {106} (\bibinfo {year} {1996})}\BibitemShut {NoStop}%
\bibitem [{\citenamefont {Nakamura}\ \emph {et~al.}(1999)\citenamefont
  {Nakamura}, \citenamefont {Hori}, \citenamefont {Goko}, \citenamefont {Uno},
  \citenamefont {Kikugawa},\ and\ \citenamefont {Fujita}}]{Nakamura1999}%
  \BibitemOpen
  \bibfield  {author} {\bibinfo {author} {\bibfnamefont {F.}~\bibnamefont
  {Nakamura}}, \bibinfo {author} {\bibfnamefont {J.}~\bibnamefont {Hori}},
  \bibinfo {author} {\bibfnamefont {T.}~\bibnamefont {Goko}}, \bibinfo {author}
  {\bibfnamefont {Y.}~\bibnamefont {Uno}}, \bibinfo {author} {\bibfnamefont
  {N.}~\bibnamefont {Kikugawa}},\ and\ \bibinfo {author} {\bibfnamefont
  {T.}~\bibnamefont {Fujita}},\ }\href
  {https://doi.org/10.1023/A:1022503213459} {\bibfield  {journal} {\bibinfo
  {journal} {J. Low Temp. Phys.}\ }\textbf {\bibinfo {volume} {117}},\ \bibinfo
  {pages} {1145} (\bibinfo {year} {1999})}\BibitemShut {NoStop}%
\bibitem [{\citenamefont {Nakamura}\ \emph {et~al.}(2000)\citenamefont
  {Nakamura}, \citenamefont {Goko}, \citenamefont {Hori}, \citenamefont {Uno},
  \citenamefont {Kikugawa},\ and\ \citenamefont {Fujita}}]{Nakamura2000}%
  \BibitemOpen
  \bibfield  {author} {\bibinfo {author} {\bibfnamefont {F.}~\bibnamefont
  {Nakamura}}, \bibinfo {author} {\bibfnamefont {T.}~\bibnamefont {Goko}},
  \bibinfo {author} {\bibfnamefont {J.}~\bibnamefont {Hori}}, \bibinfo {author}
  {\bibfnamefont {Y.}~\bibnamefont {Uno}}, \bibinfo {author} {\bibfnamefont
  {N.}~\bibnamefont {Kikugawa}},\ and\ \bibinfo {author} {\bibfnamefont
  {T.}~\bibnamefont {Fujita}},\ }\href
  {https://doi.org/10.1103/PhysRevB.61.107} {\bibfield  {journal} {\bibinfo
  {journal} {Phys. Rev. B}\ }\textbf {\bibinfo {volume} {61}},\ \bibinfo
  {pages} {107} (\bibinfo {year} {2000})}\BibitemShut {NoStop}%
\bibitem [{\citenamefont {Butko}\ \emph {et~al.}(2009)\citenamefont {Butko},
  \citenamefont {Logvenov}, \citenamefont {Bo\v{z}ovi\'{c}}, \citenamefont
  {Radovi\'{c}},\ and\ \citenamefont {Bo\v{z}ovi\'{c}}}]{Butko2009}%
  \BibitemOpen
  \bibfield  {author} {\bibinfo {author} {\bibfnamefont {V.}~\bibnamefont
  {Butko}}, \bibinfo {author} {\bibfnamefont {G.}~\bibnamefont {Logvenov}},
  \bibinfo {author} {\bibfnamefont {N.}~\bibnamefont {Bo\v{z}ovi\'{c}}},
  \bibinfo {author} {\bibfnamefont {Z.}~\bibnamefont {Radovi\'{c}}},\ and\
  \bibinfo {author} {\bibfnamefont {I.}~\bibnamefont {Bo\v{z}ovi\'{c}}},\
  }\href {https://doi.org/10.1002/adma.200803850} {\bibfield  {journal}
  {\bibinfo  {journal} {Adv. Mater.}\ }\textbf {\bibinfo {volume} {21}},\
  \bibinfo {pages} {3644} (\bibinfo {year} {2009})}\BibitemShut {NoStop}%
\bibitem [{\citenamefont {Chen}\ \emph {et~al.}(1991)\citenamefont {Chen},
  \citenamefont {Tessema},\ and\ \citenamefont {Skove}}]{Chen91}%
  \BibitemOpen
  \bibfield  {author} {\bibinfo {author} {\bibfnamefont {X.-F.}\ \bibnamefont
  {Chen}}, \bibinfo {author} {\bibfnamefont {G.}~\bibnamefont {Tessema}},\ and\
  \bibinfo {author} {\bibfnamefont {M.}~\bibnamefont {Skove}},\ }\href
  {https://doi.org/10.1016/0921-4534(91)90121-E} {\bibfield  {journal}
  {\bibinfo  {journal} {Physica C}\ }\textbf {\bibinfo {volume} {181}},\
  \bibinfo {pages} {340} (\bibinfo {year} {1991})}\BibitemShut {NoStop}%
\bibitem [{\citenamefont {Meingast}\ \emph {et~al.}(1990)\citenamefont
  {Meingast}, \citenamefont {Blank}, \citenamefont {B{\"u}rkle}, \citenamefont
  {Obst}, \citenamefont {Wolf}, \citenamefont {W{\"u}hl}, \citenamefont
  {Selvamanickam},\ and\ \citenamefont {Salama}}]{Meingast90}%
  \BibitemOpen
  \bibfield  {author} {\bibinfo {author} {\bibfnamefont {C.}~\bibnamefont
  {Meingast}}, \bibinfo {author} {\bibfnamefont {B.}~\bibnamefont {Blank}},
  \bibinfo {author} {\bibfnamefont {H.}~\bibnamefont {B{\"u}rkle}}, \bibinfo
  {author} {\bibfnamefont {B.}~\bibnamefont {Obst}}, \bibinfo {author}
  {\bibfnamefont {T.}~\bibnamefont {Wolf}}, \bibinfo {author} {\bibfnamefont
  {H.}~\bibnamefont {W{\"u}hl}}, \bibinfo {author} {\bibfnamefont
  {V.}~\bibnamefont {Selvamanickam}},\ and\ \bibinfo {author} {\bibfnamefont
  {K.}~\bibnamefont {Salama}},\ }\href
  {https://doi.org/10.1103/PhysRevB.41.11299} {\bibfield  {journal} {\bibinfo
  {journal} {Phys. Rev. B}\ }\textbf {\bibinfo {volume} {41}},\ \bibinfo
  {pages} {11299} (\bibinfo {year} {1990})}\BibitemShut {NoStop}%
\bibitem [{\citenamefont {Meingast}\ \emph {et~al.}(1991)\citenamefont
  {Meingast}, \citenamefont {Kraut}, \citenamefont {Wolf}, \citenamefont
  {W{\"u}hl}, \citenamefont {Erb},\ and\ \citenamefont
  {M{\"u}ller-Vogt}}]{Meingast91}%
  \BibitemOpen
  \bibfield  {author} {\bibinfo {author} {\bibfnamefont {C.}~\bibnamefont
  {Meingast}}, \bibinfo {author} {\bibfnamefont {O.}~\bibnamefont {Kraut}},
  \bibinfo {author} {\bibfnamefont {T.}~\bibnamefont {Wolf}}, \bibinfo {author}
  {\bibfnamefont {H.}~\bibnamefont {W{\"u}hl}}, \bibinfo {author}
  {\bibfnamefont {A.}~\bibnamefont {Erb}},\ and\ \bibinfo {author}
  {\bibfnamefont {G.}~\bibnamefont {M{\"u}ller-Vogt}},\ }\href
  {https://doi.org/10.1103/PhysRevLett.67.1634} {\bibfield  {journal} {\bibinfo
   {journal} {Phys. Rev. Lett.}\ }\textbf {\bibinfo {volume} {67}},\ \bibinfo
  {pages} {1634} (\bibinfo {year} {1991})}\BibitemShut {NoStop}%
\bibitem [{\citenamefont {Welp}\ \emph {et~al.}(1992)\citenamefont {Welp},
  \citenamefont {Grimsditch}, \citenamefont {Fleshler}, \citenamefont
  {Nessler}, \citenamefont {Downey}, \citenamefont {Crabtree},\ and\
  \citenamefont {Guimpel}}]{Welp92}%
  \BibitemOpen
  \bibfield  {author} {\bibinfo {author} {\bibfnamefont {U.}~\bibnamefont
  {Welp}}, \bibinfo {author} {\bibfnamefont {M.}~\bibnamefont {Grimsditch}},
  \bibinfo {author} {\bibfnamefont {S.}~\bibnamefont {Fleshler}}, \bibinfo
  {author} {\bibfnamefont {W.}~\bibnamefont {Nessler}}, \bibinfo {author}
  {\bibfnamefont {J.}~\bibnamefont {Downey}}, \bibinfo {author} {\bibfnamefont
  {G.}~\bibnamefont {Crabtree}},\ and\ \bibinfo {author} {\bibfnamefont
  {J.}~\bibnamefont {Guimpel}},\ }\href
  {https://doi.org/10.1103/PhysRevLett.69.2130} {\bibfield  {journal} {\bibinfo
   {journal} {Phys. Rev. Lett.}\ }\textbf {\bibinfo {volume} {69}},\ \bibinfo
  {pages} {2130} (\bibinfo {year} {1992})}\BibitemShut {NoStop}%
\bibitem [{\citenamefont {Crommie}\ \emph {et~al.}(1989)\citenamefont
  {Crommie}, \citenamefont {Liu}, \citenamefont {Zettl}, \citenamefont {Cohen},
  \citenamefont {Parilla}, \citenamefont {Hundley}, \citenamefont {Creager},
  \citenamefont {Hoen},\ and\ \citenamefont {Sherwin}}]{Crommie1989}%
  \BibitemOpen
  \bibfield  {author} {\bibinfo {author} {\bibfnamefont {M.}~\bibnamefont
  {Crommie}}, \bibinfo {author} {\bibfnamefont {A.}~\bibnamefont {Liu}},
  \bibinfo {author} {\bibfnamefont {A.}~\bibnamefont {Zettl}}, \bibinfo
  {author} {\bibfnamefont {M.}~\bibnamefont {Cohen}}, \bibinfo {author}
  {\bibfnamefont {P.}~\bibnamefont {Parilla}}, \bibinfo {author} {\bibfnamefont
  {M.}~\bibnamefont {Hundley}}, \bibinfo {author} {\bibfnamefont
  {W.}~\bibnamefont {Creager}}, \bibinfo {author} {\bibfnamefont
  {S.}~\bibnamefont {Hoen}},\ and\ \bibinfo {author} {\bibfnamefont
  {M.}~\bibnamefont {Sherwin}},\ }\href
  {https://doi.org/10.1103/PhysRevB.39.4231} {\bibfield  {journal} {\bibinfo
  {journal} {Phys. Rev. B}\ }\textbf {\bibinfo {volume} {39}},\ \bibinfo
  {pages} {4231} (\bibinfo {year} {1989})}\BibitemShut {NoStop}%
\bibitem [{\citenamefont {Bud'ko}\ \emph {et~al.}(1991)\citenamefont {Bud'ko},
  \citenamefont {Nakamura}, \citenamefont {Guimpel}, \citenamefont {Maple},\
  and\ \citenamefont {Schuller}}]{Budko91}%
  \BibitemOpen
  \bibfield  {author} {\bibinfo {author} {\bibfnamefont {S.}~\bibnamefont
  {Bud'ko}}, \bibinfo {author} {\bibfnamefont {O.}~\bibnamefont {Nakamura}},
  \bibinfo {author} {\bibfnamefont {J.}~\bibnamefont {Guimpel}}, \bibinfo
  {author} {\bibfnamefont {M.}~\bibnamefont {Maple}},\ and\ \bibinfo {author}
  {\bibfnamefont {I.~K.}\ \bibnamefont {Schuller}},\ }\href
  {https://doi.org/10.1016/0921-4534(91)91098-O} {\bibfield  {journal}
  {\bibinfo  {journal} {Physica C}\ }\textbf {\bibinfo {volume} {185-189}},\
  \bibinfo {pages} {1947} (\bibinfo {year} {1991})}\BibitemShut {NoStop}%
\bibitem [{\citenamefont {W{\"u}hl}\ \emph {et~al.}(1991)\citenamefont
  {W{\"u}hl}, \citenamefont {Benischke}, \citenamefont {Braun}, \citenamefont
  {Frank}, \citenamefont {Kraut}, \citenamefont {Ahrens}, \citenamefont
  {Br{\"a}uchle}, \citenamefont {Claus}, \citenamefont {Erb}, \citenamefont
  {Fietz}, \citenamefont {Meingast}, \citenamefont {M{\"u}ller-Vogt},\ and\
  \citenamefont {Wolf}}]{Wuhl91}%
  \BibitemOpen
  \bibfield  {author} {\bibinfo {author} {\bibfnamefont {H.}~\bibnamefont
  {W{\"u}hl}}, \bibinfo {author} {\bibfnamefont {R.}~\bibnamefont {Benischke}},
  \bibinfo {author} {\bibfnamefont {M.}~\bibnamefont {Braun}}, \bibinfo
  {author} {\bibfnamefont {B.}~\bibnamefont {Frank}}, \bibinfo {author}
  {\bibfnamefont {O.}~\bibnamefont {Kraut}}, \bibinfo {author} {\bibfnamefont
  {R.}~\bibnamefont {Ahrens}}, \bibinfo {author} {\bibfnamefont
  {G.}~\bibnamefont {Br{\"a}uchle}}, \bibinfo {author} {\bibfnamefont
  {H.}~\bibnamefont {Claus}}, \bibinfo {author} {\bibfnamefont
  {A.}~\bibnamefont {Erb}}, \bibinfo {author} {\bibfnamefont {W.}~\bibnamefont
  {Fietz}}, \bibinfo {author} {\bibfnamefont {C.}~\bibnamefont {Meingast}},
  \bibinfo {author} {\bibfnamefont {G.}~\bibnamefont {M{\"u}ller-Vogt}},\ and\
  \bibinfo {author} {\bibfnamefont {T.}~\bibnamefont {Wolf}},\ }\href
  {https://doi.org/10.1016/0921-4534(91)91601-Y} {\bibfield  {journal}
  {\bibinfo  {journal} {Physica C}\ }\textbf {\bibinfo {volume} {185-189}},\
  \bibinfo {pages} {755} (\bibinfo {year} {1991})}\BibitemShut {NoStop}%
\bibitem [{\citenamefont {Meingast}\ \emph {et~al.}(1993)\citenamefont
  {Meingast}, \citenamefont {Karpinski}, \citenamefont {Jilek},\ and\
  \citenamefont {Kaldis}}]{Meingast93}%
  \BibitemOpen
  \bibfield  {author} {\bibinfo {author} {\bibfnamefont {C.}~\bibnamefont
  {Meingast}}, \bibinfo {author} {\bibfnamefont {J.}~\bibnamefont {Karpinski}},
  \bibinfo {author} {\bibfnamefont {E.}~\bibnamefont {Jilek}},\ and\ \bibinfo
  {author} {\bibfnamefont {E.}~\bibnamefont {Kaldis}},\ }\href
  {https://doi.org/10.1016/0921-4534(93)90580-J} {\bibfield  {journal}
  {\bibinfo  {journal} {Physica C}\ }\textbf {\bibinfo {volume} {209}},\
  \bibinfo {pages} {591} (\bibinfo {year} {1993})}\BibitemShut {NoStop}%
\bibitem [{\citenamefont {Mito}\ \emph {et~al.}(2012)\citenamefont {Mito},
  \citenamefont {Imakyurei}, \citenamefont {Deguchi}, \citenamefont
  {Matsumoto}, \citenamefont {Tajiri}, \citenamefont {Hara}, \citenamefont
  {Ozaki}, \citenamefont {Takeya},\ and\ \citenamefont {Takano}}]{Mito2012}%
  \BibitemOpen
  \bibfield  {author} {\bibinfo {author} {\bibfnamefont {M.}~\bibnamefont
  {Mito}}, \bibinfo {author} {\bibfnamefont {T.}~\bibnamefont {Imakyurei}},
  \bibinfo {author} {\bibfnamefont {H.}~\bibnamefont {Deguchi}}, \bibinfo
  {author} {\bibfnamefont {K.}~\bibnamefont {Matsumoto}}, \bibinfo {author}
  {\bibfnamefont {T.}~\bibnamefont {Tajiri}}, \bibinfo {author} {\bibfnamefont
  {H.}~\bibnamefont {Hara}}, \bibinfo {author} {\bibfnamefont {T.}~\bibnamefont
  {Ozaki}}, \bibinfo {author} {\bibfnamefont {H.}~\bibnamefont {Takeya}},\ and\
  \bibinfo {author} {\bibfnamefont {Y.}~\bibnamefont {Takano}},\ }\href
  {https://doi.org/10.1143/JPSJ.81.113709} {\bibfield  {journal} {\bibinfo
  {journal} {J. Phys. Soc. Jpn.}\ }\textbf {\bibinfo {volume} {81}},\ \bibinfo
  {pages} {113709} (\bibinfo {year} {2012})}\BibitemShut {NoStop}%
\bibitem [{\citenamefont {Sidorov}\ \emph {et~al.}(2016)\citenamefont
  {Sidorov}, \citenamefont {Gavrichkov}, \citenamefont {Nikolaev},
  \citenamefont {Pchelkina},\ and\ \citenamefont {Ovchinnikov}}]{Sidorov2016}%
  \BibitemOpen
  \bibfield  {author} {\bibinfo {author} {\bibfnamefont {K.}~\bibnamefont
  {Sidorov}}, \bibinfo {author} {\bibfnamefont {V.}~\bibnamefont {Gavrichkov}},
  \bibinfo {author} {\bibfnamefont {S.}~\bibnamefont {Nikolaev}}, \bibinfo
  {author} {\bibfnamefont {Z.}~\bibnamefont {Pchelkina}},\ and\ \bibinfo
  {author} {\bibfnamefont {S.}~\bibnamefont {Ovchinnikov}},\ }\href
  {https://doi.org/10.1002/pssb.201552465} {\bibfield  {journal} {\bibinfo
  {journal} {Phys. Status Solidi B}\ }\textbf {\bibinfo {volume} {253}},\
  \bibinfo {pages} {486} (\bibinfo {year} {2016})}\BibitemShut {NoStop}%
\bibitem [{\citenamefont {Ovchinnikov}\ and\ \citenamefont
  {Sandalov}(1989)}]{Ovchinnikov89}%
  \BibitemOpen
  \bibfield  {author} {\bibinfo {author} {\bibfnamefont {S.}~\bibnamefont
  {Ovchinnikov}}\ and\ \bibinfo {author} {\bibfnamefont {I.}~\bibnamefont
  {Sandalov}},\ }\href {https://doi.org/10.1016/0921-4534(89)90397-3}
  {\bibfield  {journal} {\bibinfo  {journal} {Physica C}\ }\textbf {\bibinfo
  {volume} {161}},\ \bibinfo {pages} {607} (\bibinfo {year}
  {1989})}\BibitemShut {NoStop}%
\bibitem [{\citenamefont {Gavrichkov}\ \emph {et~al.}(2000)\citenamefont
  {Gavrichkov}, \citenamefont {Ovchinnikov}, \citenamefont {Borisov},\ and\
  \citenamefont {Goryachev}}]{Gavrichkov00}%
  \BibitemOpen
  \bibfield  {author} {\bibinfo {author} {\bibfnamefont {V.}~\bibnamefont
  {Gavrichkov}}, \bibinfo {author} {\bibfnamefont {S.}~\bibnamefont
  {Ovchinnikov}}, \bibinfo {author} {\bibfnamefont {A.}~\bibnamefont
  {Borisov}},\ and\ \bibinfo {author} {\bibfnamefont {E.}~\bibnamefont
  {Goryachev}},\ }\href {https://doi.org/10.1134/1.1311997} {\bibfield
  {journal} {\bibinfo  {journal} {J. Exp. Theor. Phys.}\ }\textbf {\bibinfo
  {volume} {91}},\ \bibinfo {pages} {369} (\bibinfo {year} {2000})}\BibitemShut
  {NoStop}%
\bibitem [{\citenamefont {Korshunov}\ \emph {et~al.}(2005)\citenamefont
  {Korshunov}, \citenamefont {Gavrichkov}, \citenamefont {Ovchinnikov},
  \citenamefont {Nekrasov}, \citenamefont {Pchelkina},\ and\ \citenamefont
  {Anisimov}}]{Korshunov05}%
  \BibitemOpen
  \bibfield  {author} {\bibinfo {author} {\bibfnamefont {M.}~\bibnamefont
  {Korshunov}}, \bibinfo {author} {\bibfnamefont {V.}~\bibnamefont
  {Gavrichkov}}, \bibinfo {author} {\bibfnamefont {S.}~\bibnamefont
  {Ovchinnikov}}, \bibinfo {author} {\bibfnamefont {I.}~\bibnamefont
  {Nekrasov}}, \bibinfo {author} {\bibfnamefont {Z.}~\bibnamefont
  {Pchelkina}},\ and\ \bibinfo {author} {\bibfnamefont {V.}~\bibnamefont
  {Anisimov}},\ }\href {https://doi.org/10.1103/PhysRevB.72.165104} {\bibfield
  {journal} {\bibinfo  {journal} {Phys. Rev. B}\ }\textbf {\bibinfo {volume}
  {72}},\ \bibinfo {pages} {165104} (\bibinfo {year} {2005})}\BibitemShut
  {NoStop}%
\bibitem [{\citenamefont {Kresin}\ \emph {et~al.}(2021)\citenamefont {Kresin},
  \citenamefont {Ovchinnikov},\ and\ \citenamefont {Wolf}}]{Kresin2021}%
  \BibitemOpen
  \bibfield  {author} {\bibinfo {author} {\bibfnamefont {V.}~\bibnamefont
  {Kresin}}, \bibinfo {author} {\bibfnamefont {S.}~\bibnamefont
  {Ovchinnikov}},\ and\ \bibinfo {author} {\bibfnamefont {S.}~\bibnamefont
  {Wolf}},\ }\href@noop {} {\emph {\bibinfo {title} {Superconducting State:
  Mechanisms and Materials}}}\ (\bibinfo  {publisher} {Oxford University
  Press},\ \bibinfo {year} {2021})\BibitemShut {NoStop}%
\bibitem [{\citenamefont {I.A.Makarov}\ and\ \citenamefont
  {Ovchinnikov}(2025)}]{Makarov2025}%
  \BibitemOpen
  \bibfield  {author} {\bibinfo {author} {\bibnamefont {I.A.Makarov}}\ and\
  \bibinfo {author} {\bibfnamefont {S.}~\bibnamefont {Ovchinnikov}},\ }\href
  {https://doi.org/10.1007/s10948-025-06928-5} {\bibfield  {journal} {\bibinfo
  {journal} {J. Supercond. Nov. Magn.}\ }\textbf {\bibinfo {volume} {38}},\
  \bibinfo {pages} {83} (\bibinfo {year} {2025})}\BibitemShut {NoStop}%
\bibitem [{\citenamefont {Raimondi}\ \emph {et~al.}(1996)\citenamefont
  {Raimondi}, \citenamefont {Jefferson},\ and\ \citenamefont
  {Feiner}}]{Raimondi96}%
  \BibitemOpen
  \bibfield  {author} {\bibinfo {author} {\bibfnamefont {R.}~\bibnamefont
  {Raimondi}}, \bibinfo {author} {\bibfnamefont {J.}~\bibnamefont
  {Jefferson}},\ and\ \bibinfo {author} {\bibfnamefont {L.}~\bibnamefont
  {Feiner}},\ }\href {https://doi.org/10.1103/PhysRevB.53.8774} {\bibfield
  {journal} {\bibinfo  {journal} {Phys. Rev. B}\ }\textbf {\bibinfo {volume}
  {53}},\ \bibinfo {pages} {8774} (\bibinfo {year} {1996})}\BibitemShut
  {NoStop}%
\bibitem [{\citenamefont {Makarov}\ \emph {et~al.}(2019)\citenamefont
  {Makarov}, \citenamefont {Gavrichkov}, \citenamefont {Shneyder},
  \citenamefont {Nekrasov}, \citenamefont {Slobodchikov}, \citenamefont
  {Ovchinnikov},\ and\ \citenamefont {Bianconi}}]{Makarov19}%
  \BibitemOpen
  \bibfield  {author} {\bibinfo {author} {\bibfnamefont {I.}~\bibnamefont
  {Makarov}}, \bibinfo {author} {\bibfnamefont {V.}~\bibnamefont {Gavrichkov}},
  \bibinfo {author} {\bibfnamefont {E.}~\bibnamefont {Shneyder}}, \bibinfo
  {author} {\bibfnamefont {I.}~\bibnamefont {Nekrasov}}, \bibinfo {author}
  {\bibfnamefont {A.}~\bibnamefont {Slobodchikov}}, \bibinfo {author}
  {\bibfnamefont {S.}~\bibnamefont {Ovchinnikov}},\ and\ \bibinfo {author}
  {\bibfnamefont {A.}~\bibnamefont {Bianconi}},\ }\href
  {https://doi.org/10.1007/s10948-018-4936-9} {\bibfield  {journal} {\bibinfo
  {journal} {J. Supercond. Nov. Magn.}\ }\textbf {\bibinfo {volume} {32}},\
  \bibinfo {pages} {1927} (\bibinfo {year} {2019})}\BibitemShut {NoStop}%
\bibitem [{\citenamefont {Schl{\"u}ter}\ \emph {et~al.}(1988)\citenamefont
  {Schl{\"u}ter}, \citenamefont {Hybertsen},\ and\ \citenamefont
  {Christensen}}]{Schluter88}%
  \BibitemOpen
  \bibfield  {author} {\bibinfo {author} {\bibfnamefont {M.}~\bibnamefont
  {Schl{\"u}ter}}, \bibinfo {author} {\bibfnamefont {M.}~\bibnamefont
  {Hybertsen}},\ and\ \bibinfo {author} {\bibfnamefont {N.}~\bibnamefont
  {Christensen}},\ }\href {https://doi.org/10.1016/0921-4534(88)90249-3}
  {\bibfield  {journal} {\bibinfo  {journal} {Physica C}\ }\textbf {\bibinfo
  {volume} {153-155}},\ \bibinfo {pages} {1217} (\bibinfo {year}
  {1988})}\BibitemShut {NoStop}%
\bibitem [{\citenamefont {Schl{\"u}ter}\ and\ \citenamefont
  {Hybertsen}(1989)}]{Schluter89}%
  \BibitemOpen
  \bibfield  {author} {\bibinfo {author} {\bibfnamefont {M.}~\bibnamefont
  {Schl{\"u}ter}}\ and\ \bibinfo {author} {\bibfnamefont {M.}~\bibnamefont
  {Hybertsen}},\ }\href {https://doi.org/10.1016/0921-4534(89)91163-5}
  {\bibfield  {journal} {\bibinfo  {journal} {Physica C}\ }\textbf {\bibinfo
  {volume} {162-164}},\ \bibinfo {pages} {583} (\bibinfo {year}
  {1989})}\BibitemShut {NoStop}%
\bibitem [{\citenamefont {Hybertsen}\ \emph {et~al.}(1989)\citenamefont
  {Hybertsen}, \citenamefont {Schl{\"u}ter},\ and\ \citenamefont
  {Christensen}}]{Hybertsen89}%
  \BibitemOpen
  \bibfield  {author} {\bibinfo {author} {\bibfnamefont {M.}~\bibnamefont
  {Hybertsen}}, \bibinfo {author} {\bibfnamefont {M.}~\bibnamefont
  {Schl{\"u}ter}},\ and\ \bibinfo {author} {\bibfnamefont {N.}~\bibnamefont
  {Christensen}},\ }\href {https://doi.org/10.1103/PhysRevB.39.9028} {\bibfield
   {journal} {\bibinfo  {journal} {Phys. Rev. B}\ }\textbf {\bibinfo {volume}
  {39}},\ \bibinfo {pages} {9028–9041} (\bibinfo {year} {1989})}\BibitemShut
  {NoStop}%
\bibitem [{\citenamefont {Hybertsen}\ \emph {et~al.}(1990)\citenamefont
  {Hybertsen}, \citenamefont {Stechel}, \citenamefont {Schl{\"u}ter},\ and\
  \citenamefont {Jennison}}]{Hybertsen90}%
  \BibitemOpen
  \bibfield  {author} {\bibinfo {author} {\bibfnamefont {M.}~\bibnamefont
  {Hybertsen}}, \bibinfo {author} {\bibfnamefont {E.}~\bibnamefont {Stechel}},
  \bibinfo {author} {\bibfnamefont {M.}~\bibnamefont {Schl{\"u}ter}},\ and\
  \bibinfo {author} {\bibfnamefont {D.}~\bibnamefont {Jennison}},\ }\href
  {https://doi.org/10.1103/PhysRevB.41.11068} {\bibfield  {journal} {\bibinfo
  {journal} {Phys. Rev. B}\ }\textbf {\bibinfo {volume} {41}},\ \bibinfo
  {pages} {11068} (\bibinfo {year} {1990})}\BibitemShut {NoStop}%
\bibitem [{\citenamefont {McMahan}\ \emph {et~al.}(1990)\citenamefont
  {McMahan}, \citenamefont {Annett},\ and\ \citenamefont {Martin}}]{Mahan90}%
  \BibitemOpen
  \bibfield  {author} {\bibinfo {author} {\bibfnamefont {A.}~\bibnamefont
  {McMahan}}, \bibinfo {author} {\bibfnamefont {J.}~\bibnamefont {Annett}},\
  and\ \bibinfo {author} {\bibfnamefont {R.}~\bibnamefont {Martin}},\ }\href
  {https://doi.org/10.1103/PhysRevB.42.6268} {\bibfield  {journal} {\bibinfo
  {journal} {Phys. Rev. B}\ }\textbf {\bibinfo {volume} {42}},\ \bibinfo
  {pages} {6268} (\bibinfo {year} {1990})}\BibitemShut {NoStop}%
\bibitem [{\citenamefont {Grant}\ and\ \citenamefont
  {McMahan}(1992)}]{Grant92}%
  \BibitemOpen
  \bibfield  {author} {\bibinfo {author} {\bibfnamefont {J.}~\bibnamefont
  {Grant}}\ and\ \bibinfo {author} {\bibfnamefont {A.}~\bibnamefont
  {McMahan}},\ }\href {https://doi.org/10.1103/PhysRevB.46.8440} {\bibfield
  {journal} {\bibinfo  {journal} {Phys. Rev. B}\ }\textbf {\bibinfo {volume}
  {46}},\ \bibinfo {pages} {8440–8455} (\bibinfo {year} {1992})}\BibitemShut
  {NoStop}%
\bibitem [{\citenamefont {Plakida}\ \emph {et~al.}(2003)\citenamefont
  {Plakida}, \citenamefont {Anton}, \citenamefont {Adam},\ and\ \citenamefont
  {Adam}}]{Plakida2003}%
  \BibitemOpen
  \bibfield  {author} {\bibinfo {author} {\bibfnamefont {N.}~\bibnamefont
  {Plakida}}, \bibinfo {author} {\bibfnamefont {L.}~\bibnamefont {Anton}},
  \bibinfo {author} {\bibfnamefont {S.}~\bibnamefont {Adam}},\ and\ \bibinfo
  {author} {\bibfnamefont {G.}~\bibnamefont {Adam}},\ }\href
  {https://doi.org/10.1134/1.1608998} {\bibfield  {journal} {\bibinfo
  {journal} {J. Exp. Theor. Phys.}\ }\textbf {\bibinfo {volume} {97}},\
  \bibinfo {pages} {331} (\bibinfo {year} {2003})}\BibitemShut {NoStop}%
\bibitem [{\citenamefont {Val’kov}\ and\ \citenamefont
  {Dzebisashvili}(2005)}]{Dzebisashvili05}%
  \BibitemOpen
  \bibfield  {author} {\bibinfo {author} {\bibfnamefont {V.}~\bibnamefont
  {Val’kov}}\ and\ \bibinfo {author} {\bibfnamefont {D.}~\bibnamefont
  {Dzebisashvili}},\ }\href {https://doi.org/10.1134/1.1901772} {\bibfield
  {journal} {\bibinfo  {journal} {J. Exp. Theor. Phys.}\ }\textbf {\bibinfo
  {volume} {100}},\ \bibinfo {pages} {608} (\bibinfo {year}
  {2005})}\BibitemShut {NoStop}%
\bibitem [{\citenamefont {Kuz'min}\ \emph {et~al.}(2020)\citenamefont
  {Kuz'min}, \citenamefont {Visotin}, \citenamefont {Nikolaev},\ and\
  \citenamefont {Ovchinnikov}}]{Kuzmin2020}%
  \BibitemOpen
  \bibfield  {author} {\bibinfo {author} {\bibfnamefont {V.}~\bibnamefont
  {Kuz'min}}, \bibinfo {author} {\bibfnamefont {M.}~\bibnamefont {Visotin}},
  \bibinfo {author} {\bibfnamefont {S.}~\bibnamefont {Nikolaev}},\ and\
  \bibinfo {author} {\bibfnamefont {S.}~\bibnamefont {Ovchinnikov}},\ }\href
  {https://doi.org/10.1103/PhysRevB.101.115141} {\bibfield  {journal} {\bibinfo
   {journal} {Phys. Rev. B}\ }\textbf {\bibinfo {volume} {101}},\ \bibinfo
  {pages} {115141} (\bibinfo {year} {2020})}\BibitemShut {NoStop}%
\bibitem [{\citenamefont {Shastry}(1989)}]{Shastry89}%
  \BibitemOpen
  \bibfield  {author} {\bibinfo {author} {\bibfnamefont {B.}~\bibnamefont
  {Shastry}},\ }\href {https://doi.org/10.1103/PhysRevLett.63.1288} {\bibfield
  {journal} {\bibinfo  {journal} {Phys. Rev. Lett.}\ }\textbf {\bibinfo
  {volume} {63}},\ \bibinfo {pages} {1288} (\bibinfo {year}
  {1989})}\BibitemShut {NoStop}%
\bibitem [{\citenamefont {Makarov}\ and\ \citenamefont
  {Ovchinnikov}(2022)}]{Makarov22}%
  \BibitemOpen
  \bibfield  {author} {\bibinfo {author} {\bibfnamefont {I.}~\bibnamefont
  {Makarov}}\ and\ \bibinfo {author} {\bibfnamefont {S.}~\bibnamefont
  {Ovchinnikov}},\ }\href {https://doi.org/10.1016/j.physleta.2022.128226}
  {\bibfield  {journal} {\bibinfo  {journal} {Phys. Lett. A}\ }\textbf
  {\bibinfo {volume} {444}},\ \bibinfo {pages} {128226} (\bibinfo {year}
  {2022})}\BibitemShut {NoStop}%
\bibitem [{\citenamefont {Korshunov}\ and\ \citenamefont
  {Ovchinnikov}(2007)}]{Korshunov07}%
  \BibitemOpen
  \bibfield  {author} {\bibinfo {author} {\bibfnamefont {M.}~\bibnamefont
  {Korshunov}}\ and\ \bibinfo {author} {\bibfnamefont {S.}~\bibnamefont
  {Ovchinnikov}},\ }\href {https://doi.org/10.1140/epjb/e2007-00179-2}
  {\bibfield  {journal} {\bibinfo  {journal} {Eur. Phys. J. B}\ }\textbf
  {\bibinfo {volume} {57}},\ \bibinfo {pages} {271} (\bibinfo {year}
  {2007})}\BibitemShut {NoStop}%
\bibitem [{\citenamefont {Gavrichkov}\ and\ \citenamefont
  {Ovchinnikov}(2008)}]{Gavrichkov2008}%
  \BibitemOpen
  \bibfield  {author} {\bibinfo {author} {\bibfnamefont {V.}~\bibnamefont
  {Gavrichkov}}\ and\ \bibinfo {author} {\bibfnamefont {S.}~\bibnamefont
  {Ovchinnikov}},\ }\href@noop {} {\bibfield  {journal} {\bibinfo  {journal}
  {Fiz. Tverd. Tela}\ }\textbf {\bibinfo {volume} {50}},\ \bibinfo {pages}
  {1037} (\bibinfo {year} {2008})}\BibitemShut {NoStop}%
\bibitem [{\citenamefont {Chao}\ \emph {et~al.}(1977)\citenamefont {Chao},
  \citenamefont {Spa{\l}ek},\ and\ \citenamefont {Ole{\'s}}}]{Chao77}%
  \BibitemOpen
  \bibfield  {author} {\bibinfo {author} {\bibfnamefont {K.}~\bibnamefont
  {Chao}}, \bibinfo {author} {\bibfnamefont {J.}~\bibnamefont {Spa{\l}ek}},\
  and\ \bibinfo {author} {\bibfnamefont {A.}~\bibnamefont {Ole{\'s}}},\ }\href
  {https://doi.org/10.1088/0022-3719/10/10/002} {\bibfield  {journal} {\bibinfo
   {journal} {J. Phys. C: Solid State Phys.}\ }\textbf {\bibinfo {volume}
  {10}},\ \bibinfo {pages} {L271} (\bibinfo {year} {1977})}\BibitemShut
  {NoStop}%
\bibitem [{\citenamefont {Harlingen}(1995)}]{VanHarlingen95}%
  \BibitemOpen
  \bibfield  {author} {\bibinfo {author} {\bibfnamefont {D.~V.}\ \bibnamefont
  {Harlingen}},\ }\href {https://doi.org/10.1103/RevModPhys.67.515} {\bibfield
  {journal} {\bibinfo  {journal} {Rev. Mod. Phys.}\ }\textbf {\bibinfo {volume}
  {67}},\ \bibinfo {pages} {515} (\bibinfo {year} {1995})}\BibitemShut
  {NoStop}%
\bibitem [{\citenamefont {Tsuei}\ and\ \citenamefont
  {Kirtley}(2000)}]{Tsuei00}%
  \BibitemOpen
  \bibfield  {author} {\bibinfo {author} {\bibfnamefont {C.}~\bibnamefont
  {Tsuei}}\ and\ \bibinfo {author} {\bibfnamefont {J.}~\bibnamefont
  {Kirtley}},\ }\href {https://doi.org/10.1103/RevModPhys.72.969} {\bibfield
  {journal} {\bibinfo  {journal} {Rev. Mod. Phys.}\ }\textbf {\bibinfo {volume}
  {72}},\ \bibinfo {pages} {969} (\bibinfo {year} {2000})}\BibitemShut
  {NoStop}%
\bibitem [{\citenamefont {Damascelli}\ \emph {et~al.}(2003)\citenamefont
  {Damascelli}, \citenamefont {Hussain},\ and\ \citenamefont
  {Shen}}]{Damascelli2003}%
  \BibitemOpen
  \bibfield  {author} {\bibinfo {author} {\bibfnamefont {A.}~\bibnamefont
  {Damascelli}}, \bibinfo {author} {\bibfnamefont {Z.}~\bibnamefont
  {Hussain}},\ and\ \bibinfo {author} {\bibfnamefont {Z.-X.}\ \bibnamefont
  {Shen}},\ }\href {https://doi.org/10.1103/RevModPhys.75.473} {\bibfield
  {journal} {\bibinfo  {journal} {Rev. Mod. Phys.}\ }\textbf {\bibinfo {volume}
  {75}},\ \bibinfo {pages} {473} (\bibinfo {year} {2003})}\BibitemShut
  {NoStop}%
\bibitem [{\citenamefont {Zaitsev}\ and\ \citenamefont
  {Ivanov}(1987)}]{Zaitsev87}%
  \BibitemOpen
  \bibfield  {author} {\bibinfo {author} {\bibfnamefont {R.}~\bibnamefont
  {Zaitsev}}\ and\ \bibinfo {author} {\bibfnamefont {V.}~\bibnamefont
  {Ivanov}},\ }\href@noop {} {\bibfield  {journal} {\bibinfo  {journal} {Fizika
  Tverdogo Tela}\ }\textbf {\bibinfo {volume} {29}},\ \bibinfo {pages} {2554}
  (\bibinfo {year} {1987})}\BibitemShut {NoStop}%
\bibitem [{\citenamefont {Zaitsev}\ and\ \citenamefont
  {Ivanov}(1988)}]{Zaitsev88}%
  \BibitemOpen
  \bibfield  {author} {\bibinfo {author} {\bibfnamefont {R.}~\bibnamefont
  {Zaitsev}}\ and\ \bibinfo {author} {\bibfnamefont {V.}~\bibnamefont
  {Ivanov}},\ }\href {https://doi.org/10.1016/0921-4534(88)90288-2} {\bibfield
  {journal} {\bibinfo  {journal} {Physica C}\ }\textbf {\bibinfo {volume}
  {153-153}},\ \bibinfo {pages} {1295} (\bibinfo {year} {1988})}\BibitemShut
  {NoStop}%
\bibitem [{\citenamefont {Zaitsev}(1990)}]{Zaitsev88_2}%
  \BibitemOpen
  \bibfield  {author} {\bibinfo {author} {\bibfnamefont {R.}~\bibnamefont
  {Zaitsev}},\ }\href {https://doi.org/10.1016/0375-9601(88)90821-3} {\bibfield
   {journal} {\bibinfo  {journal} {Phys. Lett. A}\ }\textbf {\bibinfo {volume}
  {134}},\ \bibinfo {pages} {199} (\bibinfo {year} {1990})}\BibitemShut
  {NoStop}%
\end{thebibliography}%
\end{document}